\documentclass[twocolumn,twocolappendix]{aastex63}

\usepackage{ragged2e}
\usepackage[]{natbib}
\setcitestyle{comma}
\usepackage{amsmath}
\usepackage[nameinlink]{cleveref}

\usepackage[utf8]{inputenc}
\usepackage[T1]{fontenc}

\newcommand{\Eq}[1]{Equation\,(\ref{#1})}

\newcommand{\Sec}[1]{Section~\ref{#1}}

\newcommand{\Fig}[1]{Figure~\ref{#1}}

\newcommand{\Tab}[1]{Table \ref{#1}}

\def\K{\; {\rm K}}

\shorttitle{Patchy clouds on ultra-hot Jupiters}
\shortauthors{T. Komacek \& X. Tan, P. Gao \& E. Lee}
\graphicspath{{./}{figs/}}

\begin{document}

\title{Patchy nightside clouds on ultra-hot Jupiters: General Circulation Model simulations with radiatively active cloud tracers}

\correspondingauthor{Thaddeus D. Komacek}
\email{tkomacek@umd.edu}
\author[0000-0002-9258-5311]{Thaddeus D. Komacek}
\altaffiliation{T.D.K. and X.T. contributed equally to this work.}
\affiliation{Department of Astronomy, University of Maryland, College Park, MD 20742, USA}
\author[0000-0003-2278-6932]{Xianyu Tan}
\altaffiliation{T.D.K. and X.T. contributed equally to this work.}
\affiliation{Atmospheric, Oceanic and Planetary Physics, Department of Physics, University of Oxford, OX1 3PU, UK}
\author[0000-0002-8518-9601]{Peter Gao}
\altaffiliation{P.G. and E.K.H.L. contributed equally to this work.}
\affiliation{Earth and Planets Laboratory, Carnegie Institution for Science, 5241 Broad Branch Road, NW, Washington, DC 20015, USA}
\author[0000-0002-3052-7116]{Elspeth K.H. Lee}
\altaffiliation{P.G. and E.K.H.L. contributed equally to this work.}
\affiliation{Center for Space and Habitability, University of Bern, Gesellschaftsstrasse 6, CH-3012 Bern, Switzerland}




\begin{abstract}
The atmospheres of ultra-hot Jupiters have been characterized in detail through recent phase curve and low- and high-resolution emission and transmission spectroscopic observations. 
Previous numerical studies have analyzed the effect of the localized recombination of hydrogen on the atmospheric dynamics and heat transport of ultra-hot Jupiters, finding that hydrogen dissociation and recombination lead to a reduction in the day-to-night contrasts of ultra-hot Jupiters relative to previous expectations. In this work, we add to previous efforts by also considering the localized condensation of clouds in the atmospheres of ultra-hot Jupiters, their resulting transport by the atmospheric circulation, and the radiative feedback of clouds on the atmospheric dynamics. To do so, we include radiatively active cloud tracers into the existing \texttt{MITgcm} framework for simulating the atmospheric dynamics of ultra-hot Jupiters. We take cloud condensate properties appropriate for the high-temperature condensate corundum from \texttt{CARMA} cloud microphysics models. We conduct a suite of GCM simulations with varying cloud microphysical and radiative properties, and we find that partial cloud coverage is a ubiquitous outcome of our simulations. This patchy cloud distribution is inherently set by atmospheric dynamics in addition to equilibrium cloud condensation, and causes a cloud greenhouse effect that warms the atmosphere below the cloud deck. Nightside clouds are further sequestered at depth due to a dynamically induced high-altitude thermal inversion. We post-process our GCMs with the Monte Carlo radiative transfer code \texttt{gCMCRT} and find that the patchy clouds on ultra-hot Jupiters do not significantly impact transmission spectra but can affect their phase-dependent emission spectra. 
\end{abstract}



\section{Introduction} 
\label{sec:intro}
\indent Ultra-hot Jupiters are a novel class of substellar object with atmospheres that lie in a regime between the cooler hot Jupiters and those of late-type stars. These gaseous exoplanets orbit extremely close-in to their host star, with zero albedo full-redistribution equilibrium temperatures in excess of $2200~\mathrm{K}$. The enormous incident stellar flux that ultra-hot Jupiters receive along with their likely tidally synchronized rotational state cause a large day-to-night temperature contrast akin to cooler hot Jupiters, which in turn is predicted to generate planetary-scale waves that drive an eastward equatorial jet (\citealp{Showman_Polvani_2011}, see recent reviews of the atmospheric circulation of extrasolar gas giant planets by \citealp{Showman:2020rev,Zhang:2020rev,Fortney:2021aa}). As a result, the present theoretical understanding of ultra-hot Jupiters relies upon extensions of numerical general circulation models (GCMs) developed to understand their cooler cousins (e.g., \citealp{Parmentier:2018aa,Tan:2019aa,May:2021ab,Beltz:2021aa}). However, a variety of processes that are either inactive or weak in the atmospheres of standard hot Jupiters are expected to affect the atmospheric structure and dynamics of ultra-hot Jupiters, necessitating a coupled framework to develop further understanding. 

The most prominent difference between the atmospheres of hot and ultra-hot Jupiters is expected to be the thermal dissociation of molecular species \citep{Parmentier:2018aa,Lothringer:2018aa,Kitzmann:2018aa}, for which there is observational evidence in a wide range of low-resolution transmission spectra, emission spectra, and phase curves to date \citep{Stevenson:2014tx,Haynes:2015,Beatty:2017aa,Evans:2017aa,Sheppard:2017,Arcangeli:2018aa,Kreidberg:2018aa,Mansfield:2018aa,Baxter:2020aa,Gandhi:2020ws,Mansfield:2020aa,Mikal-Evans:2020aa,Wilson:2020uc,Wong:2019aa,Fu:2021aa,Mansfield:2021tz,Changeat:2021vl,Mikal-Evans:2022wq}. Thermal dissociation also affects the primary atmospheric constituent of ultra-hot Jupiters, molecular hydrogen, causing dissociation of molecular hydrogen to atomic form on the hot dayside and recombination of molecular hydrogen on the cooler limbs and nightside \citep{Bell:2018aa,Komacek:2018aa,Tan:2019aa,Gandhi:2020aa,Roth:2021un}. The thermodynamic impact of hydrogen dissociation and recombination shapes the atmospheric structure and dynamics of ultra-hot Jupiters, reducing day-night temperature contrasts and affecting the planetary scale standing wave pattern of ultra-hot Jupiters, leading to a reduction in the speed of the equatorial jet \citep{Bell:2018aa,Tan:2019aa}. 


A variety of other processes along with molecular dissociation conspire to set the atmospheric structure and dynamics of ultra-hot Jupiters. This includes absorption of incident stellar radiation by atomic metals in the planetary atmosphere that can lead to ``inverted'' temperature-pressure profiles which increase in temperature with decreasing pressure \citep{Fortney:2008,Lothringer:2018aa,Kitzmann:2018aa,Gandhi:2019ve,Malik:2019vr} and can be especially strong for ultra-hot Jupiters that orbit early-type stars \citep{Lothringer:2019uo,Fu:2022ud}. Recent high spectral resolution observations of ultra-hot Jupiters have found a wealth of metallic species along with evidence for thermal inversions \citep{Nugroho:2017aa,Hoeijmakers:2018aa,Jensen:2018aa,Seidel:2019aa,Cabot:2020aa,Ehrenreich:2020aa,Hoeijmakers:2020aa,Nugroho:2020vx,Pino:2020aa,Yan:2020ui,Kasper:2021aa,Kesseli:2021ab,Prinoth:2021aa,Tabernero:2020ab,Yan:2022uz}. The hot daysides of ultra-hot Jupiters should be sufficiently ionized that magnetohydrodynamic mechanisms can affect their atmospheric circulation \citep{Perna_2010_1,Menou:2012fu,batygin_2013,Rauscher_2013,Rogers:2020,Rogers:2014,Hindle:2019aa,Beltz:2022aa}, potentially causing large-amplitude time-variability due to induced atmospheric magnetic fields \citep{Rogers:2017,Rogers:2017a,Hindle:2021ti,Hindle:2021wm}. Additionally, many ultra-hot Jupiters are highly inflated, implying a significant internal heat flux \citep{Thorngren:2018,Thorngren:2019aa,Sarkis:2021aa} that can affect their deep atmospheric circulation and mixing \citep{Tremblin:2017,Sainsbury-Martinez:2019aa,Carone:2019aa,Baeyens:2021uo}. 


Along with molecular dissociation, atomic metal absorption, thermal ionization and magnetohydrodynamic effects, and internal heating, one other process likely acts to shape the emergent properties of ultra-hot Jupiters: aerosol coverage. Unlike the other processes at work in ultra-hot Jupiter atmospheres, aerosols and their radiative feedback on the circulation likely has a more minor effect on the atmospheric dynamics of ultra-hot Jupiters than for hot Jupiters due to the high temperatures preventing condensation of many mineral species (for a comprehensive recent review of exoplanet aerosols, see \citealp{Gao:2021tq}). Given the high temperatures on the daysides of both hot and ultra-hot Jupiters and large horizontal temperature contrasts, it is expected that their aerosol coverage is non-uniform, with enhanced condensate cloud coverage on the cooler western limb and nightside and with haze distributions dependent on particle size \citep{Helling:2016,Parmentier16,Kempton:2017aa,Wakeford:2017,Mendonca:2018,Powell:2018aa,Powell:2019aa,Gao:2020aa,Adams:2021ud,Helling:2021aa,Parmentier:2021tt,Roman:2021wl,Steinrueck:2021aa,Robbins-Blanch:2022ue}. There is observational evidence of such non-uniform aerosol distributions from the reflected light signature in Kepler phase curves \citep{Demory_2013,Esteves:2015,Hu:2015,Schwartz:2015,Parmentier16}, and both low and high resolution transmission spectra have been suggestive of non-uniform aerosol coverage that changes with local atmospheric temperature \citep{Line2016,Sing:2015a,Ehrenreich:2020aa}. 

Though hot Jupiters have non-uniform dayside aerosol distributions, they are expected to have a thick nightside condensate cloud deck that acts to reduce their outgoing longwave radiation, producing the weak observed trend in nightside infrared brightness temperature with increasing equilibrium temperature \citep{Beatty19,Keating:2019aa,Bell:2021aa}. This ``flat nightside temperature'' trend may be analogous to the ``fixed anvil temperature'' hypothesis proposed to explain the independence of outgoing longwave radiation on the global-mean temperature of our warming Earth \citep{Hartmann:2002tj,Kuang:2007uu,Zelinka:2010uo}, as both rely on an increase in the cloud top altitude with increasing planetary-mean temperature to mute the changes in outgoing longwave radiation \citep{Gao:2021vp}. The observed flat nightside trend breaks and abruptly steepens in the ultra-hot Jupiter regime, potentially due to the dissipation of the uniform nightside cloud deck \citep{Parmentier:2021tt,Roman:2021wl} or changes in the heat transport properties of ultra-hot Jupiters, which may be analogous to the more complex interplay between clouds, circulation, and climate found in modern cloud resolving models of Earth  \citep{Seeley:2019um,Wing:2020ul} than expected in the framework of a fixed anvil cloud temperature. 

The observational evidence for non-uniform aerosol distributions in hot and ultra-hot Jupiter atmospheres necessitates three-dimensional models of their coupled atmospheric circulation and aerosol distribution. A range of models have been developed that incorporate the radiative feedback of aerosols on the atmospheric circulation of hot Jupiters \citep{Lee:2016,Lines:2018,lines:2019,Roman:2019aa,Parmentier:2021tt,Roman:2021wl,Christie:2021tu}. Models that include cloud-radiative feedback are especially critical for understanding the behavior of patchy clouds on atmospheric circulation. This is because patchy clouds lead to spatially inhomogeneous cloud radiative forcing, with a cloud greenhouse effect caused by the weak outgoing longwave radiation at cloud tops and enhanced cooling in cloud-free regions. 

A substellar regime in which patchy cloud formation, cloud-radiative feedback, and vertical mixing of both clouds and chemical species have been studied in detail is that of brown dwarfs and directly imaged giant planets \citep{Ackerman2001,Freytag:2010,Morley2012,Showman:2013,Morley:2014wr,Bordwell:2018tz,Tan:2019tj,Tan:2021wv,Tan:2021vx,Tremblin:2021ub}. Notably, \cite{Tan:2021wv,Tan:2021vx} studied the effect of cloud-radiative feedback on the atmospheric dynamics of brown dwarfs, finding that cloud-radiative feedback can be a key driver of the atmospheric dynamics and inhomogeneous cloud structures as expected from analytic  theory of cloud-radiative instability \citep{Gierasch:1973aa}. Models of the global circulation of brown dwarfs \citep{Showman:2019us,Tan:2021vx,Tan:2022bd} predict significant variability due to a combination of gravity waves generated by interaction with the convective interior and cloud-radiative feedback. Such variability has been observed in a range of brown dwarfs to date (e.g., \citealp{Gelino:2002aa,Artigau:2009tx,Radigan:2012we,Biller:2013tq,Crossfield:2014tm,Faherty:2014td,Karalidi:2016vj,Lew:2016up,Apai:2017to,Allers:2020aa,Vos:2022aa}), providing evidence that cloud patchiness and large-scale wave motions induce the observed variability. Additionally, the observational characterization of highly irradiated brown dwarfs orbiting white dwarfs (e.g., \citealp{Casewell:2018tf,Casewell:2020vk,Lew:2022ur,Zhou:2022to}) has recently motivated numerical models of their atmospheric circulation \citep{Lee:2020vv,Tan:2020va,Sainsbury-Martinez:2021us}. Their atmospheric dynamics are expected to be unique given their placement in parameter space as hot and high-gravity objects with large internal heat fluxes, which motivates further understanding as a population analogous to ultra-hot Jupiters \citep{Showman:2020rev}.

Nascent studies of the effect of patchy clouds on the observable properties of ultra-hot Jupiters have leveraged improved instrumental capabilities to probe planetary atmospheres with high temporal and spatial resolution and at short wavelengths. Recent near-ultraviolet (NUV) spectra of ultra-hot Jupiters have been studied to probe the temperature at which clouds condense at the terminator of hot Jupiters \citep{Lothringer:2020aa}, with the hotter WASP-121b and WASP-178b showing evidence of a cloud-free limb \citep{Sing:2019tq,Lothringer:2022aa} but spectra of the cooler HAT-P-41b \citep{Lewis:2020ue,Wakeford:2020tu} signaling the presence of clouds. Contemporaneously, time-resolved high spectral resolution transit observations have found evidence for non-uniform absorption during the transits of WASP-76b \citep{Ehrenreich:2020aa,Kesseli:2021ab,Kesseli:2021uk,Seidel:2021tm} and WASP-121b \citep{Borsa:2021wr}. Due to the influence of atmospheric climate dynamics on the shape and wavelength of spectral lines \citep{Kempton:2012vk,showman_2013_doppler,Kempton:2014,Zhang:2017b,Seidel:2020vi}, high spectral resolution observations require three-dimensional models to fully extract the information embedded in the spectrum about the planetary atmosphere \citep{Flowers:2018aa,Beltz:2021aa}. Three-dimensional climate dynamics and radiative transfer modeling has recently been applied to study the observed time-resolved high-resolution transmission spectrum of WASP-76b \citep{Wardenier:2021td,Savel:2021aa}, and demonstrate the influence of the atmospheric temperature, winds, and aerosol coverage on the resulting observable spectrum. 


Recent TESS discoveries and Spitzer characterization have enabled the study of hot and ultra-hot Jupiters as a population (see Figure 17 of \citealp{Wong:2021td}). These discoveries include TOI-1431b, which is a young ($\approx 0.3$ Ga) ultra-hot Jupiter with an equilibrium temperature of $\approx 2370~\mathrm{K}$ that was discovered and characterized by \cite{Addison:2021aa} and \cite{Stangret:2021aa}. TOI-1431b lies in the regime just hotter than the transition point from hot to ultra-hot Jupiters, and similar to HAT-P-7b \citep{Bell:2021aa} it has a surprisingly small day-to-night temperature contrast due to an observed high nightside brightness temperature in the TESS bandpass. As a result, TOI-1431b is an ideal case study of the impact of the confluence of molecular dissociation, cloud coverage, and youth on the atmospheric circulation of ultra-hot Jupiters. In this work, we leverage TOI-1431b as a test case to study the combination of the thermodynamic effect of molecular dissociation and recombination and cloud-radiative feedback on the atmospheric circulation, cloud transport, and observable properties of ultra-hot Jupiters. To do so, we add radiatively active condensate cloud tracers to the existing ultra-hot Jupiter \texttt{MITgcm} with thermodynamically active atomic hydrogen tracers \citep{Tan:2019aa}. We utilize \texttt{CARMA} cloud microphysics simulations \citep{Gao:2020aa,Gao:2021vp} to determine microphysical properties relevant for both the mixing of cloud tracers and their radiative feedback on the atmosphere. We then apply the \texttt{gCMCRT} Monte Carlo radiative transfer code \citep{Lee:2021uv} to post-process our GCM simulations and make predictions for the impact of high-temperature cloud condensates on the observable properties of ultra-hot Jupiters. 


The outline of this work is as follows. Section \ref{sec:setup} describes the setup of the GCM simulations conducted in this work, including evolutionary calculations to determine the internal heat flux, implementation of radiatively active cloud tracers, and the GCM parameter sweep that we carry out. We describe the resulting atmospheric dynamics and cloud coverage from our GCMs in \Sec{sec:results} along with the effects of clouds on emergent spectra and phase curves of ultra-hot Jupiters. We discuss the potential implications of our model results for the understanding of the effect of clouds on observations of ultra-hot Jupiters in \Sec{sec:disc}, and describe present limitations of our model along with possible improvements to our current modeling framework. 
Lastly, we denote conclusions in \Sec{sec:conc}.

\section{Model Setup}
\label{sec:setup}
\subsection{\texttt{MESA} evolutionary models}
\label{sec:mesa}
In order to determine the temperature structure near the bottom of our GCM domain, we conducted a suite of planetary evolution calculations with the \texttt{MESA} stellar and planetary evolution code \citep{Paxton:2011,Paxton:2013,Paxton:2015,Paxton:2018aa,Paxton:2019aa}. These planetary evolution calculations have an setup similar to that in \cite{Komacek:2017a} and \cite{Komacek:2020ab}, and solve the stellar structure equations \citep{Chandra39,Kippenhahn:2012}, including mass conservation, hydrostatic equilibrium, energy conservation, and energy transport. The key modifications to the stellar structure equations in our model framework are the inclusion of an external irradiation by adding an energy generation rate in a specified outer column mass of the atmosphere, and the addition of an extra heating term in the interior spread as a Gaussian with a standard deviation of one-half a pressure scale height centered at a given pressure (see also \citealp{Wu:2013,Millholland:2019aa,Lous:2020aa,Glanz:2021aa}). These models do not include a central heavy element core, and as a result prescribe upper limits on planetary radius \citep{bode01,Guillot_2002,Thorngren16} for a given planetary mass, irradiation, and amount of deposited heating in the interior. 

\begin{figure*}
    \centering
    \includegraphics[width=0.32\textwidth]{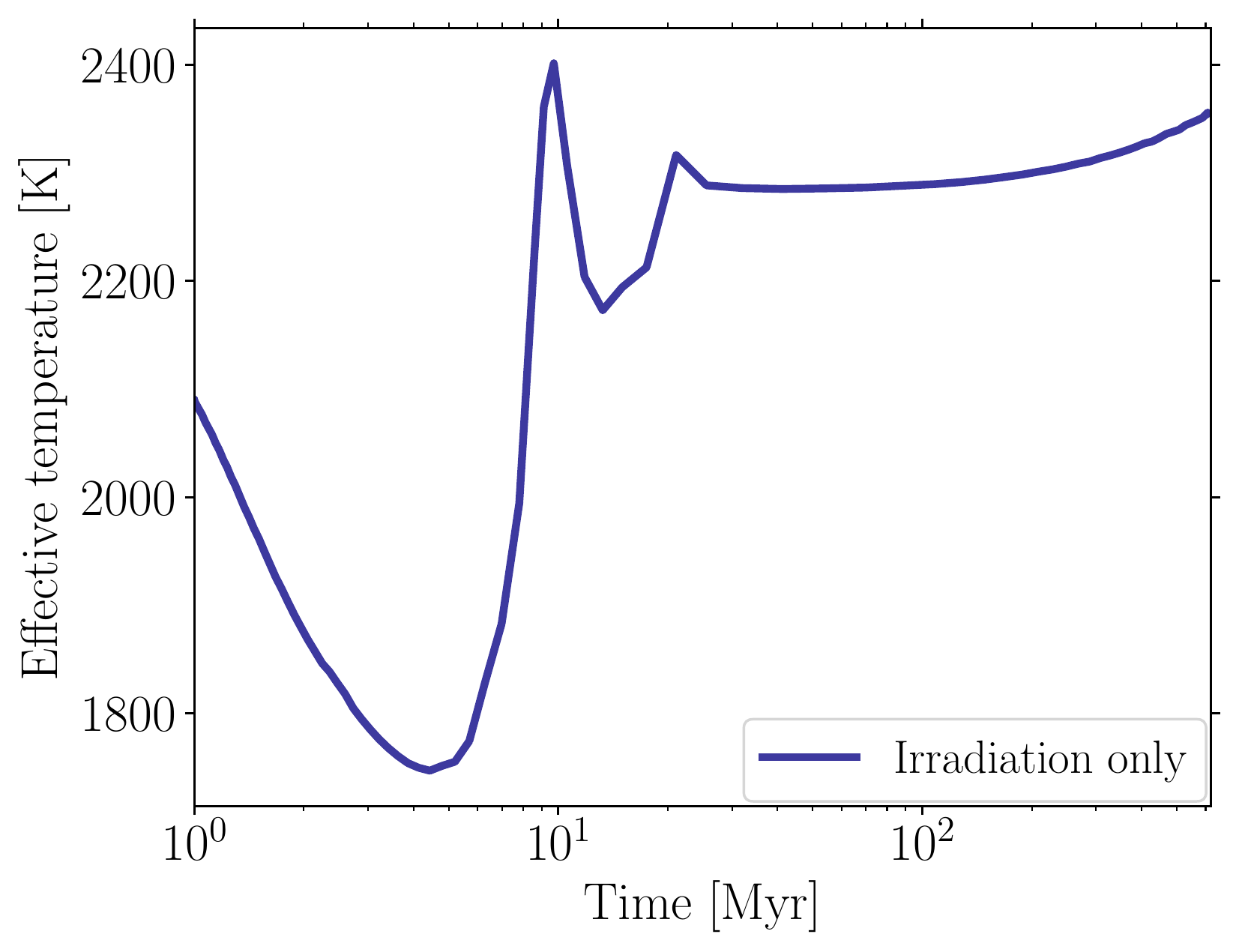}
    \includegraphics[width=0.32\textwidth]{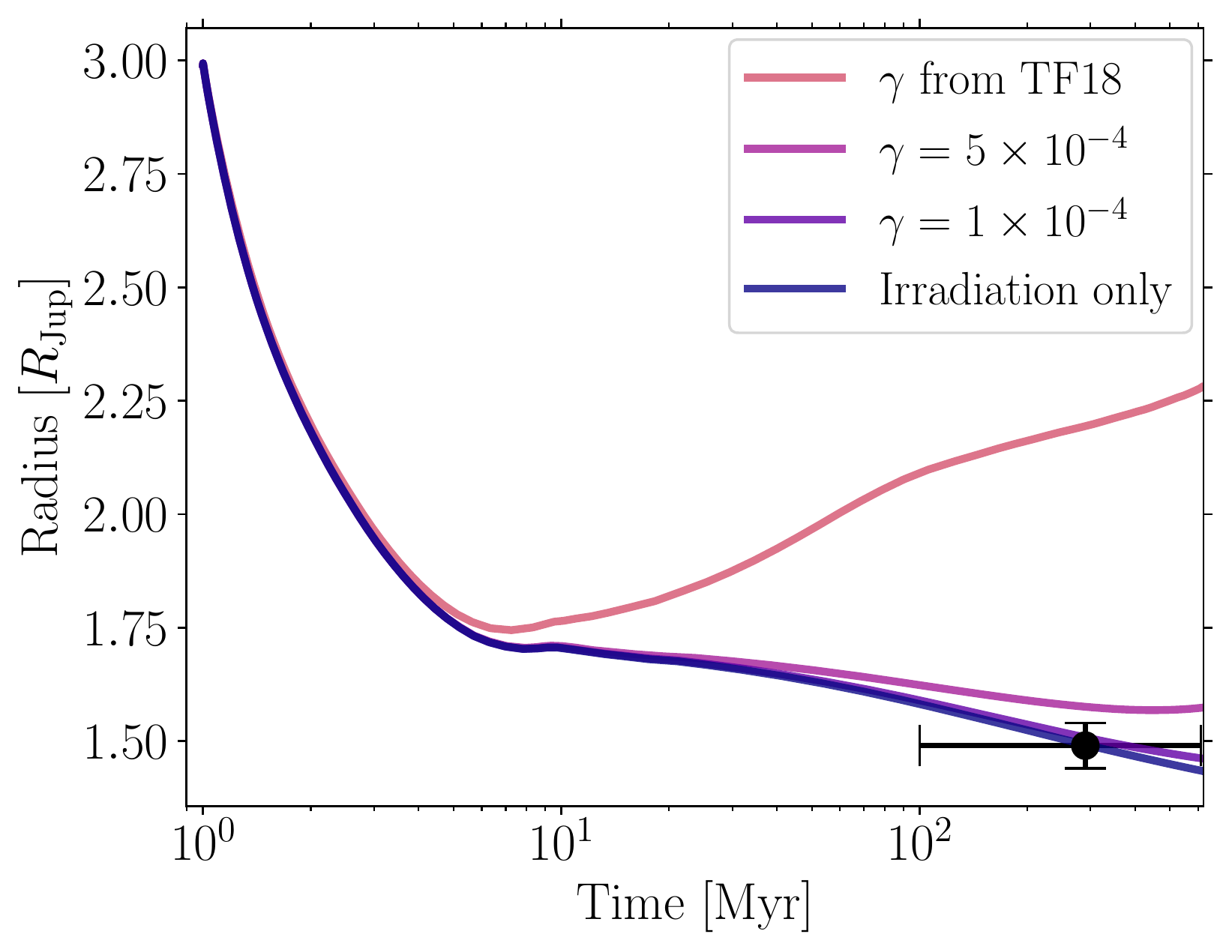}
    \includegraphics[width=0.32\textwidth]{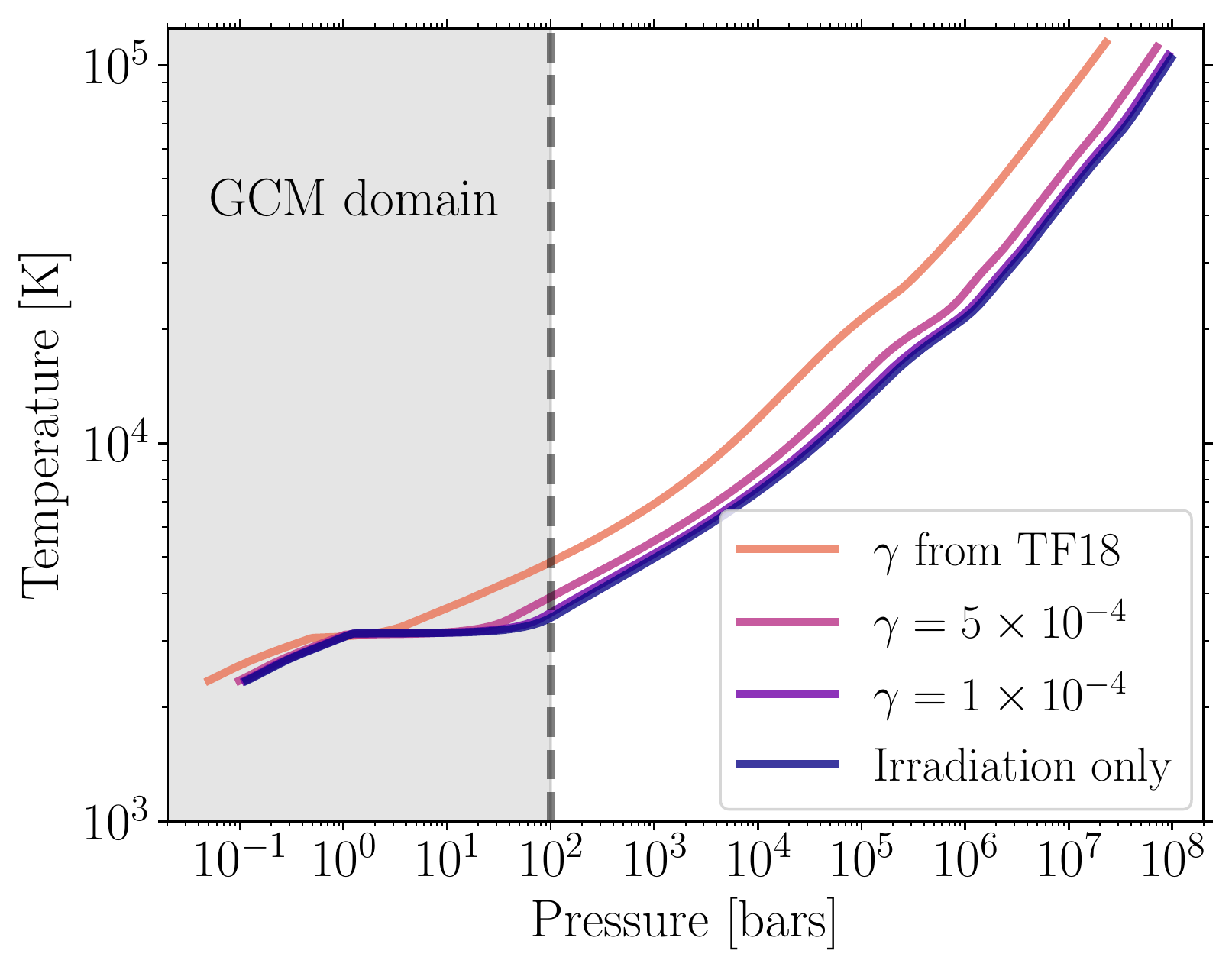}
    \caption{Evolution and final internal temperature structure for TOI-1431b models undergoing irradiation and varying levels of deposited heating. The left-hand panel shows the time-evolution of effective temperature, the central panel the radius evolution, and the right-hand panel the temperature-pressure profile for varying assumptions of the fraction of incident stellar flux converted to deposited heat, $\gamma$. The lightest colored line shows the case with a $\gamma$ dependent on the incident stellar flux as derived from the full sample of hot Jupiters \citep{Thorngren:2018}. The observed radius and age of TOI-1431b \citep{Addison:2021aa} is over-plotted on the radius evolution track. Models with weak or no deposited heating are consistent with the present-day radius of TOI-1431b.}
    \label{fig:evolution}
\end{figure*}

As in \cite{Komacek:2020ab}, we include the time-varying irradiation due to stellar evolution in our planetary evolution calculations. We take the time-dependent stellar luminosity of TOI-1431 from MIST \citep{Choi:2016aa,Dotter:2016aa} models with an initial mass of $1.895~M_\odot$, metallicity of $[\mathrm{Fe}/\mathrm{H}] = 0.43$, and a rotation rate of $v/v_{\mathrm{crit}} = 0.4$. We apply irradiation in an outer mass column of $300~\mathrm{g}~\mathrm{cm}^{-2}$, taking the semi-major axis of TOI-1431b to be $0.0471~\mathrm{au}$. We assume full heat redistribution, allowing our planetary evolution calculations to represent the global-mean energy budget. The left-hand panel of \Fig{fig:evolution} displays the evolution of the effective temperature of TOI-1431b, showing the impacts of the early evolution of TOI-1431 on the irradiation received by TOI-1431b. Note that we show results from each \texttt{MESA} model to an age of $0.61~\mathrm{Gyr}$, consistent with the $1\sigma$ upper limit on the system age \citep{Addison:2021aa}. 

We include a fraction of the irradiation power, $\gamma$ as deposited heating in the deep interior of the planet. As a result, the total amount of heat deposited in the interior of the planet, $\Gamma$, depends on $\gamma$ and the incident stellar power as
\begin{equation}
    \Gamma = \gamma L_\mathrm{irr} = \pi \gamma R_p^2 F_\star \mathrm{,}
\end{equation}
where $L_\mathrm{irr}$ is the incident stellar power, $R_p$ is the planetary radius, and $F_\star$ is the incident stellar flux. Note that we include the positive feedback between planetary radius and deposited heat due to the increasing cross-sectional area of the planet \citep{Batygin_2011}. We conduct a suite of models with varying $\gamma$ from 0 to $5 \times 10^{-4}$, along with a simulation with time-varying $\gamma$ following the dependence of internal heating on incident stellar flux derived from the full suite of hot Jupiters (Equation 34 of \citealp{Thorngren:2018}). For all cases, we assume that the maximum of heat deposition occurs at the very center of the planet, in line with our motivation described above to set upper limits on the predicted planetary radius for a given heating rate in our model. We assume a planet mass of $3.12~M_\mathrm{Jup}$ \citep{Addison:2021aa} for all cases, and we vary the initial planet radius from $2.5 - 4~R_\mathrm{Jup}$ given the unknown initial entropy post-formation \citep{Marleau:2014,Berardo:2017aa}. We find that the assumed initial radius does not affect the present-day properties of TOI-1431b due to its short ($\lesssim 10~\mathrm{Myr}$) Kelvin-Helmoltz contraction timescale \citep{Ginzburg:2015,Lous:2020aa} -- as a result, we only display results from the case with an intermediate initial radius of $3~R_\mathrm{Jup}$. 

\Fig{fig:evolution} shows the radius evolution and final temperature structures from evolutionary calculations of TOI-1431b with varied assumptions for the deposited heat in the interior. As in \cite{Addison:2021aa}, we find that TOI-1431b is under-inflated for its age, with a present-day radius that can be fit either with only irradiation acting to slow cooling, a weak amount of incident stellar flux converted to heat in the deep interior, or if the planet did not form at its current location, but migrated inward \citep{Lous:2020aa}. The efficiency of conversion of incident stellar power to deposited heat ($\gamma$) expected from the suite of hot Jupiters over-predicts the present-day radius, either implying that the mechanism that inflates the majority of hot Jupiters is weaker than expected in TOI-1431b or that it does not apply over the full evolution of TOI-1431b. TOI-1431b has a deep radiative envelope in models with weak or zero deposited internal heating, with radiative-convective boundaries between $104.4 - 43.5~\mathrm{bars}$ with increasing $\gamma$ from $0 - 5 \times 10^{-4}$. We further constrain the 100-bar temperature at the end of our evolutionary calculations to lie between $3459 - 3894~\mathrm{K}$ with varying $\gamma$ from $0 - 5 \times 10^{-4}$. Given the uncertainty in system age and heating rate, we take an intermediate value for the 100 bar temperature of $3750~\mathrm{K}$ to prescribe the bottom boundary condition in our GCM simulations, as described below. 
\subsection{\texttt{MITgcm} simulations}
\subsubsection{Double-gray GCM framework with radiatively active cloud tracers}
For the GCM simulations conducted in this work, we solve the three-dimensional primitive equations of meteorology with the \texttt{MITgcm} \citep{Adcroft:2004} including the thermodynamic impact of hydrogen dissociation and recombination \citep{Tan:2019aa,Mansfield:2020aa,May:2021ab}. These include the equations of horizontal momentum, local hydrostatic equilibrium, mass conservation, energy conservation, and the ideal gas law as an equation of state, as follows:
\begin{equation}
\label{eq:mom}
    \frac{d{\bf v}}{dt} = - f \hat{\bf k} \times {\bf v} - \nabla_p \Phi + \mathcal{F}_\mathrm{drag} + \mathcal{D}_\mathrm{S} \mathrm{,}
\end{equation}
\begin{equation}
    \frac{\partial \Phi}{\partial p} = -\frac{1}{\rho} \mathrm{,}
\end{equation}
\begin{equation}
\label{eq:masscont}
    \nabla_p \cdot {\bf v} + \frac{\partial \omega}{\partial p} = 0 \mathrm{,}
\end{equation}
\begin{equation}
\label{eq:thermo}
    \frac{d\theta'}{dt} = \frac{\theta'}{\bar{c}_pT}\left(g\frac{\partial F}{\partial p} + \frac{L_H \delta q_H}{\tau_\mathrm{relax}}\right) +  \mathcal{H}_\mathrm{drag} + \mathcal{E}_\mathrm{S} \mathrm{,}
\end{equation}
\begin{equation}
\label{eq:state}
    p = \rho \bar{R} T \mathrm{.}
\end{equation}

Symbols used in Equations (\ref{eq:mom}) - (\ref{eq:state}) above include the pressure $p$, the horizontal velocity vector on isobars ${\bf v}$, the vertical velocity in pressure coordinates $\omega = dp/dt$, the horizontal gradient on an isobar $\nabla_p$, the total (material) derivative $d/dt = \partial/\partial t + {\bf v} \cdot \nabla_p + \omega \partial/\partial p$, the Coriolis parameter $f = 2 \Omega~\mathrm{sin}\phi$, where $\Omega$ is planetary rotation rate and $\phi$ is latitude, the local vertical unit vector $\hat{\bf k}$, the geopotential $\Phi = gz$, where $g$ is gravity and $z$ is altitude, the mean gas density $\rho$, the mass-weighted specific gas constant and heat capacity $\bar{R}$ and $\bar{c}_p$, the temperature $T$, the modified potential temperature $\theta' = T \left(p_0/p\right)^{\bar{R}/\bar{c}_p}$ with a reference pressure $p_0$, the net radiative flux $F$, the recombination energy of hydrogen $L_H$, the mass mixing ratio of atomic hydrogen relative to the total air mass $q_H$, and the relaxation timescale of atomic hydrogen $\tau_\mathrm{relax}$. The change in atomic hydrogen mass mixing ratio due to H-H$_2$ conversion $\delta q_H$ is detailed in \Eq{eq:deltaqh}. Note that model parameter choices are shown in Table \ref{table:params}. 

The $\mathcal{F}_\mathrm{drag}$ term in \Eq{eq:mom} corresponds to a frictional Rayleigh drag applied throughout the atmosphere as 
\begin{equation}
    \mathcal{F}_\mathrm{drag} = -\frac{{\bf v}}{\tau_\mathrm{drag}} \mathrm{,}
\end{equation}
where $\tau_\mathrm{drag}$ is a height-independent drag timescale. The kinetic energy dissipated via frictional drag is converted back into thermal energy as indicated by the $\mathcal{H}_\mathrm{drag}$ term in \Eq{eq:thermo}. The $\mathcal{D}_\mathrm{S}$ term in \Eq{eq:mom} represents a high-order Shapiro filter that acts to damp the momentum of sub-grid scale flow and prevent kinetic energy build up at large wavenumbers, and the $\mathcal{E}_\mathrm{S}$ term in \Eq{eq:thermo} represents the conversion of kinetic to thermal energy by the Shapiro filter. 

We couple the primitive equations of motion to tracer equations for the transport of condensible vapor and cloud condensate \citep{Tan:2021wv,Tan:2021vx}:
\begin{equation}
\label{eq:vapor}
    \frac{dq_v}{dt} = (1 - s)\frac{\mathrm{min}(q_s - q_v, q_c)}{\tau_\mathrm{c}} - s\frac{\left(q_v - q_s\right)}{\tau_\mathrm{c}} - \frac{q_v - q_\mathrm{deep}}{\tau_\mathrm{deep}} \mathrm{,} 
\end{equation}
\begin{equation}
\label{eq:condensate}
    \frac{dq_c}{dt} = s \frac{\left(q_v-q_s\right)}{\tau_\mathrm{c}} - (1-s) \frac{\mathrm{min}(q_s - q_v, q_c)}{\tau_\mathrm{c}} - \frac{\partial\left(q_c \langle V_\mathrm{p}\rangle \right)}{\partial p} \mathrm{.}
\end{equation}
Symbols used in Equations (\ref{eq:vapor}) and (\ref{eq:condensate}) include the mass mixing ratio of condensible vapor relative to the background hydrogen-helium air $q_v$, the mass mixing ratio of cloud condensate particles $q_c$, the mass mixing ratio of condensible vapor at saturation $q_s$, the supersaturation indicator $s$, which is set to one when vapor is super-saturated and zero if vapor is sub-saturated, the cloud and condensible vapor tracer relaxation timescale $\tau_\mathrm{c}$, the deep vapor mass mixing ratio $q_\mathrm{deep}$, the deep vapor replenishment timescale $\tau_\mathrm{deep}$, and the terminal settling velocity in pressure coordinates that is properly averaged over the particle size distribution $\langle V_\mathrm{p}\rangle$. In this work, we take $\tau_\mathrm{c}$ to be $15~\mathrm{s}$, slightly longer than the time step of the dynamical core and the same as the relaxation timescale $\tau_\mathrm{relax}$ for the atomic hydrogen tracer. We do so because microphysical inter-conversion (evaporation and nucleation) timescales are generally much shorter than dynamical timescales in gas giant atmospheres \citep{Helling14,Gao:2018tz,Powell:2018aa}. However, our model does not include condensational growth, which can occur on timescales that are comparable to or longer than the dynamical timescale in hot Jupiter atmospheres \citep{Powell:2018aa}. Additionally, note that we do not include the thermodynamic impact of latent heat release from cloud condensation given its negligible magnitude in the ultra-hot Jupiter regime \citep{Tan:2017aa}. 

The first two terms on the right hand side of Equations (\ref{eq:vapor}) and (\ref{eq:condensate}) represent the source and sink of condensible vapor and cloud condensate, respectively. The final term on the right hand side of Equation (\ref{eq:vapor}) relaxes the condensible vapor mass mixing ratio at pressures higher than $p_\mathrm{deep}$ (here taken to be $50~\mathrm{bars}$, see Table \ref{table:params}) toward a deep source mixing ratio $q_\mathrm{deep}$ calculated assuming solar composition \citep{Lodders03}. As discussed further in \Sec{sec:suite}, we assume that Al$_2$O$_3$ (corundum) is the dominant condensible species. We take the equilibrium condensation curve of corundum from Equation (4) of \cite{Wakeford:2017}. If the temperature is below the condensation temperature at any time and any location in the model, all local vapor is assumed to be supersaturated and therefore $q_s$ is nearly zero; otherwise all condensates are relaxed back to vapor.

The final term on the right hand side of Equation (\ref{eq:condensate}) represents the gravitational settling of cloud condensate at the terminal velocity $V_\mathrm{p}$. Its form in altitude coordinates is
\begin{equation}
\label{eq:term}
    V_\mathrm{term} = \frac{2 \beta r^2 g \left(\rho_c - \rho\right)}{9\eta} \mathrm{,}
\end{equation}
where $r$ is the cloud condensate particle size, $\rho_c$ is the cloud condensate density, $\beta$ is the Cunningham factor, which accounts for kinetic effects when the mean free path is larger than particle size, and $\eta$ is the gas viscosity. We convert $V_\mathrm{term}$ in height coordinates to $V_\mathrm{p}$ in pressure coordinates assuming hydrostatic balance. As in \cite{parmentier_2013} and \cite{Komacek:2019aa}, we parameterize the Cunningham factor as a function of the Knudsen number $\mathrm{Kn}$, the ratio of the mean free path to cloud particle size,
\begin{equation}
\label{eq:kn}
   \mathrm{Kn} = \frac{k_B T}{\sqrt{2} \pi r d^2 p} \mathrm{,}
\end{equation}
as 
\begin{equation}
  \beta = 1 + \mathrm{Kn}\left(1.256 + 0.4\mathrm{e}^{-1.1/\mathrm{Kn}}\right) \mathrm{,}
\end{equation} 
and we parameterize the molecular viscosity as a function of temperature as:
\begin{equation}
\label{eq:visc}
 \eta = \frac{5}{16} \frac{\sqrt{\bar{m} k_B T}}{\sqrt{\pi}d^2} \frac{\left(k_B T/\epsilon\right)^{0.16}}{1.22} \mathrm{.}
\end{equation}
Symbols used in Equations (\ref{eq:kn}) - (\ref{eq:visc}) above include the Boltzmann constant $k_B$, the diameter of hydrogen gas $d = 2.827 \times 10^{-10}~\mathrm{m}$, the mean molecular mass $\bar{m}$, and the depth of the H$_2$ potential well $\epsilon = 59.7 k_b~\mathrm{K}$. The expressions in Equations (\ref{eq:kn}) - (\ref{eq:visc}) above are valid for temperatures ranging from 300 to 3000 K and at pressures less than 100 bars. As described below, condensate clouds do not form in high temperature regions in our model. As a result, cloud settling is confined to relatively cool regions with a low mixing ratio of atomic hydrogen, ensuring that the condensation settling scheme above is a valid sink of cloud condensate in our GCM simulations. 

As in \cite{Tan:2019aa}, we further couple the primitive equations to a tracer equation for the transport of the mass mixing ratio of atomic hydrogen relative to the total gas:
\begin{equation}
    \frac{dq_H}{dt} = -\frac{\delta q_H}{\tau_\mathrm{relax}} \mathrm{.}
\end{equation}
The change of the atomic hydrogen mass mixing ratio due to tracer transport $\delta q_H$ is
\begin{equation}
\label{eq:deltaqh}
    \delta q_H = \frac{q_H  - q_{H,eq}}{1 + \frac{L_H}{\bar{c}_p}\frac{\partial q_{H,eq}}{\partial T}} \mathrm{,}
\end{equation}
where the equilibrium mass mixing ratio of atomic hydrogen is $q_{H,eq}$ (see the following paragraph for further details). In the GCM, local changes in $q_H$ affect the dynamics mainly through heating and cooling due to hydrogen recombination and dissociation, as expressed by the $\frac{\theta'}{\bar{c}_pT} \frac{L_H \delta q_H}{\tau_\mathrm{relax}}$ term on the right hand side of \Eq{eq:thermo}. Other slightly minor dynamical effects include the change of mean molecular weight and specific heat.

We improve upon the tracer scheme for atomic hydrogen from \cite{Tan:2019aa} by incorporating the presence of helium in the mass budget. Given a molar fraction of atomic hydrogen relative to the total gas $\chi_H$ calculated from the Saha equation of the H$_2$-H system \citep{Berardo:2017aa,Bell:2018aa} and a prescribed molar ratio of helium to hydrogen $x_{He/H}$ (assumed to be Solar), we calculate the equilibrium mass mixing ratio of atomic hydrogen as
\begin{equation}
    q_{H,eq} = \frac{\chi_H}{2 - \chi_H + 4 x_{He/H} (2-\chi_H)} \mathrm{.}
\end{equation}
This sets an upper limit on the equilibrium mass mixing ratio of atomic hydrogen of $q_{H,max} = 0.759$ if all of the hydrogen present at a given location is expected to be in atomic form. We also take into account helium gas in our formulation of the mean molecular weight and thus the mass-weighted specific gas constant and heat capacity:
\begin{equation}
  \bar{R} = R_H q_{H} + R_{H_2} \left(q_{H,max} - q_H\right) + R_{He} \left(1- q_{H,max}\right) \mathrm{,}
\end{equation}
\begin{equation}
    \bar{c}_p = c_{p,H} q_{H} + c_{p,H_2} \left(q_{H,max} - q_H\right) + c_{p,He} \left(1- q_{H,max}\right) \mathrm{.}
\end{equation}
Our assumed values for specific heat capacity and gas constant of each component of the atmosphere (H, H$_2$, He) are shown in Table \ref{table:params}. Note that our choices of the specific heat capacity for each component result in an average $c_p=13000~\mathrm{J}~\mathrm{kg}^{-1}~\mathrm{K}^{-1}$ at low temperature (i.e., with $q_H = 0$), consistent with that used in previous \texttt{MITgcm} simulations of the atmospheric dynamics of hot and ultra-hot Jupiters (e.g., \citealp{Showmanetal_2009,Liu:2013,kataria_2013,Tan:2019aa,Steinrueck:2021aa}). Though the inclusion of inert helium gas reduces the thermodynamic effect of hydrogen dissociation and recombination, we show in \Sec{sec:Hdisc} that the thermal impact of hydrogen dissociation and recombination has a strong effect on the thermal structure predicted by our GCMs. 

As in \cite{Tan:2019aa} and \cite{May:2021ab}, we couple a modified version of the \texttt{DISORT} \texttt{TWOSTR} plane-parallel two-stream double-gray radiative transfer scheme with the multiple scattering approximation \citep{Stamnes:2027,Kylling:1995} to our dynamical core. Gas opacities are considered to be purely absorptive, but cloud opacities include multiple scattering. We set the visible and infrared background gas opacities to be functions of pressure alone, with the gas opacity in the visible band parameterized as
\begin{equation}
    \kappa_\mathrm{g,vis} = 10^{0.0478(\mathrm{log}_{10}p[\mathrm{Pa}])^2 - 0.1366(\mathrm{log}_{10}p[\mathrm{Pa}]) - 3.2095}\mathrm{m}^2~\mathrm{kg}^{-1} \mathrm{,}
\end{equation}
and the gas opacity in the infrared band set to
\begin{equation}
\begin{split}
    \kappa_\mathrm{g,IR} = \mathrm{max}\big(& 10^{0.0498(\mathrm{log}_{10}p[\mathrm{Pa}])^2 - 0.1329(\mathrm{log}_{10}p[\mathrm{Pa}]) - 2.9457},  \\
    & 10^{-3}\big)~\mathrm{m}^2~\mathrm{kg}^{-1} \mathrm{.}
    \end{split}
\end{equation}
As in \cite{May:2021ab,Tan:2021wv,Tan:2021vx}, we include a constant minimal opacity given that the use of a Rosseland-mean opacity in the double-gray framework would otherwise cause low pressures in the model to become optically thin and radiatively inactive, where in reality narrow absorption bands drive radiative heating/cooling. We choose a minimal opacity $\kappa_\mathrm{min}$ of $10^{-3}~\mathrm{m}^2~\mathrm{kg}^{-1}$, consistent with the thermal opacity used in previous double-gray models of hot Jupiter atmospheres \citep{Guillot:2010}. To incorporate the current evolutionary state of TOI-1431b modeled in \Sec{sec:mesa}, as in \cite{May:2021ab} we prescribe a ``surface'' temperature of 3750 K at 100 bars in our radiative transfer scheme. Our simulations do not include a dry convective adjustment scheme due to the inclusion of hydrogen dissociation and recombination, which enforces the lapse rate to be smaller than the lapse rate of H$_2$ alone. As a result, the thermal structure is not prescribed at any atmospheric level, and instead is controlled by dynamics and radiative transfer.

We include the cloud opacity as follows. We calculate the cloud extinction opacity by scaling from the gas opacity in both the visible and thermal band as
\begin{equation}
\begin{split}
    & \kappa_\mathrm{c,vis} = \frac{q_c}{q_\mathrm{deep}} c_{\kappa,\mathrm{cld}} \kappa_{\mathrm{g,vis}} \mathrm{,} \\
    &\kappa_\mathrm{c,IR} = \frac{q_c}{q_\mathrm{deep}} c_{\kappa,\mathrm{cld}} \kappa_{\mathrm{g,IR}} \mathrm{,}
\end{split}
\end{equation} 
where $c_{\kappa,\mathrm{cld}}$ is a scaling factor for the cloud extinction opacity relative to the gas opacity in both the visible and thermal bands. Here, $q_{\rm deep}$ is a fixed parameter in each simulation, and the local cloud opacity depends on the time-dependent and spatially inhomogeneous distribution of clouds which are coupled to the dynamics. From \texttt{CARMA} microphysics simulations as described below, clouds on the nightside of hot Jupiters are usually optically thick and  highly scattering. We prescribe a fixed cloud asymmetry parameter ($g_\mathrm{cld}$) and single scattering albedo ($\varpi_{0,\mathrm{cld}}$) to determine the cloud optical properties, discussed further in \Sec{sec:suite}. These assumed cloud optical properties are used in both the double-gray multiple scattering radiative transfer scheme in the GCM along with the \texttt{gCMCRT} code used in post-processing.

Note that our cloud extinction opacity has the same scaling to the gas opacity in both the visible and optical bands. This simplification is justified in the regime of ultra-hot Jupiters due to the lack of significant cloud mass on the dayside and thus a minimal cloud-radiative effect in the visible wavelength band. There are other schemes to parameterize cloud opacity in semi-grey GCMs of hot Jupiters and brown dwarfs (e.g., \citealp{Roman:2021wl,Tan:2021vx}) that specify the cloud particle size or number density. Our current method is simpler and provides a clear setup to understand how the atmospheric circulation and cloud structure are affected by the radiative effect of clouds.

The total atmospheric opacity at a given layer for each wavelength band is simply set by a weighted sum of the gas and cloud opacities,
\begin{equation}
\begin{split}
  & \kappa_{\mathrm{vis}} = c_{\kappa,\mathrm{vis}} \kappa_{\mathrm{g,vis}} + \kappa_{\mathrm{c,vis}} \mathrm{,} \\
  & \kappa_{\mathrm{IR}} = \kappa_{\mathrm{g,IR}} + \kappa_{\mathrm{c,IR}} \mathrm{.}
  \end{split}
\end{equation}
We include an additional scaling factor for the visible-band gas opacity ($c_{\kappa,\mathrm{vis}}$) in order to mimic the effects of visible absorbers (e.g., atomic metal species) on the temperature structure by considering an increased visible band opacity. We formulate the total opacity in each band in this manner in order to have only two free radiative transfer parameters in our model suite: $c_{\kappa,\mathrm{vis}}$ and $c_{\kappa,\mathrm{cld}}$. Our two-stream, double-gray radiative transfer scheme including cloud extinction and scattering is then coupled to the dynamical core through the vertical divergence of net radiative flux leading to heating and cooling, as shown by the term $\frac{g \theta'}{\bar{c}_pT} \frac{\partial F}{\partial p}$ on the right hand side of \Eq{eq:thermo}. 


\begin{table}
\vspace{0.5cm}
\setlength{\tabcolsep}{0pt}
\footnotesize
\begin{center}
\begin{tabular}{ l l }
\hline
{\bf Parameter} & {\bf Value} \\
\hline
\hspace{1.25cm} Planetary properties \\
\hline
Radius ($R$) & 1.54 $R_\mathrm{Jup,eq}$ \\
Gravity ($g$) & 32.75 m~s$^{-2}$\\
Rotation period ($P_{rot}$) &  2.65 Earth days \\
\hline
\hspace{1.25cm} Thermodynamic quantities \\
\hline
Specific heat capacity of H$_2$ ($c_{p,H_2}$) & $1.548 \times 10^4$ J~kg$^{-1}$~K$^{-1}$ \\
Specific heat capacity of H ($c_{p,H}$) & $2.079 \times 10^4$ J~kg$^{-1}$~K$^{-1}$   \\
Specific heat capacity of He ($c_{p,He}$)& $5193$ J~kg$^{-1}$~K$^{-1}$  \\
He/H molar ratio ($x_{He/H}$) & 0.0793 \\ 
Specific gas constant of H$_2$ ($R_{H_2}$)& 4124 J~kg$^{-1}$~K$^{-1}$ \\
Specific gas constant of H ($R_{H}$)& 8248 J~kg$^{-1}$~K$^{-1}$ \\
Specific gas constant of He ($R_{He}$)& 2062 J~kg$^{-1}$~K$^{-1}$ \\
Specific recombination energy of H$_2$ ($L_H$)& $2.18 \times 10^8$ J~kg$^{-1}$ \\ 
\hline 
\hspace{1.25cm} Radiative transfer parameters  \\
\hline
Irradiation temperature ($T_\mathrm{irr}$)& 3348.85  K \\
Interior upward heat flux ($T_{100}$) & 3750 K \\
Minimum thermal opacity ($\kappa_\mathrm{IR,min}$) & $10^{-3}$ m$^2$~kg$^{-1}$ \\ 
Visible opacity scaling factor ($c_{\kappa,\mathrm{vis}}$) & [{\bf 1}, 10] \\
Cloud opacity scaling factor ($c_{\kappa,\mathrm{cld}}$) & [1, {\bf 10}] \\
Cloud asymmetry parameter ($g_\mathrm{cld}$) & 0.8 \\
Cloud single scattering albedo ($\varpi_{0,\mathrm{cld}}$) & 0.95 \\ 
Clear thermal photosphere ($p_{\mathrm{phot,IR}}$) & $142.5~\mathrm{mbars}$ \\ 
Clear visible photosphere ($p_{\mathrm{phot,vis}}$) & $[{\bf 251.3}, 39.58]~\mathrm{mbars}$ \\

\hline 
\hspace{1.25cm} Cloud properties \\
\hline
Condensate vapor deep  & $1.13 \times 10^{-4}$ kg~kg$^{-1}$ \\
\hspace{0.25cm} mixing ratio ($q_\mathrm{deep}$) & \\
Condensate vapor source  & $5 \times 10^6$ Pa \\
\hspace{0.25cm} pressure ($p_\mathrm{deep}$) & \\ 
Condensate vapor deep relaxation & $10^3$ s \\ 
\hspace{0.25cm} timescale ($\tau_\mathrm{deep}$) & \\
Condensate density ($\rho_c$) & 3950 kg~m$^{-3}$ \\ 
Mean particle size ($r_0$) & [2, {\bf 5}] $\mu\mathrm{m}$ \\ 
Lognormal distribution width ($\sigma$) & 1 \\
Minimum particle bin radius ($r_\mathrm{min}$) & 0.1 $\mu\mathrm{m}$ \\
Maximum particle bin radius ($r_\mathrm{max}$) & 100 $\mu\mathrm{m}$ \\
\hline 
\hspace{2cm} Numerical parameters \\
\hline
Drag timescale ($\tau_\mathrm{drag}$) & $10^7$ s\\
Horizontal resolution & C48 \\
Vertical resolution & 70 layers\\
Lower boundary & 100 bars \\
Upper boundary & 10 $\mu$bars \\
Reference pressure ($p_0$) & 1 bar \\ 
Shapiro filter order & 4 \\
Shapiro filter timescale & 40 s \\ 
Dynamical time step & 5 s, {\bf 10 s} \\
Radiative time step & 15 s, {\bf 30 s} \\
H tracer relaxation timescale $\tau_\mathrm{relax}$ & 15 s \\
Condensate tracer relaxation & 15 s \\ 
\hspace{0.25cm} timescale ($\tau_\mathrm{c}$) & \\ 
\hline
\end{tabular}%
\caption{Planetary and atmospheric properties assumed for the suite of GCM simulations presented in this work. Brackets indicate parameter sensitivity tests over visible-band gas opacity, cloud opacity, and cloud particle size, while bold indicates baseline values for those parameters in our full suite of models.}
\label{table:params}
\end{center}
\end{table}

\subsubsection{Cloud microphysical and radiative properties}
\label{sec:suite}
We parameterize cloud microphysical and radiative properties in our GCM using predicted cloud particle size distributions and optical properties from the \texttt{CARMA} cloud microphysical simulations of \cite{Gao:2021vp}. \texttt{CARMA} has been recently applied to  study the cloud microphysics of hot Jupiters by \cite{Powell:2018aa,Powell:2019aa,Gao:2020aa,Gao:2021vp}, and it models a range of microphysical processes including heterogeneous and homogeneous nucleation, condensational growth and evaporation, coagulation, and vertical lofting and settling (see Appendix A of \citealp{Gao:2018aa} for details). 

Specifically, we utilize the cloud microphysical and optical properties from the $T_\mathrm{eq} = 2100~\mathrm{K}$ model of \cite{Gao:2021vp}, which considers the heterogeneous nucleation of a variety of species, most notably forsterite (Mg$_2$SiO$_4$) and corundum (Al$_2$O$_3$), on titanium dioxide (TiO$_2$) grains, along with the homogeneous nucleation of TiO$_2$ and Fe, among other species. The resulting cloud particle size distribution for each condensate species from these \texttt{CARMA} simulations is roughly log-normal, with characteristic particle sizes ranging from $\sim 1-10~\mu\mathrm{m}$. As a result, we parameterize a log-normal cloud particle size distribution in our GCM as
\begin{equation}
\label{eq:cldparsize}
    \frac{dn_c}{dr} = \frac{n_\mathrm{c}}{\sqrt{2\pi}\sigma r} e^{-\left[\mathrm{ln}(r/r_0)\right]^2/(2\sigma^2)} \mathrm{,}
\end{equation}
where $n_c$ is the cloud particle number per dry air mass, $r_0$ is the mean particle size, $\sigma$ is the lognormal distribution width, and $n_\mathrm{c}$ is the number of cloud particles per dry air mass, calculated as
\begin{equation}
    n_\mathrm{c} = \frac{3q_c}{4\pi \rho_c r_0^3 e^{(4.5\sigma^2)}} \mathrm{.}
\end{equation}
Given that our cloud optical properties are set to a fixed value independent of particle size (see below), our assumed cloud particle size distribution only impacts the vertical settling of particles through the sink term for cloud condensate on the right hand side of \Eq{eq:condensate}, $- \frac{\partial\left(q_c \langle V_\mathrm{p}\rangle\right)}{\partial p}$, in which the mean terminal velocity  is obtained from proper averaging over the size distribution in \Eq{eq:cldparsize}. 

In this work, we include a single cloud condensate tracer along with a single tracer for condensible vapor, using cloud condensate and condensible vapor properties appropriate for corundum. We isolate the local condensation and transport of corundum in this work because it is the highest-temperature condensate expected to contribute significantly to the cloud opacity of ultra-hot Jupiters \citep{Gao:2021vp}. As a result, by considering corundum condensation alone in this work we constrain how dynamics, cloud tracer transport, and cloud-radiative feedback set the greatest possible extent of the cloud deck on ultra-hot Jupiters. We assume the density of corundum clouds from \cite{Roman:2021wl}, and take the condensation temperature and deep condensible vapor mixing ratio appropriate for corundum from \cite{Wakeford:2017} and \cite{Lodders03}. We choose baseline cloud optical properties, including the cloud asymmetry parameter, single scattering albedo, and cloud opacity scaling factor, from their characteristic values in the near-infrared at the $10-100~\mathrm{mbar}$ level in the $T_\mathrm{eq} = 2100~\mathrm{K}$ \texttt{CARMA} simulation of \cite{Gao:2021vp} where there is a thick corundum cloud deck in the model. The cloud single scattering albedo and asymmetry parameter are unchanged throughout the suite of GCMs conducted here, and instead the only free parameter that directly affects the cloud-radiative feedback is the cloud opacity scaling factor $c_{\kappa,\mathrm{cld}}$. Table \ref{table:params} shows all of our parameter choices in the model, with free parameters expressed in brackets and their baseline value in bold face. 

\subsubsection{Suite of GCMs with cloud tracers}
\begin{table}
\setlength{\tabcolsep}{2pt}
\footnotesize
\begin{center}
\begin{tabular}{ l  l }
\hline
{\bf Simulation ID} & {\bf Parameter modifications} \\
\hline
Baseline & None \\
No cloud RT & Radiatively inactive clouds \\ 
Reduced cloud opacity (Red. $\kappa_\mathrm{cld}$) & $c_{\kappa,\mathrm{cld}} = 1$ \\
Reduced cloud particle size (Red. $r_0$) & $r_0 = 2~\mu\mathrm{m}$ \\
Enhanced visible opacity (Enh. $\kappa_\mathrm{vis}$) & $c_{\kappa,\mathrm{vis}} = 10$ \\
\hline
\end{tabular}%
\caption{List of GCM simulations conducted for this work, along with the parameters modified from the baseline model for each case. We conduct five GCM simulations in total in order to study the effect of varying the properties and existence of cloud-radiative feedback and of varying the visible-to-infrared gas opacity ratio (i.e., $\kappa_\mathrm{g,vis}/\kappa_\mathrm{g,IR}$, \citealp{Guillot:2010}) on the atmospheric dynamics of ultra-hot Jupiters. }
\label{table:sims}
\end{center}
\end{table}

We conduct a suite of GCMs modifying three free parameters individually: the cloud opacity scaling factor $c_{\kappa,\mathrm{cld}}$, the visible gas opacity scaling factor $c_{\kappa,\mathrm{vis}}$, and the mean cloud particle size $r_0$. We additionally conduct a simulation with the cloud-radiative feedback turned off, resulting in no cloud extinction or scattering in the radiative transfer module. Along with our baseline case with radiatively active clouds, this results in a total of five GCMs in our main model suite. Table \ref{table:sims} shows the simulation identification and parameter modification from the baseline case for each simulation in our GCM suite.  

Each GCM in our model suite has a cubed-sphere resolution of C48, approximately equivalent to a horizontal resolution of $192 \times 96$ in longitude and latitude. Each GCM uses $70$ vertical layers, spaced logarithmically in pressure from 100 bars to $10~\mu\mathrm{bars}$. We take planetary parameters (radius, gravity, rotation period, irradiation temperature) appropriate for TOI-1431b from \cite{Addison:2021aa}. We use a weak height-independent frictional drag in all simulations, characterized by a drag timescale $\tau_\mathrm{drag} = 10^7~\mathrm{s}$. This weak drag ensures that our simulations reach an equilibrated end-state, and \cite{Komacek:2015} and \cite{Komacek:2017} have previously shown that varying the drag timescale from $10^7~\mathrm{s}$ to $\infty$ does not significantly affect the resulting flow structure. The GCM further includes a fourth-order Shapiro filter in order to prevent sub-grid scale momentum build-up without qualitatively affecting the large-scale flow. Our standard time step in the dynamical core is $10~\mathrm{s}$, with the radiative time step set equal to three times the dynamical time step. In the case with an enhanced visible opacity ($c_{\kappa,\mathrm{vis}} = 10$), we reduce the dynamical time step to $5~\mathrm{s}$ for numerical stability. We set the hydrogen and condensation tracer relaxation timescale $\tau_\mathrm{relax} = \tau_\mathrm{c} = 15~\mathrm{s}$, slightly larger than a dynamical time step. Each simulation is initialized from rest with an isothermal temperature profile of $T = 2368~\mathrm{K}$. Each case is then continued to at least 2,500 Earth days, after which we confirm that each GCM simulation has reached an equilibrated state in both the domain-integrated kinetic and thermal energy\footnote{Note that this may be only the first time at which the model would reach kinetic and thermal energy ``equilibrium,'' as long-timescale (tens to hundreds of thousands of Earth day) integrations of gas giant and sub-Neptune GCMs have found significant time-evolution in both the domain-integrated energetics and qualitative predictions for atmospheric dynamics \citep{Mayne:2017,Sainsbury-Martinez:2019aa,Young:2019aa,Mendonca:2020aa,Wang:2020aa,Schneider2022}.}. As expected from previous studies of the atmospheric dynamics of hot Jupiters \citep{Menou03,Rauscher07,Dobbs-Dixon:2010aa,Komacek:2020aa,Cho:2021wb}, our simulations display time-variability in temperature, winds, and cloud mass mixing ratio. However, in this work we focus on the impact of clouds on the climate of ultra-hot Jupiters, and leave more detailed studies of meteorology to future work. As a result, all GCM results shown are time-averages over the final 500 Earth days of simulated time.

\section{Results}
\label{sec:results}

\subsection{Atmospheric dynamics and cloud coverage}
\subsubsection{Baseline case}
\begin{figure*}
    \centering
    \includegraphics[height=0.95\textheight]{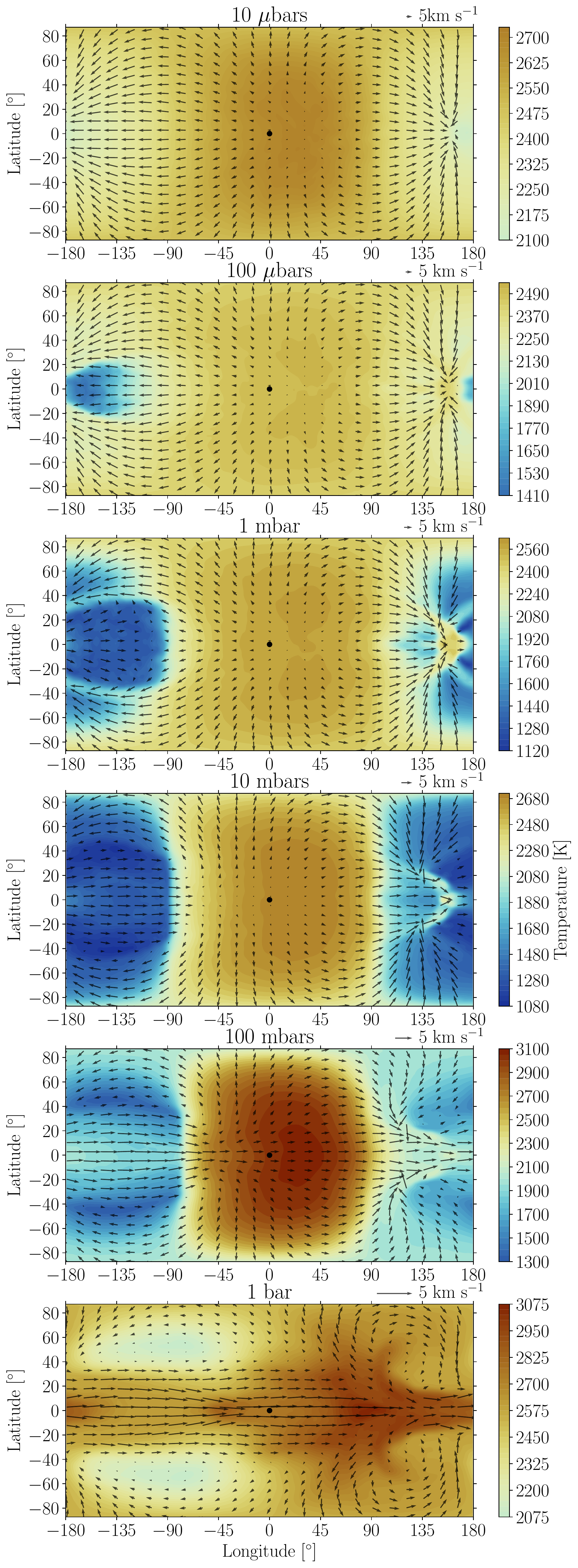}
    \includegraphics[height=0.95\textheight]{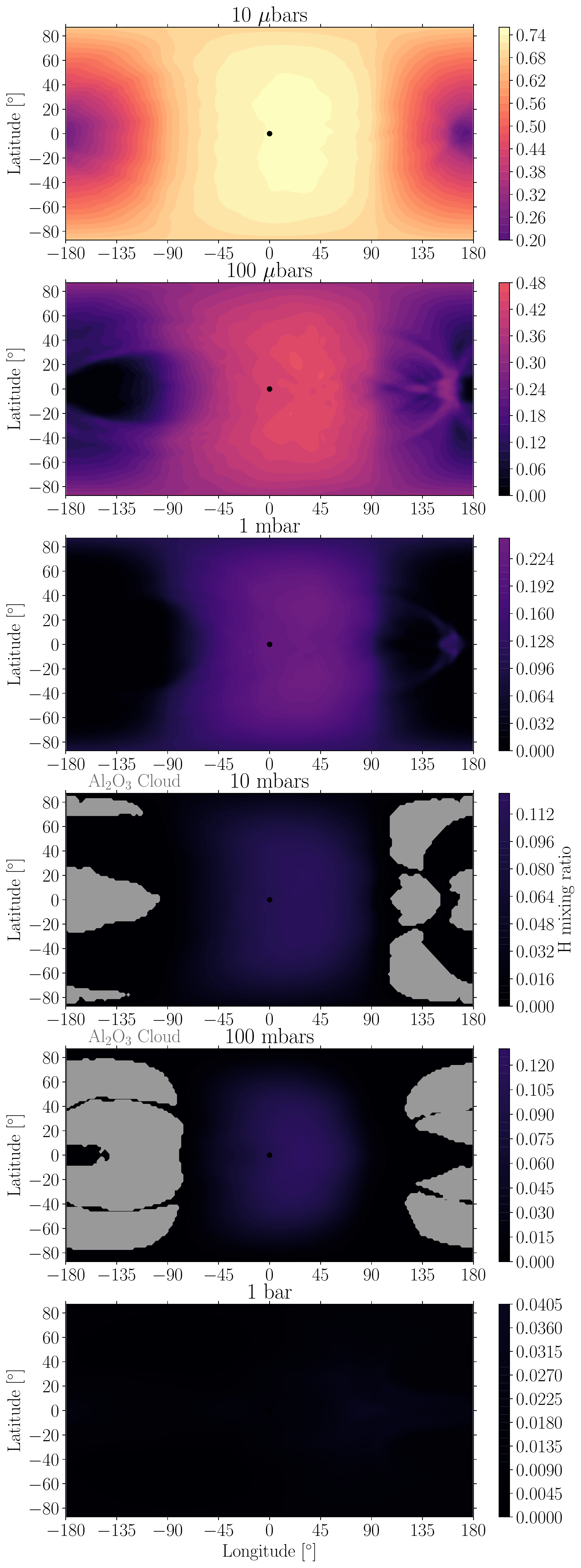}
    \caption{Temperature maps with overlaid wind arrows (left) and atomic hydrogen mass mixing ratio (right, colors) with overlaid cloud tracer distributions (gray regions show where the cloud mass mixing ratio is $\ge 5 \times 10^{-5}~\mathrm{kg}~\mathrm{kg}^{-1}$) plotted on isobars logarithmically spaced from 10 $\mu\mathrm{bars}$ to $1~\mathrm{bar}$ from the baseline GCM simulation with radiatively active cloud tracers. All panels in each respective column share a color scale. The upper atmosphere has significant atomic hydrogen on the dayside, and hydrogen on the nightside is fully recombined at pressures $\gg 1~\mathrm{mbar}$. Patchy clouds occur on the nightside at pressures of 10 and 100 mbars.}
    \label{fig:tempwindp}
\end{figure*}
Before comparing models with varying cloud microphyscial and radiative assumptions, we first study the emergent behavior in our baseline model with radiatively active clouds. \Fig{fig:tempwindp} shows the temperature and wind patterns along with the atomic hydrogen mixing ratio and cloud coverage from the baseline model on isboars ranging from 1 bar to 10 $\mu$bars. Note that temperature, wind, atomic hydrogen mixing ratio, and cloud coverage maps on isobars from the entire suite of GCMs are shown in Appendix \ref{sec:appendixtemp}. 

The temperature and wind patterns near the infrared photosphere (which lies at a pressure of $142.5~\mathrm{mbars}$, see Table \ref{table:params}) in the baseline case are similar to those expected from the broad range of previous studies of hot and ultra-hot Jupiters \citep{Heng:2014b,Showman:2020rev}. At these pressures, the atmosphere is characterized by a large day-to-night temperature contrast \citep{perna_2012,Perez-Becker:2013fv,Komacek:2015} that triggers a planetary-scale equatorial wave pattern \citep{Showman_Polvani_2011,showman_2013_doppler,Tsai:2014,Hammond:2018aa,Penn:2018ws,Pierrehumbert:2019vk,Hammond:2021aa}. The superposition of mid-latitude westward shifted Rossby modes and equatorially trapped eastward Kelvin modes leads to phase tilts that act to pump momentum toward the equator, causing a superrotating equatorial jet shown by the eastward direction of winds throughout the equatorial region \citep{Showman_Polvani_2011} at pressures $\ge 10~\mathrm{mbars}$ in \Fig{fig:tempwindp}. This superrotating equatorial jet then Doppler shifts the planetary-scale wave pattern, leading to an eastward hot spot offset due to the eastward shift of the equatorially trapped Kelvin wave component \citep{Penn:2017tp,Hammond:2018aa}. The Rossby wave component of the planetary-scale wave pattern induces cyclonic motions (``Rossby gyres'') poleward of the mid-latitude flanks of the equatorial jet on the nightside \citep{Cho:2021wb}. Relevant for cloud condensation, the coldest regions in the simulation occur at the intersection of the equatorial jet and Rossby gyres, at which convergence induces a narrow band of cold downwelling air in the mid-latitudes. 

The atmospheric circulation at pressures $\le 1~\mathrm{mbar}$ is markedly different than that at higher pressures. This is because the radiative timescale scales linearly with the overlying column mass of atmosphere and thus the pressure itself \citep{showman_2002}. The short radiative timescale at low pressures leads to strong radiative damping of the planetary-scale wave pattern, preventing the equatorward momentum transport required to sustain the equatorial jet \citep{showman_2013_doppler}. As a result, the circulation at low pressures is characterized by substellar-to-antistellar flow, rather than a superrotating eqatorial jet and Rossby gyres as for higher pressures. Low pressures also promote the dissociation of molecular hydrogen \citep{Bell:2018aa}, leading to an increase in the mass mixing ratio of atomic hydrogen on the dayside with decreasing pressure. As we display further in \Sec{sec:temp} and demonstrate in \Sec{sec:Hdisc}, the increased atomic hydrogen mass mixing ratio at low pressures on the dayside leads to a thermodynamically induced thermal inversion on the nightside due to the heat release during the recombination of atomic to molecular hydrogen. This causes a decrease in the day-to-night temperature contrast at the lowest pressures in the model domain relative to the day-night contrasts near the infrared photosphere. 

\begin{figure*}
    \centering
    \includegraphics[width=0.32\textwidth]{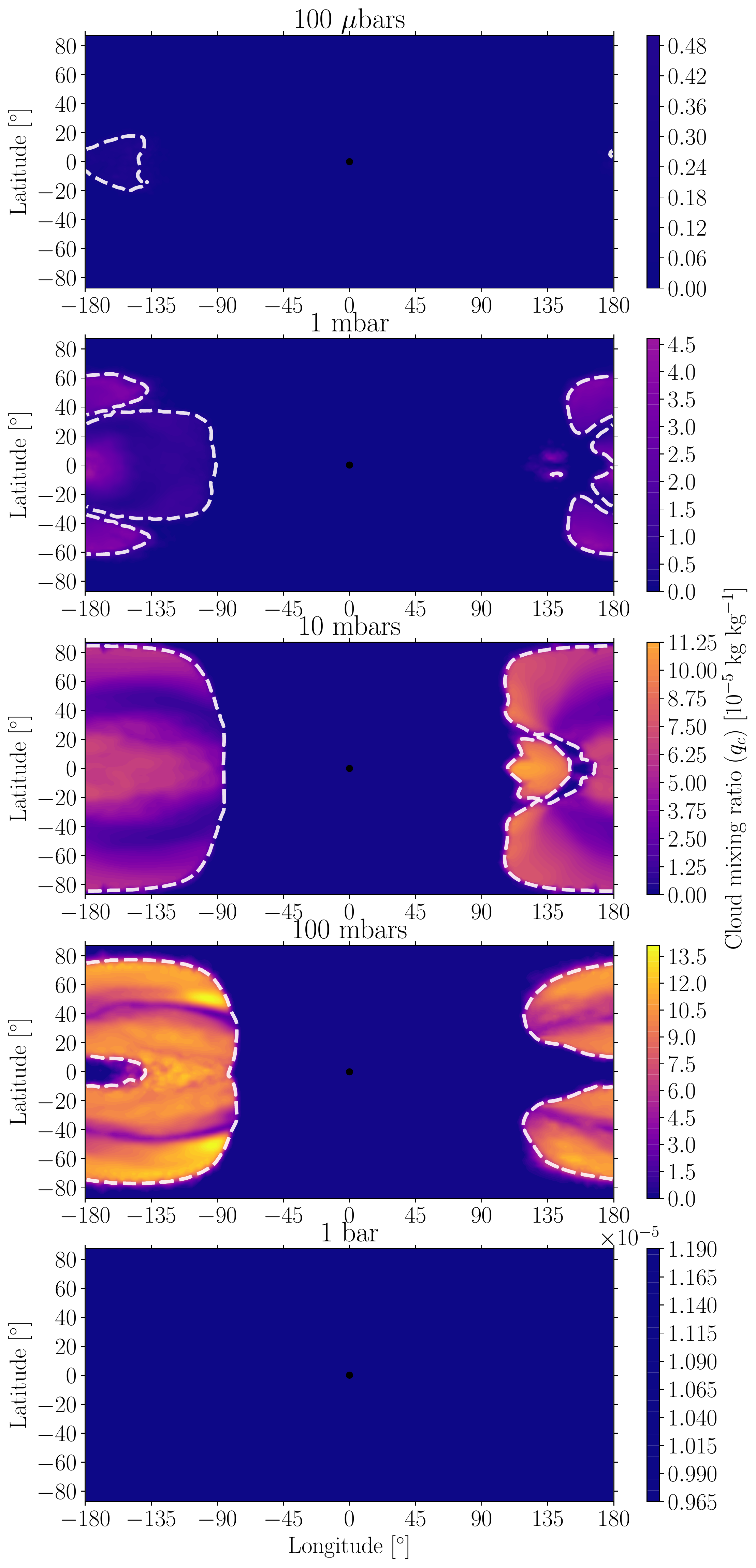}
    \includegraphics[width=0.32\textwidth]{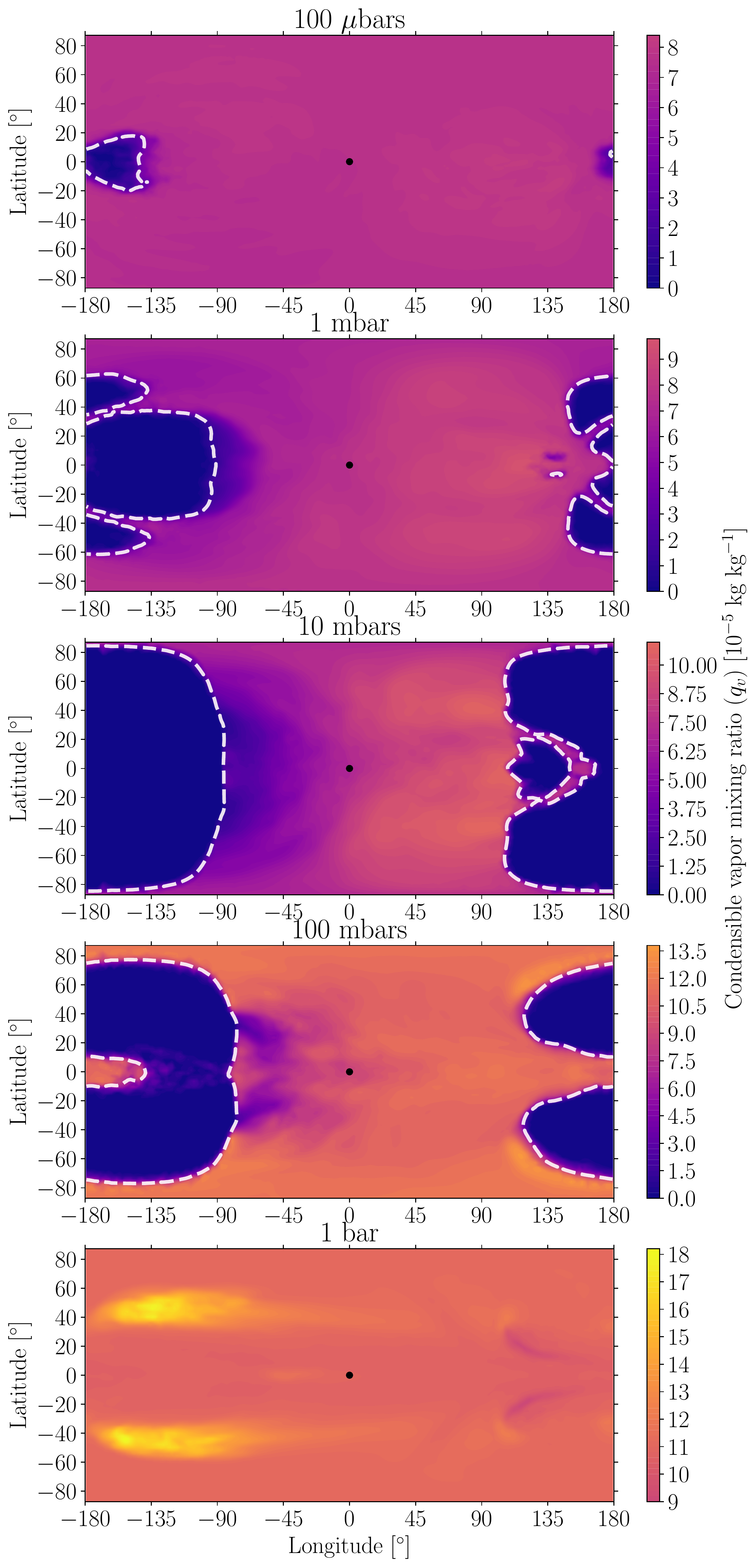}
    \includegraphics[width=0.316\textwidth]{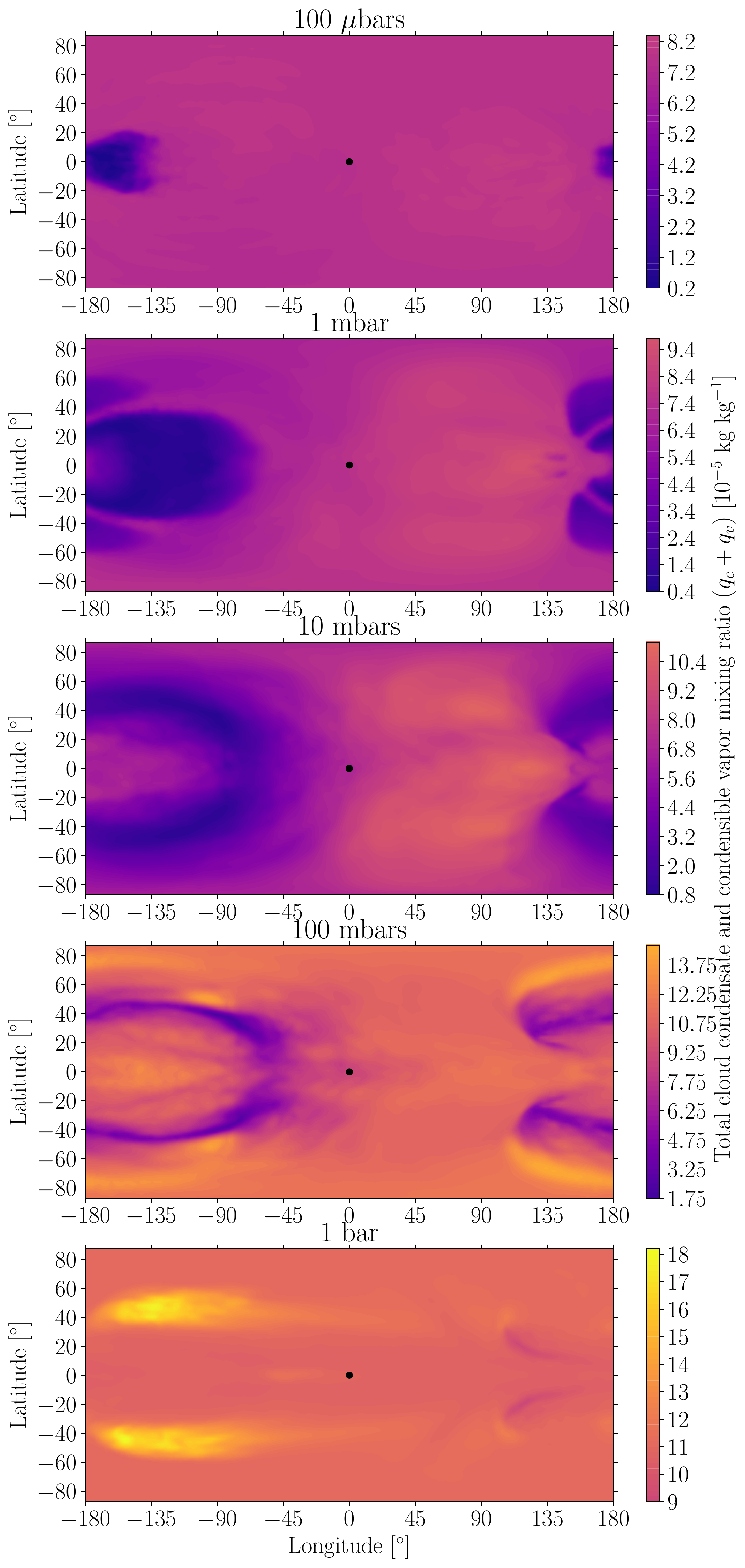}
    \caption{Maps of the cloud condensate and condensible vapor tracer distribution from our baseline GCM simulation with radiatively active cloud tracers. Shown are the cloud mass mixing ratio (left), condensible vapor mass mixing ratio (middle), and the sum of cloud and condensible vapor mass mixing ratio (right) on isobars logarithmically spaced from $100~\mu\mathrm{bars}$ to 1 bar. All panels in a given column share a color scale, and all mass mixing ratios are shown in units of $10^{-5}~\mathrm{kg}~\mathrm{kg}^{-1}$. The snow white dashed contour in the cloud and condensible vapor maps displays where the gas temperature is equal to the corundum condensation temperature on each isobar. Patchy clouds form on the nightside and western limb, but are confined between pressure levels of $\sim$ mbar to a bar. The cloud mass mixing ratio is largest in the mid-latitudes at the 100 mbar level.}
    \label{fig:cloudbaseline}
\end{figure*}

The regions with significant corundum condensate cloud coverage are shown by the gray area on the right-hand column of \Fig{fig:tempwindp}. Clouds with a local mass mixing ratio $\ge 5 \times 10^{-5}~\mathrm{kg}~\mathrm{kg}^{-1}$ only persist at pressure levels between 10 and 100 mbars in this baseline simulation. \Fig {fig:cloudbaseline} shows maps of the cloud and vapor mass mixing ratios from the baseline case with radiatively active cloud tracers, which display that significant cloud mass does persist to the 1 mbar level. Note that Appendix \ref{sec:cldapp} shows maps of cloud and vapor mass mixing ratio from the suite of models with varying cloud microphysical and radiative properties. As discussed in \Sec{sec:cloud}, we do not find significant cloud coverage at pressures $\ll 1~\mathrm{mbar}$ in any case in our model suite, as there is a decrease of almost two orders of magnitude in cloud mass mixing ratio from the peak at 100 mbars to the 100 $\mu$bar level. 
We further find that the cloud distribution is inherently non-uniform in both the horizontal and vertical directions. We find that the dayside is generally cloud-free, except for a thin region at depth near the western terminator. The cloud coverage on the nightside is patchy, with a maximum in cloud mass mixing ratio at the 100 mbar level in the mid-latitudes and local minima in cloud mass mixing ratio at latitudes of $\approx 40^\circ$ and at localized regions near the equator. 

Notably, the cloud coverage does not align solely with the atmospheric temperature structure -- the coldest regions at 10 and 100 mbars are not uniformly cloudy. This is especially true for the regions on the mid-latitude nightside where there is a transition from the eastward superrotating jet at lower latitudes to cyclonic flow in the Rossby gyres at higher latitudes. As we will demonstrate in \Sec{sec:mixeff}, this is due to the dependence of vertical mixing on the correlation between the vertical velocity and cloud tracer distribution \citep{Holton:1986,Zhang:2018tp,Zhang:2018te}. The convergence between the superrotating jet and Rossby gyres implies downwelling motions through mass continuity in the primitive equations of motion (Equation \ref{eq:masscont}, see also \citealp{Holton:2013,vallis2017}), which then causes net downward transport of cloud tracers. Cloud tracers are then either advected into upwelling regions or converted into vapor at the warmer higher pressure levels. 


\begin{figure}
    \centering
    \includegraphics[width=0.5\textwidth]{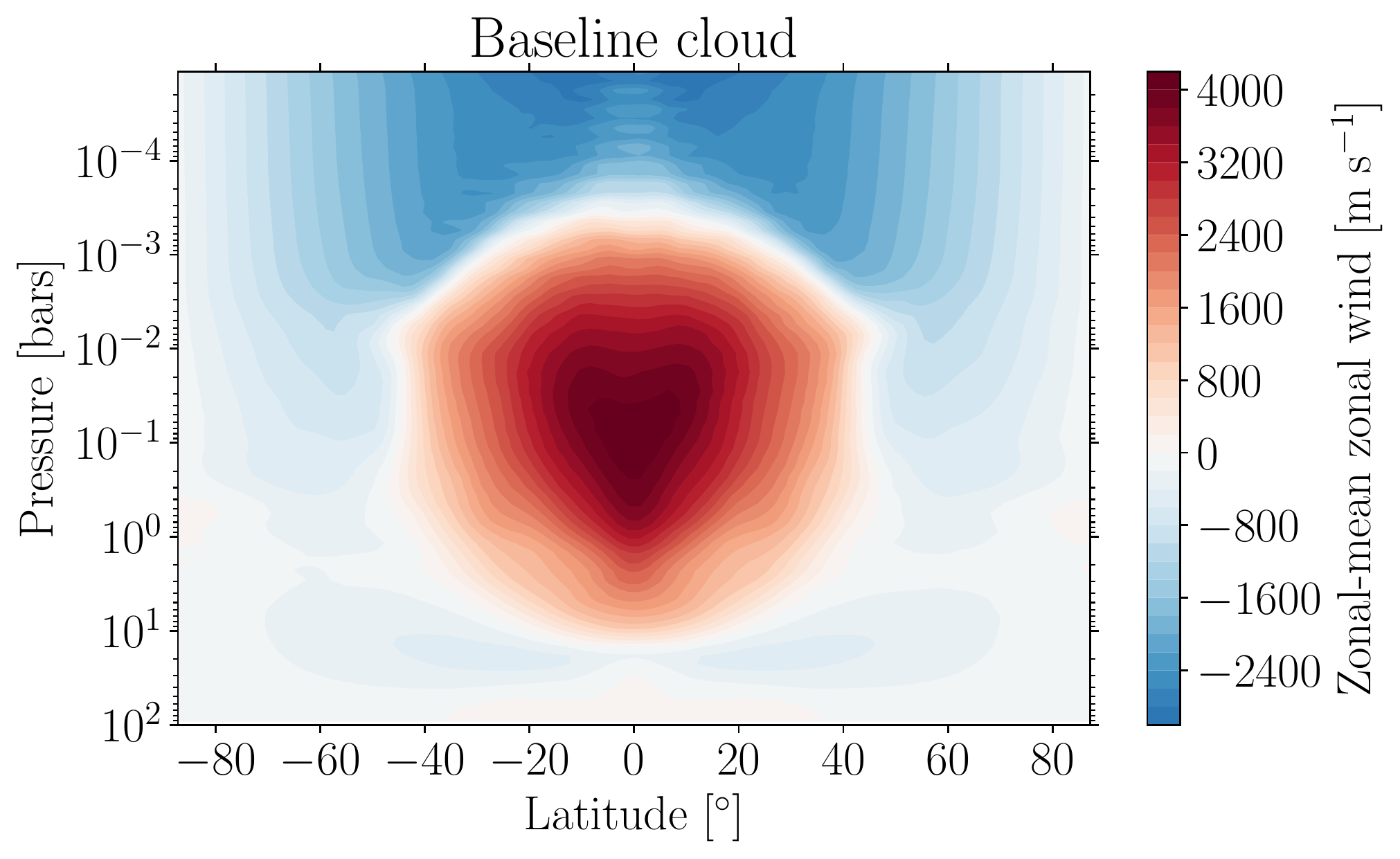}
    \caption{Zonal-mean zonal wind from the baseline GCM simulation with active cloud tracers. There is a superrotating equatorial jet at pressures $\gtrsim 1~\mathrm{mbar}$ that extends to the $\sim 10~\mathrm{bar}$ level along with a transition from superrotation to sub-rotation with decreasing pressure.}
    \label{fig:uzonal}
\end{figure}

\Fig{fig:uzonal} shows the zonal-mean zonal wind (east-west average of the east-west wind) from our baseline case with active cloud tracers. Note that the zonal-mean zonal wind speeds for the remaining cases in our model suite are shown in Appendix \ref{sec:appendixwind}. In the baseline case, we find a maximum zonal-mean zonal wind speed just above $4~\mathrm{km}~\mathrm{s}^{-1}$. As demonstrated in \cite{Tan:2019aa}, this maximum wind speed is smaller than in equivalent cases without the thermodynamic effect of hydrogen dissociation and recombination due to the muted planetary-scale wave pattern. 

Similar to the high-temperature cases with a fixed rotation period of 2.43 Earth days (close to our assumed rotation period of 2.65 Earth days) in \cite{Tan:2019aa}, we further find that the direction of the zonal-mean equatorial flow, $\bar{u}$, reverses from superrotating (eastward) to sub-rotating (westward) with decreasing pressure. This reversal coincides with the transition between superrotation and day-to-night flow shown in \Fig{fig:tempwindp}. It also coincides with pressures at which atomic hydrogen persists on the nightside, causing direct thermodynamic effects on the atmospheric circulation. The thermodynamic impact of hydrogen dissociation and recombination along with the strong radiative cooling at low pressures combine to disrupt the planetary-scale equatorial wave pattern and thus the wave-mean flow interactions that drive the equatorial jet, preventing the superrotating equatorial jet from extending to lower pressures. 

Additional insights of this vertical flow reversal come from the thermal wind balance, whose  form at low latitudes can be written as (e.g., \citealp{Holton:2013}, Chapter 12.6):
\begin{equation}
\begin{split}
    & \frac{\partial\bar{u}}{\partial\ln p} = \frac{1}{\beta}\frac{\partial^2\overline{\bar{R}T}}{\partial y^2} = \frac{R_{H_2}}{\beta}\times \\
    & \frac{\partial^2 \overline{\left [\frac{R_H}{R_{H_2}} q_{H} +  \left(q_{H,max} - q_H\right) + \frac{R_{He}}{R_{H_2}} \left(1- q_{H,max}\right)\right]T}}{\partial y^2},
    \end{split}
    \label{eq.thermalwind}
\end{equation}
where $y$ is distance that increases northward, $\beta=df/dy$ is the meridional gradient of the Coriolis parameter $f$ at the equator, and $\overline{A}$ denotes the zonal-mean of the quantity $A$. The horizontal wind patterns at low pressures preferentially transport atomic H from the dayside deeply into the nightside at high latitudes but less so from the equator. Meanwhile, low latitudes on the nightside are colder and have more H$_2$ than high latitudes at low pressures (see Figure \ref{fig:tempwindp}). By thermal wind balance as in Equation (\ref{eq.thermalwind}), both the meridional temperature and composition gradient at low pressures  favour a strong positive $\partial\bar{u}/\partial\ln p$, giving rise to the flow reversal seen in Figure \ref{fig:uzonal}. Note that we do not imply that the thermal wind balance fully holds on slowly rotating hot Jupiters. In the weak frictional drag regime, the horizontal force balance on hot Jupiters is expected to be set by a combination of the pressure gradient force, Coriolis force, and the nonlinear advection term \citep{showman_2013_doppler}. However, the strong equator-to-pole thermal and compositional gradients at low pressures still place an important constraint on the vertical shear of the zonal-mean flow.


\subsubsection{Cloud coverage}
\label{sec:cloud}
\begin{figure}
    \centering
    \includegraphics[width=0.5\textwidth]{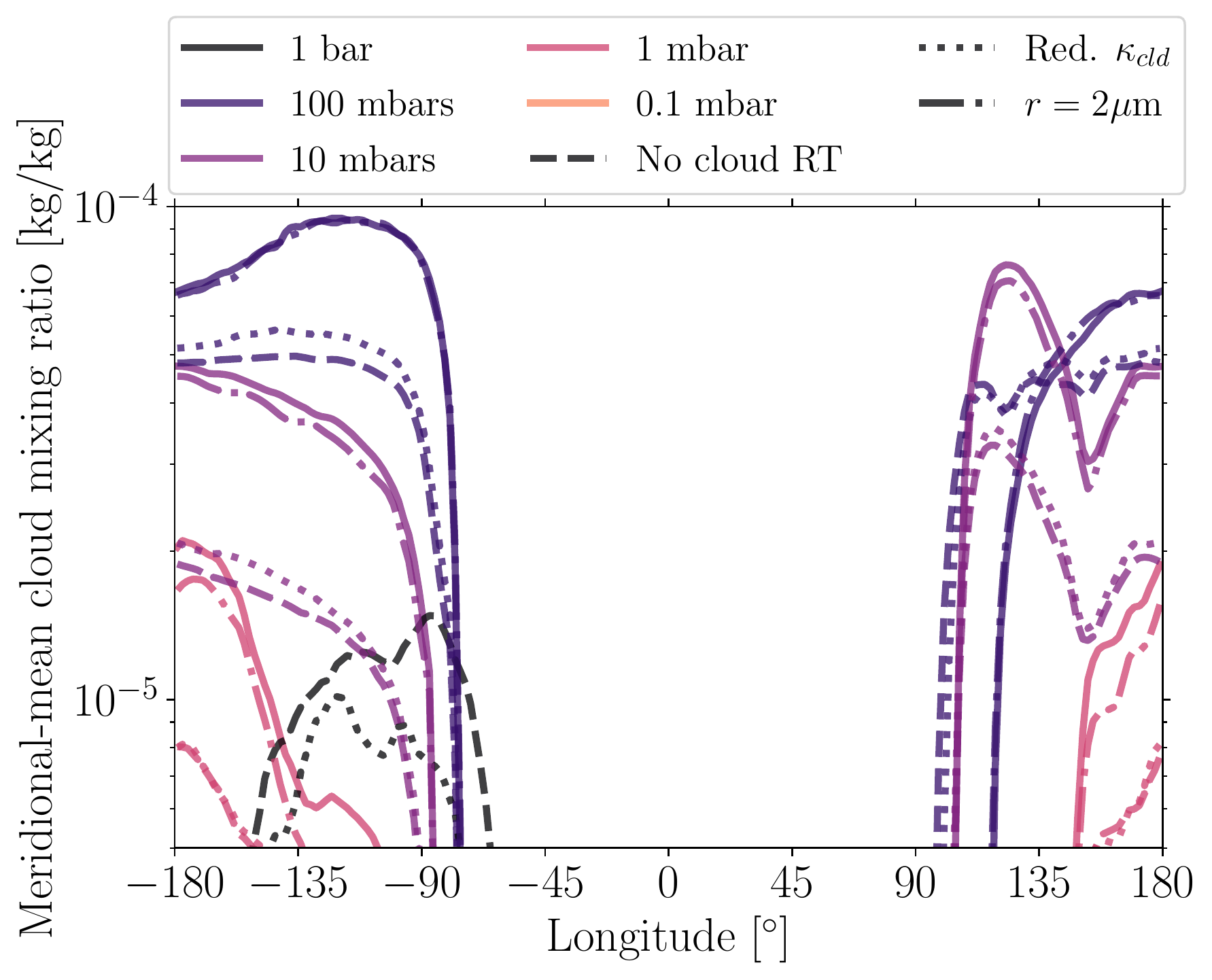}
    \caption{Meridional mean of cloud mass mixing ratio as a function of longitude at different pressures from GCM simulations with varying cloud assumptions. Clouds are largely confined to pressures $\gtrsim 10~\mathrm{mbars}$ on the western limb, and to $\gtrsim 1 ~\mathrm{mbar}$ on the nightside. Though the western terminator is cloudy at $100~\mathrm{mbars}$ in all cases, the eastern terminator is always cloud-free. There are no clouds at $0.1~\mathrm{mbar}$ with a mass mixing ratio $\gtrsim 10^{-6}~\mathrm{kg}~\mathrm{kg}^{-1}$. Clouds form at high pressures of $1~\mathrm{bar}$ only in the case without cloud-radiative feedback, as otherwise the cloud greenhouse effect warms the deep atmosphere, preventing cloud condensation at depth. }
    \label{fig:cloudlong}
\end{figure}

In all cases with cloud condensation, we find that the cloud coverage in the atmosphere of our simulated ultra-hot Jupiter is patchy, being both horizontally and vertically non-uniform. \Fig{fig:cloudlong} shows the meridional-mean condensate cloud tracer mass mixing ratio on isobars from 1 bar to 0.1 mbar as a function of longitude for four simulations: the baseline case, the case without cloud-radiative feedback, the case with reduced cloud opacity, and the case with a reduced characteristic cloud particle size. Results are not shown from the case with an enhanced visible opacity because only a small mass of condensate cloud persists at depth. We find that in the cases with cloud condensation, clouds only form near the western limb and on the nightside. Further, clouds are sequestered at depth, with no cloud condensate persisting at pressures significantly less than $1~\mathrm{mbar}$ on the nightside and western limb. The cloud decks are further prevented from extending to depth, as they only persist to the 1 bar level in the cases without significant cloud-radiative feedback. As described in \Sec{sec:temp}, cloud-radiative feedback leads to a cloud greenhouse effect that acts to warm the atmosphere at higher pressures than the cloud deck, preventing cloud condensation at depth. 

\begin{figure}
    \centering
    \includegraphics[width=0.44\textwidth]{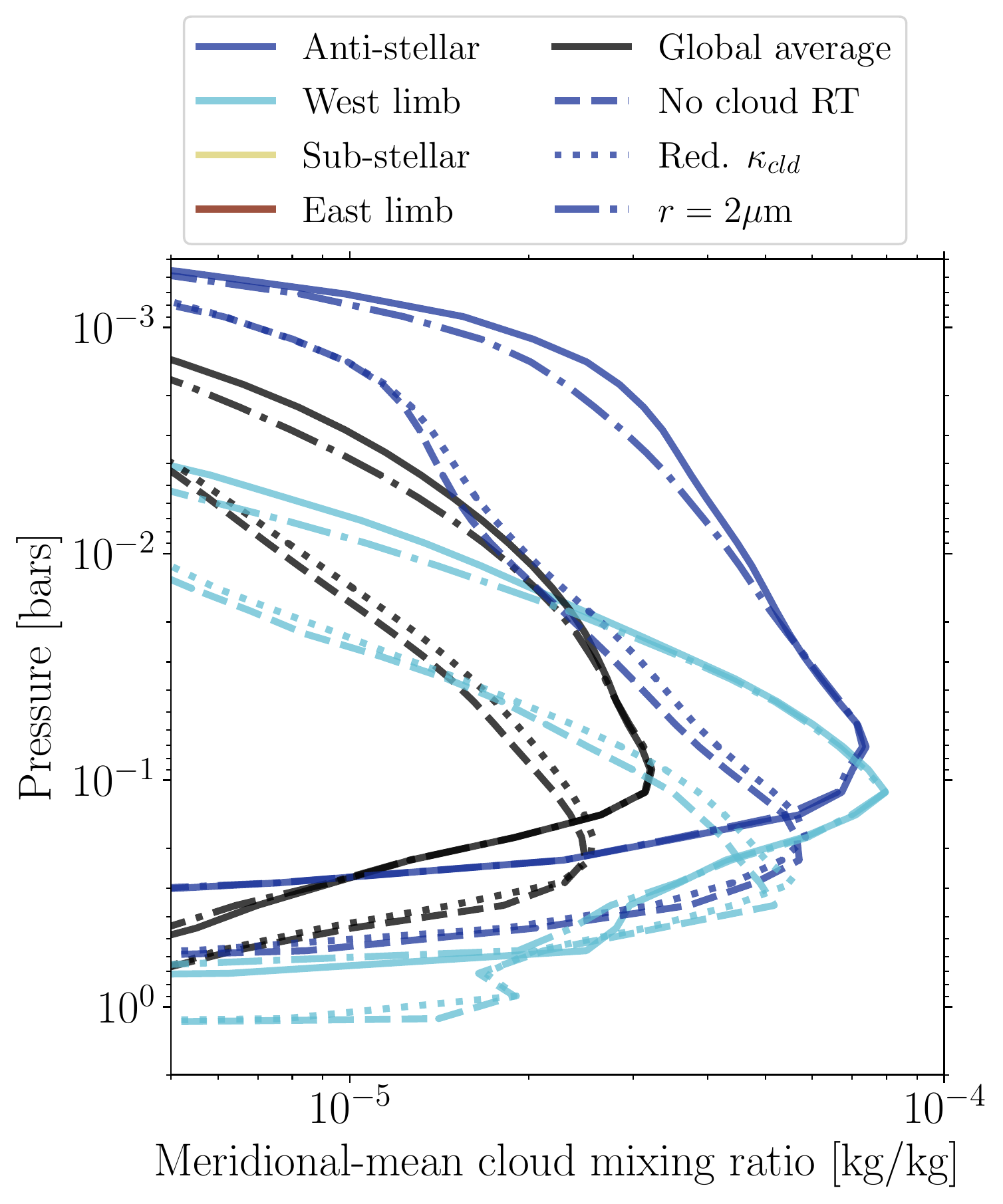}
     \includegraphics[width=0.44\textwidth]{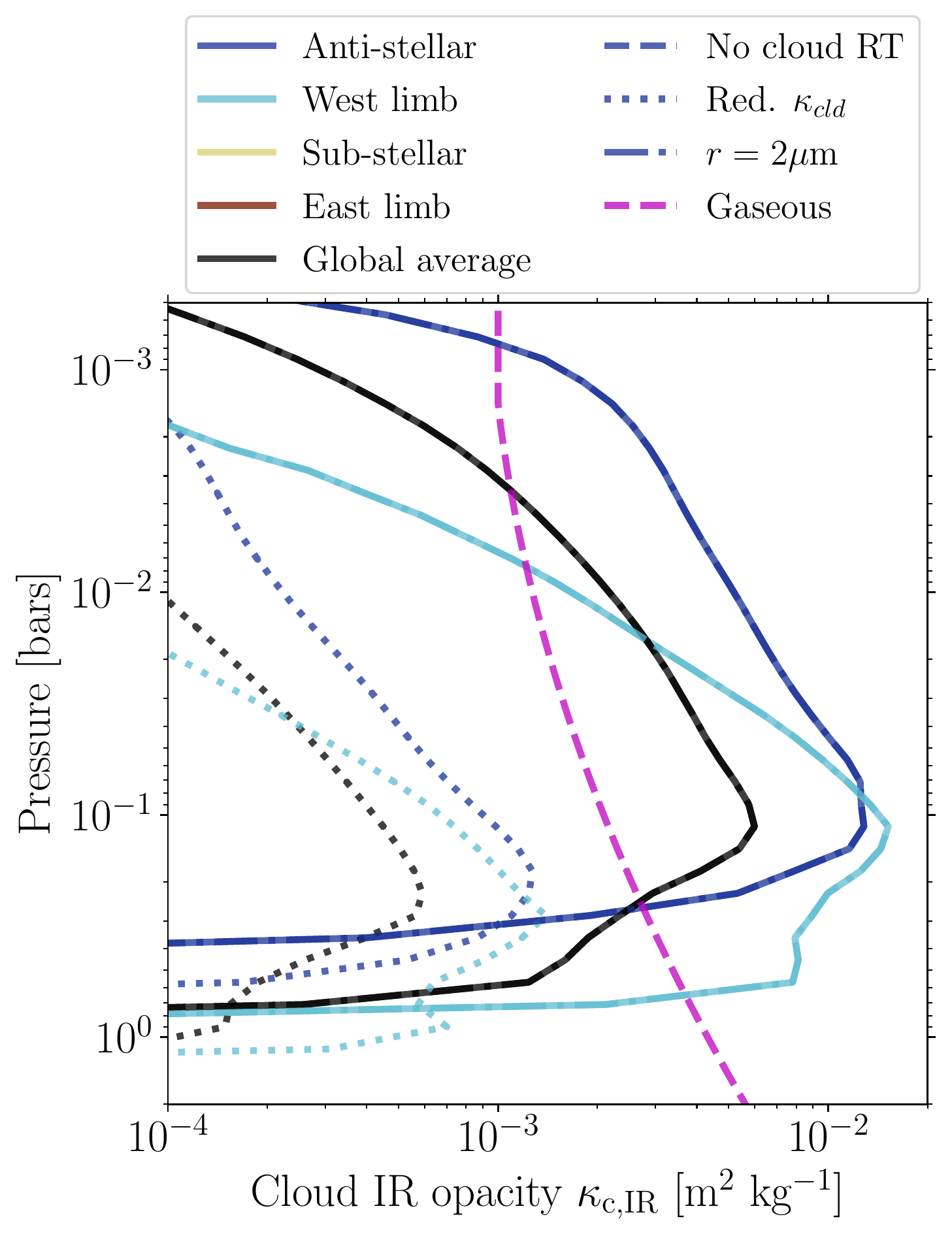}
    \caption{Top: Vertical profiles of the meridional mean of cloud mixing ratio at various longitudes (colors) and the respective global horizontal mean (black lines) from GCM simulations with varying cloud assumptions. Bottom: Vertical profiles of the corresponding IR cloud extinction opacity. There is no significant cloud mass and resulting opacity at any pressure at the sub-stellar point or eastern limb. We find that the cloud top pressure is lower in cases with higher cloud extinction opacity and resulting stronger cloud-radiative feedback. Clouds on the nightside and western limb are sequestered at depth, with the cloud opacity much smaller than the gas opacity at pressures $\ll 1$ mbar.}
    \label{fig:cloudp}
\end{figure}

In all models considered, we find that clouds are sequestered between isobars of approximately 1 bar and several hundreds of $\mu$bars. \Fig{fig:cloudp} shows pressure profiles of the meridional-mean cloud mass mixing ratio and infrared cloud extinction opacity at the anti-stellar point, west limb, and on a global average in the four cases where cloud condensate forms. Note that no significant cloud mass forms at the sub-stellar point or east limb for any of our assumed cloud microphysical and radiative properties. The pressure range over which the cloud deck extends depends on both the longitude at which the cloud deck forms and assumptions about cloud radiative properties. In all cases, the cloud top occurs at a lower pressure on the anti-stellar point than at the western limb. 
The cloud top also occurs at lower pressures in cases that include a strong cloud-radiative feedback (i.e., the baseline and reduced cloud particle size cases) relative to those with a weak (reduced cloud opacity) or zero cloud-radiative feedback, both at the anti-stellar point and western limb individually and on a global average. This is due to the radiative feedback of the cloud deck on the thermal structure, as the cloud greenhouse effect warms the underlying atmosphere, causing the cloud deck to move to lower pressures in cases with a strong cloud-radiative feedback. However, the cloud extinction opacity is much smaller than the gas opacity at low pressures $\ll~1~\mathrm{mbar}$. Near the base of the cloud deck, the cloud extinction opacity on the nightside and western limb can be larger than the infrared gas opacity alone. As we discuss in \Sec{sec:gcmcrt}, this implies that clouds have a minor impact on transmission spectroscopy but a significant impact on nightside emission and phase curve observations in the cases considered here.

\subsubsection{Vertical mixing}
\label{sec:mixeff}
\begin{figure*}
    \centering
    \includegraphics[width=0.324\textwidth]{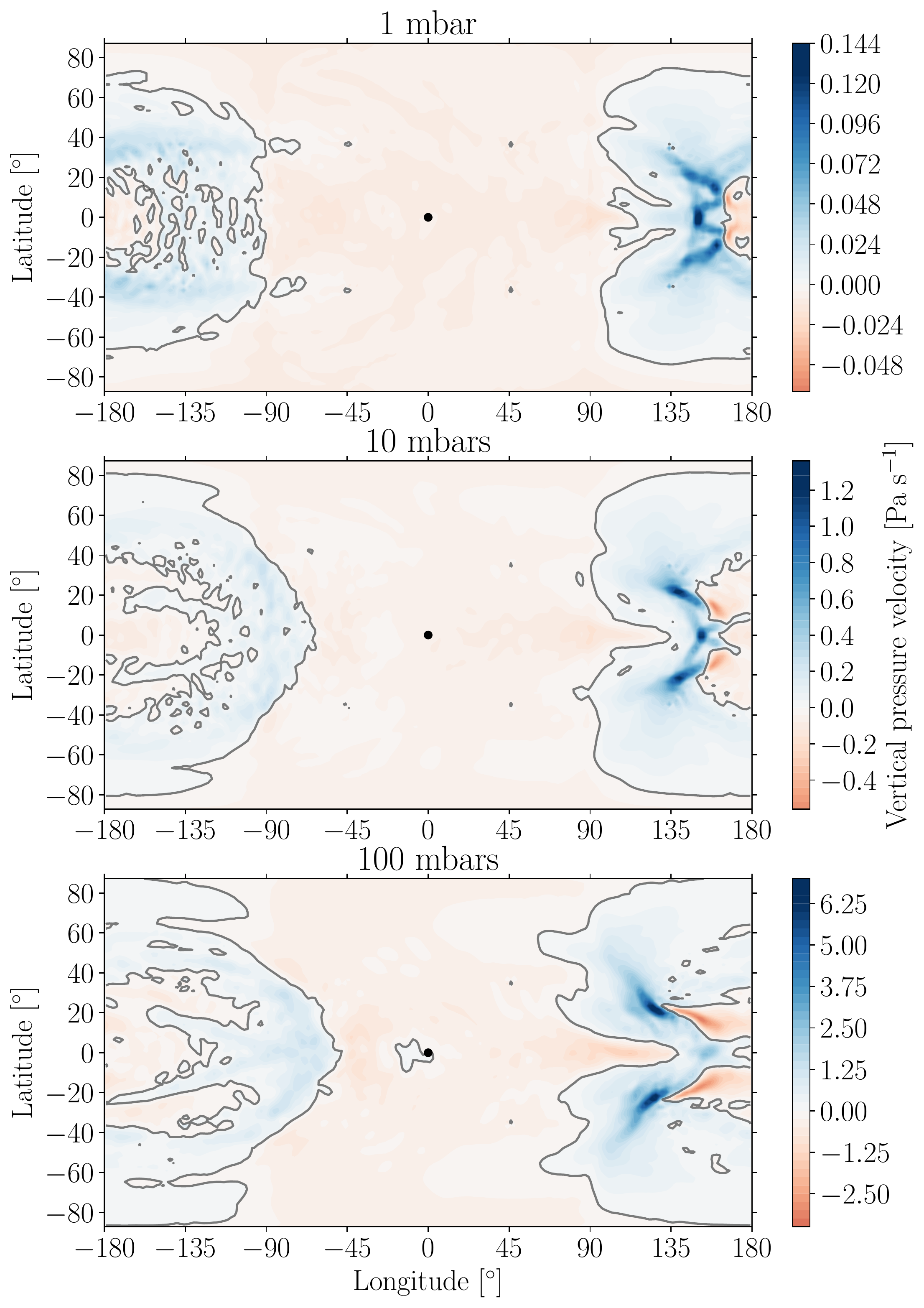}
    \includegraphics[width=0.32\textwidth]{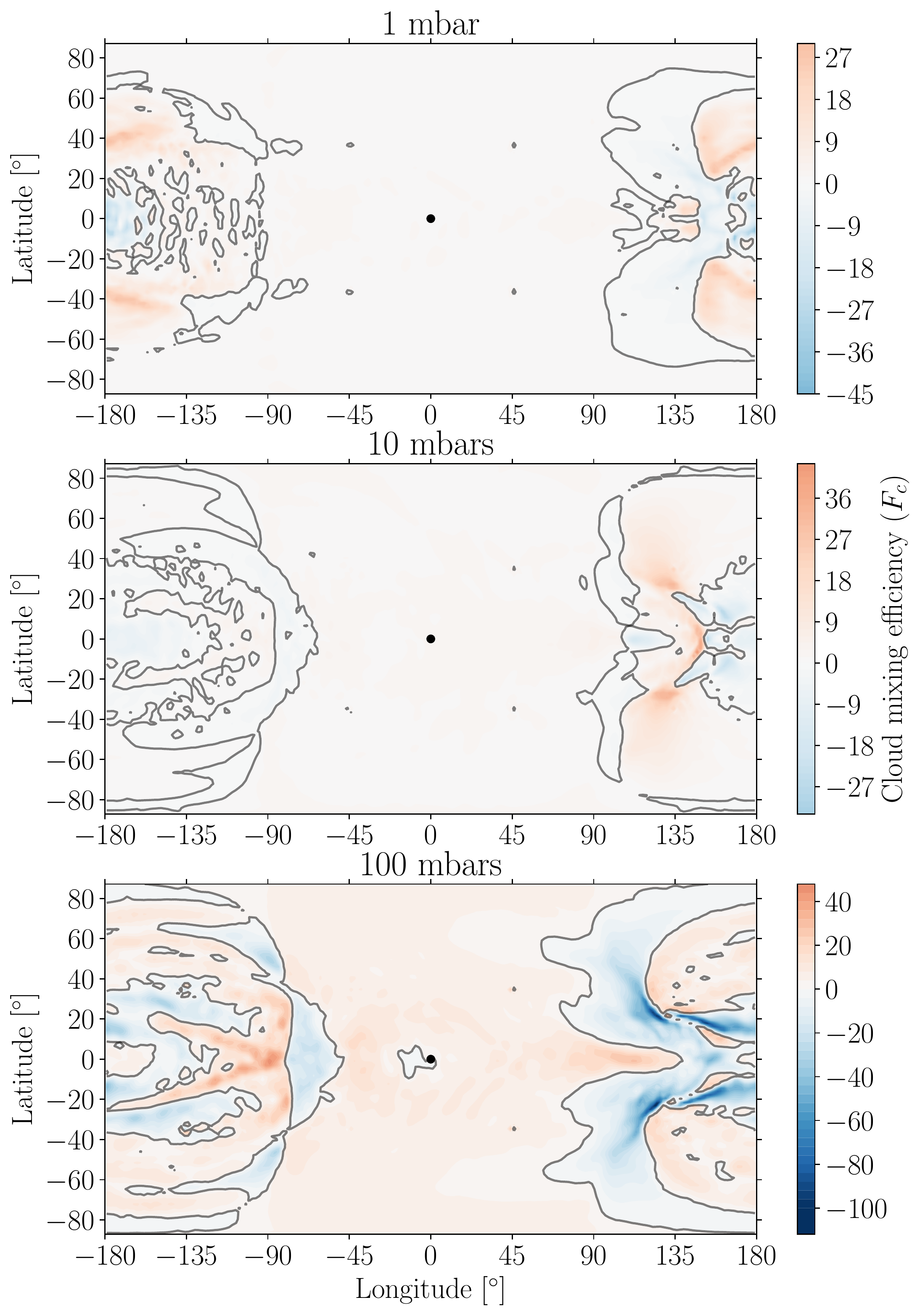}
    \includegraphics[width=0.32\textwidth]{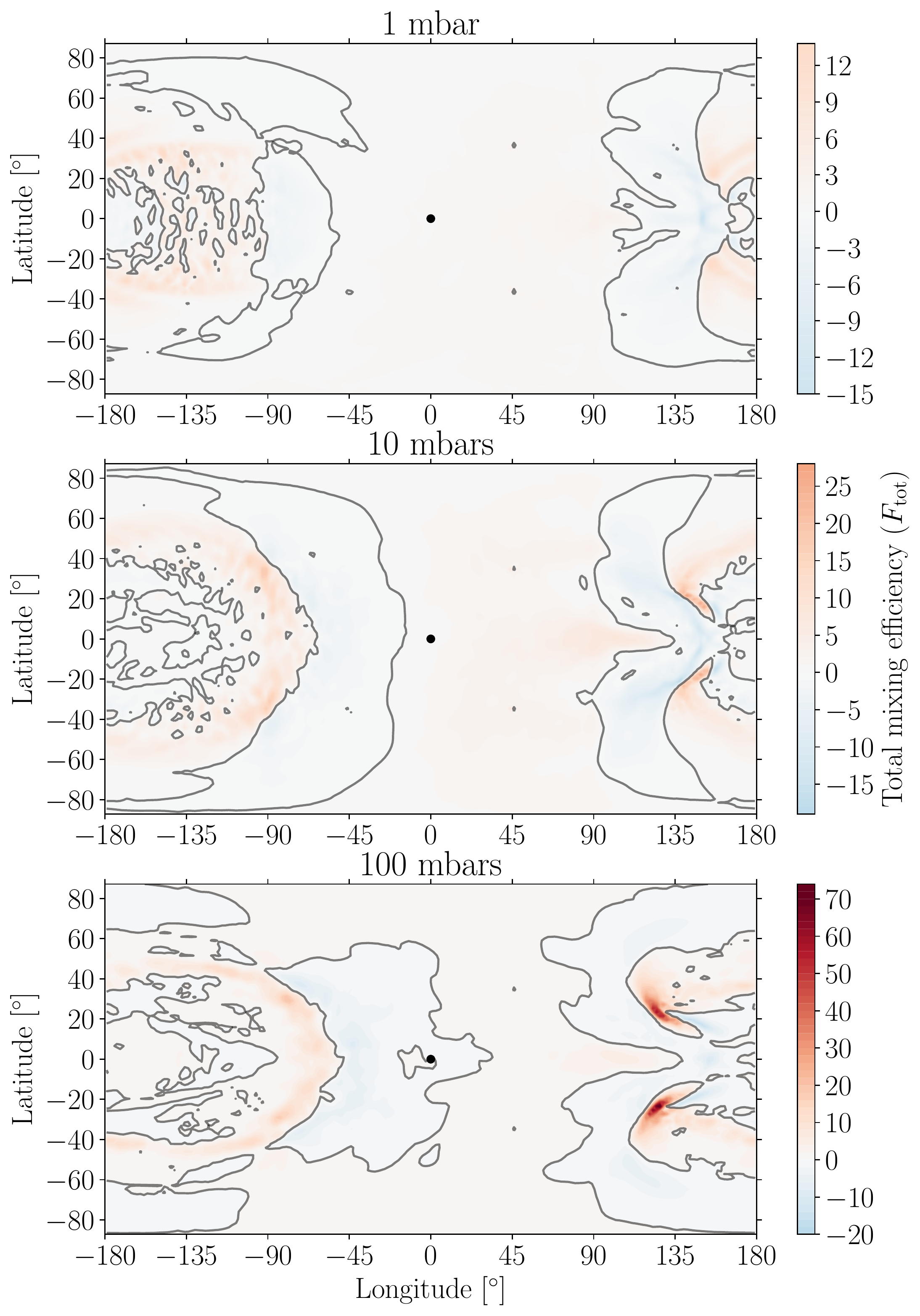}
    \caption{Contour maps of the vertical pressure velocity (left), cloud tracer mixing efficiency (middle), and total (cloud plus vapor) mixing efficiency (right) on the 1, 10, and 100 mbar isobars from the baseline simulation. The solid gray contour denotes the zero vertical velocity or mixing efficiency lines. All panels in each respective column share a color scale. Note that a negative vertical pressure velocity corresponds to upwelling motions and a positive vertical pressure velocity corresponds to downwelling. The mixing efficiency is defined in \Eq{eq:mixeff}, where $q = q_c$ for the cloud mixing efficiency (middle panels) and $q = q_c + q_v$ for the total mixing efficiency (right-hand panels). Positive mixing efficiency corresponds to local net upward tracer transport, while negative mixing efficiency corresponds to downward net tracer transport. As a result, in both columns red corresponds to upward motion or mixing while blue corresponds to downward motion or mixing. The strong spatial variations in vertical mixing on the nightside promote the patchy nature of nightside cloud coverage.}
    \label{fig:mixeffisobars}
\end{figure*}

To study how the vertical mixing of cloud tracer sets the patchy cloud distribution, we calculate the vertical mixing efficiency from our suite of models. The mixing efficiency is defined as \citep{parmentier_2013}:
\begin{equation}
\label{eq:mixeff}
F = \frac{\omega(q-\left<q\right>)}{\left<\omega q \right>} \mathrm{.}
\end{equation}
In \Eq{eq:mixeff}, $\omega$ is the vertical velocity in pressure coordinates ($\omega=dp/dt$), $q$ is the tracer mass mixing ratio (of cloud condensate and/or condensible vapor), and brackets represent the mean on an isobar. Note that the actual mixing efficiency could be higher than that estimated in our GCM framework due to turbulence generated by shear instabilities \citep{Menou:2019aa,Menou:2021vh}. The mixing efficiency encapsulates a key property of tracer mixing: net vertical transport of tracer across isobars can only occur at locations where there is a correlation between the tracer abundance and the vertical velocity (for a schematic depiction, see Figure 2 of \citealp{Zhang:2018tp}). In \Eq{eq:mixeff}, the numerator $\omega(q-\left<q\right>)$ represents this correlation between vertical velocity and tracer mixing ratio, as a net upward vertical transport of tracer will occur if $\omega(q-\left<q\right>) < 0$, and a net downward vertical transport will occur if $\omega(q-\left<q\right>) > 0$ locally\footnote{Note the reversed sign from standard vertical velocity with dimensions of length rather than pressure.}. The denominator in \Eq{eq:mixeff} ($\left<\omega q \right>$) is related to the mean upward flux of material across isobars, and normalizes the mixing efficiency $F$ such that it represents the local contribution to the total upward flux on an isobar. 

Positive mixing efficiency thus corresponds to net upward tracer transport on an isobar, and negative mixing efficiency corresponds to net downward tracer transport on an isobar. Importantly, local positive mixing efficiency can correspond to one of two possibilities: upward winds are transporting air enhanced in tracer across isobars \textit{or} downward winds are transporting air depleted in tracer across isobars. Both of these scenarios create the correlation between vertical velocity and tracer mixing ratio required for net upward tracer transport. For net downward tracer transport, which corresponds to negative mixing efficiency, the opposite is true: either downward winds are transporting air enhanced in tracer across isobars, or upward winds are transporting air depleted in tracer across isobars. 

 We define two separate mixing efficiencies: a ``cloud mixing efficiency'' ($F_c$) for the cloud condensate alone, with $q = q_c$ in \Eq{eq:mixeff}, and a ``total mixing efficiency'' ($F_\mathrm{tot}$) for the total tracer including both cloud condensate and condensible vapor, with $q = q_c + q_v$ in \Eq{eq:mixeff}. We do so because though the cloud mixing efficiency represents the vertical mixing of cloud species, this is affected by the inhomogeneous temperature structure, most notably the large day-to-night contrast that prevents cloud condensation throughout much of the dayside. Meanwhile, the total mixing efficiency more closely displays the direct effect of transport on tracer mixing. 

The mixing efficiencies at pressure levels of 1 mbar, 10 mbars, and 100 mbars from our baseline case with radiatively active cloud tracers is shown in \Fig{fig:mixeffisobars} along with vertical pressure velocity contours on the same isobars. Additionally, the vertical velocities and mixing efficiencies at the 100 mbar isobar from cases with varying cloud microphysical and radiative assumptions are further shown in Appendix \ref{sec:appendixmix}.  Note that the mixing efficiency is not undefined in any of these locations, given that all pressures considered have at least a minuscule amount of cloud and condensible vapor tracer present in the numerical scheme. We find that there is strong spatial inhomogeneity in the pattern of both the cloud and total mixing efficiency on the nightside and limbs, which drives the emergent patchy cloud behavior. At the equator, local changes in mixing efficiency generally correspond to changes in the speed of the equatorial jet, which drives convergence or divergence that causes downwelling or upwelling, respectively \citep{parmentier_2013}. The eastward flow near the equator causes a horizontal asymmetry in the background vapor (with more condensible vapor eastward of the substellar point, see \Fig{fig:cloudbaseline}), which at pressures of 10-100 mbars causes net total tracer lofting (i.e., positive total mixing efficiency) eastward of the substellar point and net tracer settling westward of the substellar point. One specific region with strong variations in both cloud and total mixing efficiency is the descending region eastward of the eastern terminator where the equatorial jet slows, which has been shown to lead to local adiabatic warming in previous simulations of ultra-hot Jupiters \citep{Beltz:2022aa}. The downward vertical velocity in this region and resulting negative mixing efficiency at 100 mbars transports cloud condensate downward and causes a local minimum in cloud mixing ratio on the nightside that at 100 mbars can extend eastward of the anti-stellar point. 

\Fig{fig:mixeffisobars} shows that along with the variations in mixing along the equator, there are also significant changes in vertical pressure velocity and mixing efficiency with latitude and pressure. These changes are most noticeable on the mid-latitude eastern nightside at latitudes near $40^\circ$ where the flank of the superrotating equatorial jet converges with the nightside high-latitude cyclonic flow. This convergence causes downwelling motion that leads to a negative cloud mixing efficiency in regions where cloud condensate is enhanced and a positive cloud mixing efficiency where cloud condensate is depleted. As a result, the equatorward flank of this region corresponds to negative cloud mixing efficiency and the poleward flank has a positive cloud mixing efficiency at the 100 mbar level. This leads to an enhancement of cloud tracers at higher latitudes and a local depletion in the mid-latitudes. Importantly, this local depletion occurs where the atmosphere is coldest, and the corresponding peak in cloud condensate mass mixing ratio occurs at higher latitudes than the coldest region. As a result, one-dimensional models or three-dimensional dynamical models with equilibrium cloud condensation schemes but without cloud tracers would predict the atmosphere to be cloudy at the cold mid-latitude regions where we find a local minimum in cloud condensate mass mixing ratio. In general, we find that three-dimensional mixing processes set the spatially inhomogeneous cloud condensate and condensible vapor tracer distribution in our GCM simulations.




\subsubsection{Three-dimensional temperature structure with varying cloud assumptions}
\label{sec:temp}

\begin{figure}
    \centering
    \includegraphics[width=0.45\textwidth]{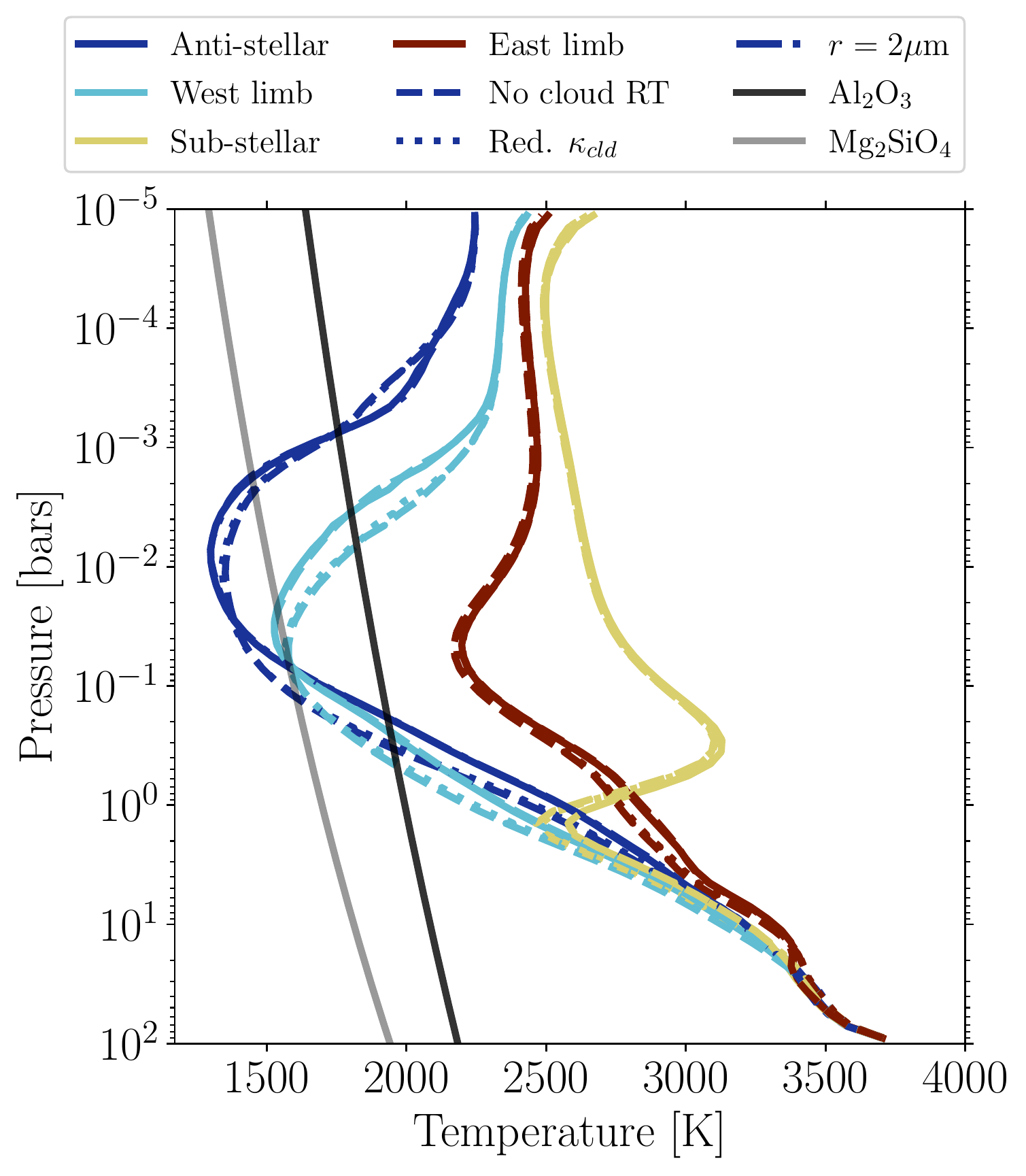}
    \caption{Meridional-mean temperature-pressure profiles from GCM simulations with varying cloud assumptions. Condensation curves for corundum \citep{Wakeford:2017} and forsterite \citep{Visscher10} given Solar metallicity are over-plotted. We find that varying cloud assumptions impact the three-dimensional temperature structure, and this effect is largest on the nightside and western limb where corundum cloud condensation can occur. Most notably, cases with significant cloud opacity (baseline and $r = 2~\mu\mathrm{m}$) have hotter nightside atmospheres below the cloud deck than cases with weak or no cloud opacity (No cloud RT, Red. $\kappa_\mathrm{cld}$) due to the cloud greenhouse effect.}
    \label{fig:tp}
\end{figure}

The dynamic interaction between the atmospheric circulation, clouds, and radiation sets the three-dimensional temperature structure in our GCMs. \Fig{fig:tp} shows profiles of the meridional-mean of temperature as a function of pressure at the anti-stellar point, west limb, sub-stellar point, and east limb from the cases with varying cloud microphysical and radiative parameters. In all cases, there is a significant change in the temperature profile from day to night, due to the large day-night temperature contrasts at pressures $\lesssim 1~\mathrm{bar}$. The eastern limb is everywhere hotter than the western limb due to the eastward advection of warm dayside air by the superrotating equatorial jet. Note that the temperature contrasts between the limbs decrease with decreasing pressure as the flow transitions from an eastward equatorial jet at depth to day-night flow at lower pressures. We also find a thermal inversion at low pressures on the anti-stellar point and west limb in all cases. This thermal inversion causes the conversion of cloud condensate to vapor at low pressures, setting the cloud top pressure at the anti-stellar point and west limb shown in \Fig{fig:cloudp}. In \Sec{sec:Hdisc}, we isolate the thermodynamic impact of hydrogen dissociation and recombination to demonstrate that this thermal inversion is generated by the heat release from recombination of atomic to molecular hydrogen on the nightside and limb. 


\begin{figure}
    \centering
    \includegraphics[width=0.45\textwidth]{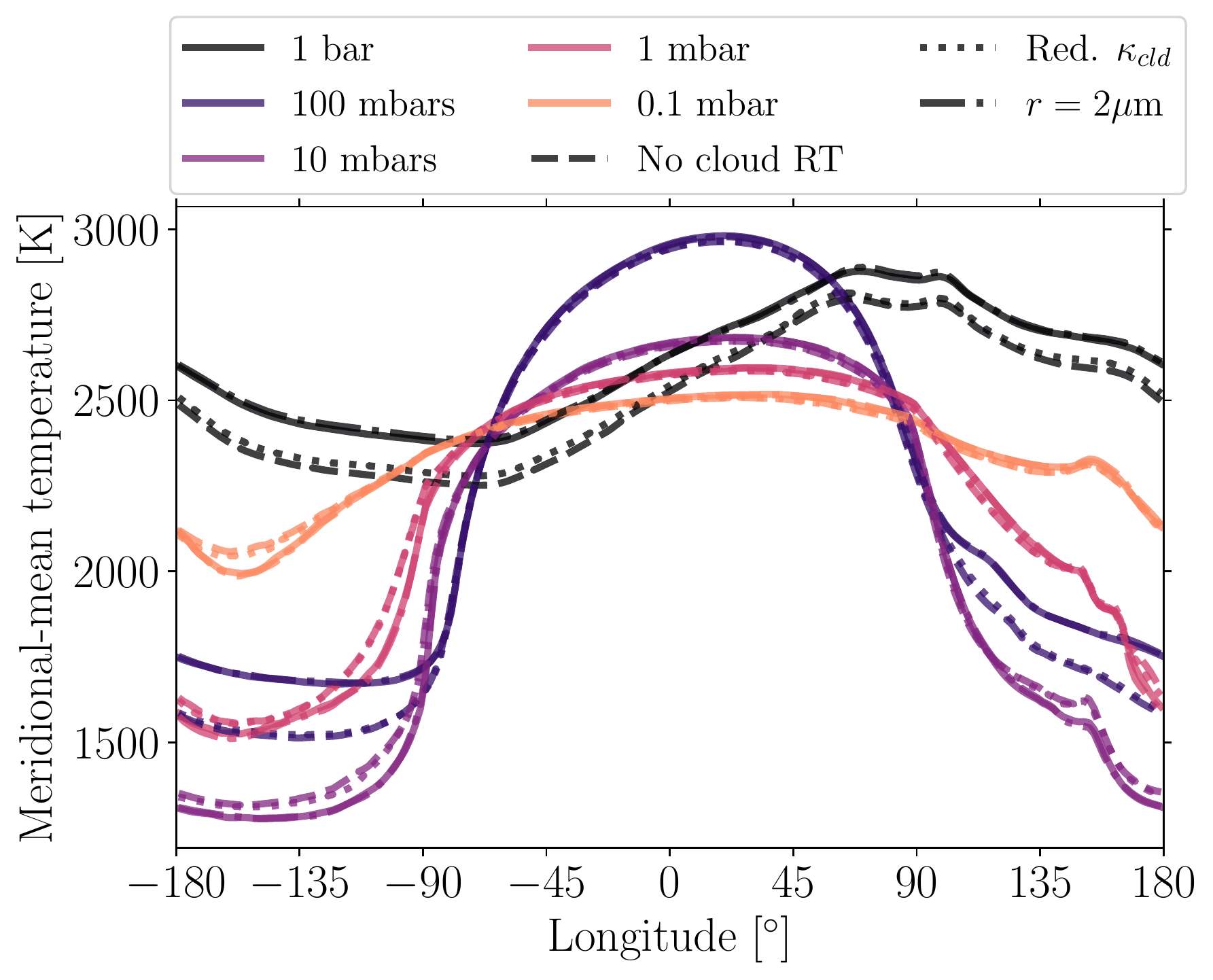}
    \caption{Meridional mean of temperature as a function of longitude at different pressures from GCM simulations with varying cloud assumptions. We find that the largest differences between cases with large and small or non-existent cloud opacity occur at pressures $\gtrsim 100~\mathrm{mbars}$, due to a cloud greenhouse effect that warms the atmosphere below the nominal cloud deck.}
    \label{fig:templong}
\end{figure}

We find significant differences in the three-dimensional temperature structure between cases with different cloud radiative assumptions. The deep atmosphere at pressures $\gtrsim 100~\mathrm{mbars}$ is warmer in the cases with a strong cloud-radiative feedback (the baseline and reduced cloud particle size cases) than in the cases with a weak or zero cloud-radiative feedback (the reduced cloud opacity and no cloud radiative feedback cases). \Fig{fig:templong} shows the meridional-mean temperature as a function of longitude on isobars from 1 bar to 0.1 mbar for the four cases with varying cloud microphysical and radiative parameters. In cases with cloud-radiative feedback, the reduced outgoing longwave radiation from the colder cloud tops relative to the deeper underlying atmosphere causes a cloud greenhouse effect that warms the underlying air. This cloud greenhouse effect is strongest on the nightside within the cloud deck (shown in the 100 mbar isobar on \Fig{fig:templong}), and further causes a global increase in temperature below the cloud deck (see the 1 bar level in \Fig{fig:templong}). Conversely, the cloud greenhouse effect leads to a slight cooling of the nightside atmosphere above the bulk of the cloud deck in cases with strong cloud-radiative feedback, as shown at the 10 and 1 mbar levels in \Fig{fig:templong}. The warming of the deep atmosphere in cases with strong cloud-radiative feedback inhibits cloud condensation at depth, causing the cloud deck to move upward in cases with strong cloud-radiative feedback as shown in \Fig{fig:cloudp}. 

\subsection{Emergent spectra and phase curves}
\label{sec:gcmcrt}
We post-process our GCM simulations with the state-of-the-art \texttt{gCMCRT} code to predict the effect of cloud microphysical and optical properties on the observable properties of ultra-hot Jupiters. \texttt{gCMCRT} is a publicly available\footnote{\url{https://github.com/ELeeAstro}} hybrid Monte Carlo Radiative Transfer (MCRT) and raytracing radiative transfer code. The model is described in detail in \cite{Lee:2021uv}, and builds upon the MCRT code developed by \cite{Hood:2008aa} to study a range of exoplanet atmospheres \citep{Lee:2017aa,Lee:2019aa}. \texttt{gCMCRT} can natively compute albedo, transmission, and emission spectra at both low and high spectral resolution. Due to its ray tracing capabilities, \texttt{gCMCRT} takes into account the three-dimensional nature of transmission spectra, given that heterogenities across the limb are expected to impact transit observations of ultra-hot Jupiters \citep{caldas2019,Pluriel:2020aa,Wardenier:2022aa}. \texttt{gCMCRT} uses custom k-tables which take cross-section data from both HELIOS-K \citep{Grimm:2021aa} and EXOPLINES \citep{Gharib-Nezhad:2021aa}. 

In this work, we apply \texttt{gCMCRT} to compute low-resolution emission spectra, phase curves, and transmission spectra at $R \approx 100$ from our GCM simulations. We use the three-dimensional temperature, atomic hydrogen and condensate cloud tracer mixing ratio, and specific gas constant from the time-averaged end-state of each case. We assume the same cloud particle size distribution as our GCM (see Equation \ref{eq:cldparsize}). We discuss the effect of cloud assumptions in the GCM on the phase-dependent emission spectra in \Sec{sec:emission}, predicted phase curves in \Sec{sec:phasecurve}, and resulting transmission spectra in \Sec{sec:transmission}.

\subsubsection{Phase-dependent emission spectra}
\label{sec:emission}
\begin{figure*}
    \centering
    \includegraphics[width=0.5\textwidth]{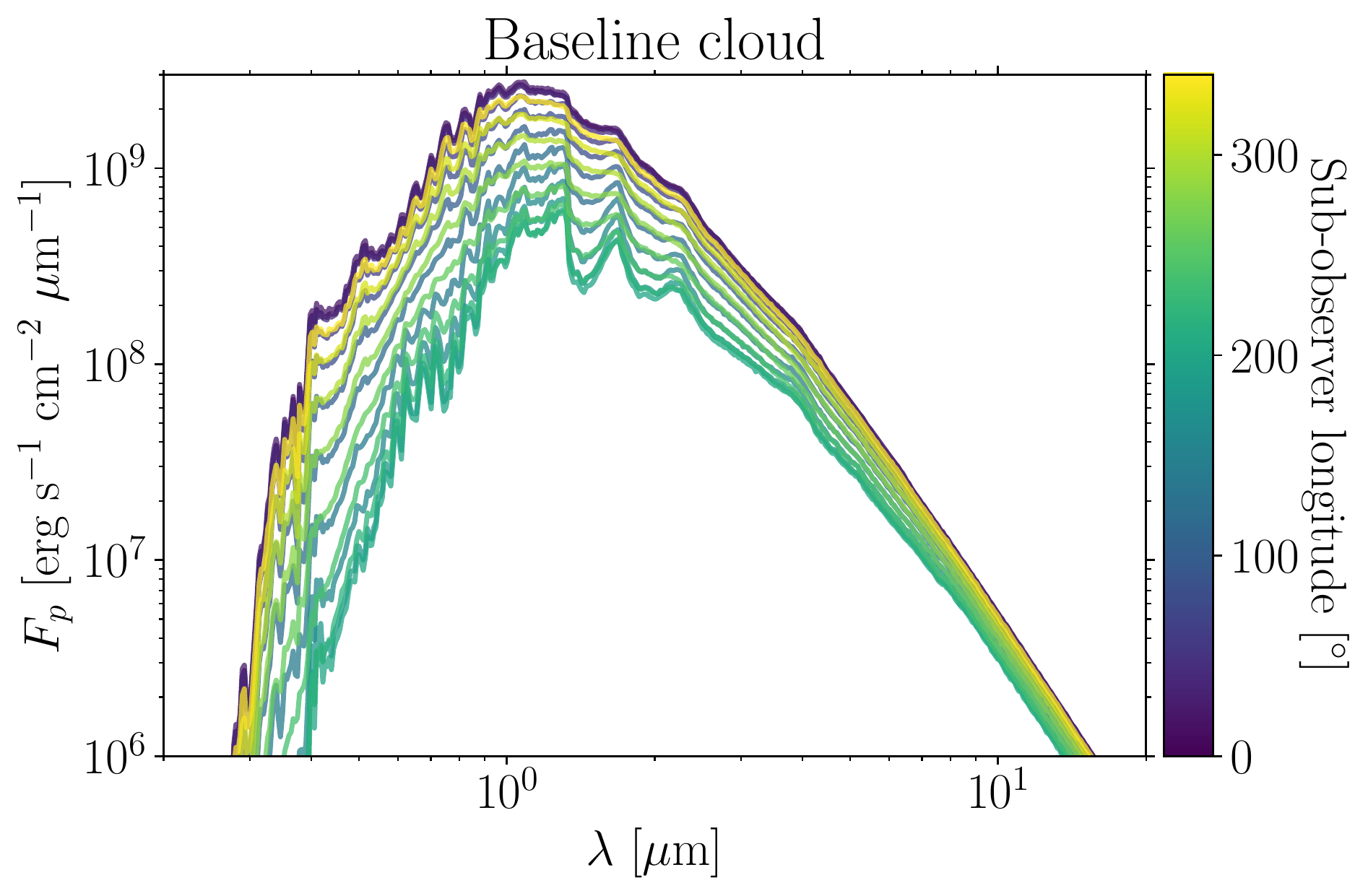}
    \includegraphics[width=0.45\textwidth]{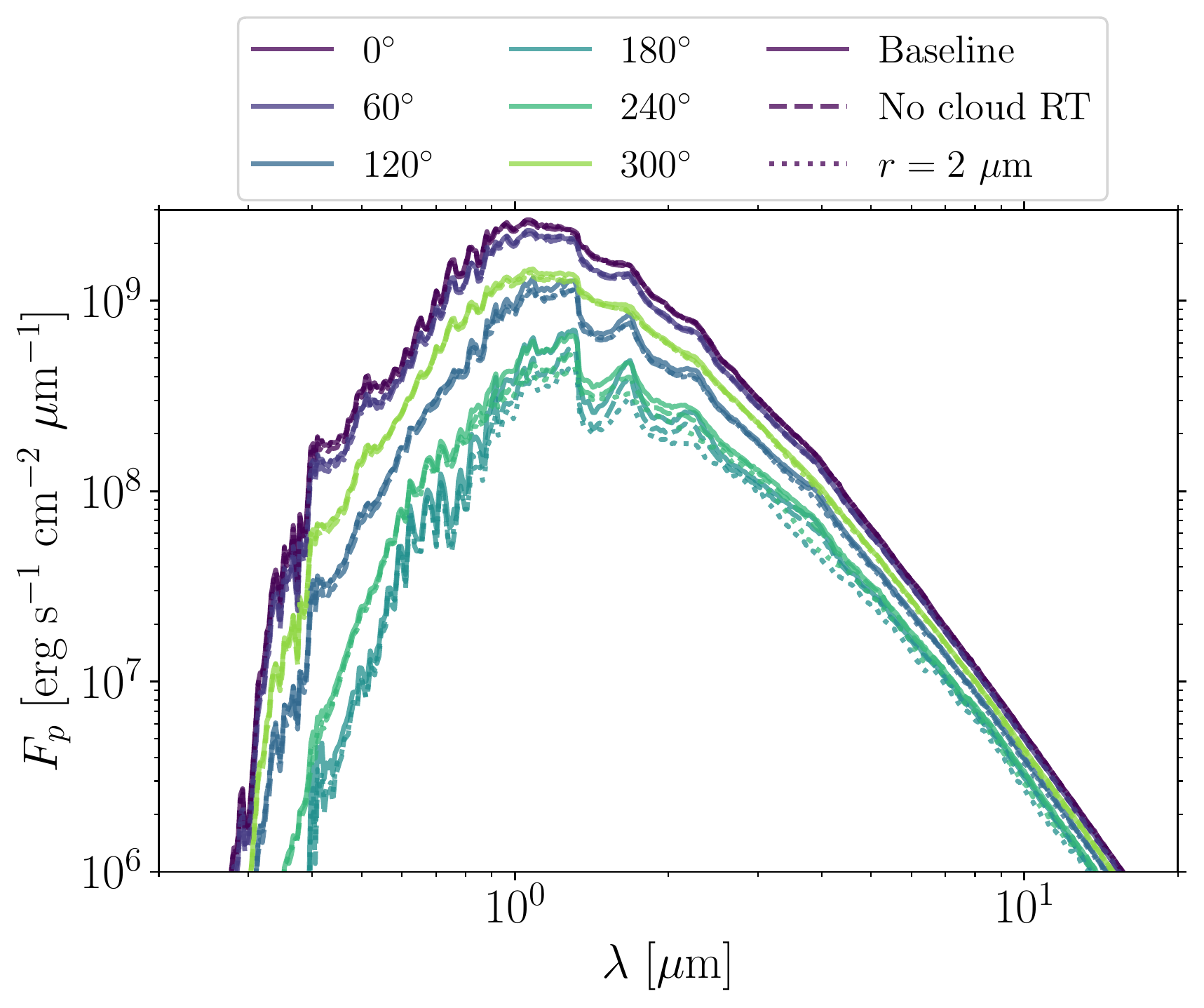}
    \caption{Left: Planetary emergent spectra as a function of sub-observer longitude from the baseline case with active cloud tracers. Right: Comparison of the emergent spectra at six sub-observer longitudes between cases with different cloud assumptions. A sub-observer longitude of $180^\circ$ corresponds to transit and $0^\circ$ corresponds to secondary eclipse. Clouds have largest impact on the top-of-atmosphere flux on the nightside in the near infrared.}
    \label{fig:emergentphase}
\end{figure*}

The planetary emergent spectrum in all of our simulations with cloud tracers is strongly phase-dependent and longitudinally asymmetric due to east-west variations in the partial cloud coverage. \Fig{fig:emergentphase} shows the planetary emission spectrum as a function of sub-observer longitude from our baseline case and for a subset of sub-observer longitudes from cases with varying cloud microphysical and radiative properties. We do not show results from the reduced $\kappa_\mathrm{cld}$ case because the behavior of the emergent spectra is similar to the case without cloud-radiative feedback in the GCM. There is a significant east-west asymmetry in the planetary emission spectrum in all cases, with the eastern dayside and limb appearing brighter than the western dayside and limb. This difference is largest near the peak of the planetary Planck spectrum in the near-infrared. This is expected from previous GCMs of hot Jupiters (e.g., \citealp{Showmanetal_2009,Heng:2011a,Rauscher_2012,Dobbs-Dixon:2013,Mayne:2014,kata16}) and occurs because of the strong longitudinal temperature asymmetry driven by the confluence of the planetary-scale wave pattern and superrotating equatorial jet. 

In tandem with the longitudinal asymmetry in flux, there is also a longitudinal asymmetry in the depth of absorption features in the planetary spectrum. In all cases, sub-observer longitudes where the planetary disk is centered on the eastern hemisphere have greater absorption feature depths than those centered on the western hemisphere (e.g., compare the $60^\circ$ and $300^\circ$ or $120^\circ$ and $240^\circ$ lines in the right-hand panel). The lapse rate at the near-infrared photosphere is larger on the western hemisphere (see Figure \ref{fig:tp}), which would imply stronger absorption features on the cooler western limb due to the larger vertical temperature contrast. However, the cloud deck extends onto the dayside on the western hemisphere, with a global maximum in cloud mass on the nightside near the western limb (see Figure \ref{fig:cloudlong}). The patchy cloud deck reduces the temperature of the continuum level in emission, resulting in reduced absorption features on the western hemisphere. In general, the combination of temperature structure and cloud coverage acts to shape the emergent spectra in our simulations with cloud tracers. 

\begin{figure}
    \centering
    \includegraphics[width=0.45\textwidth]{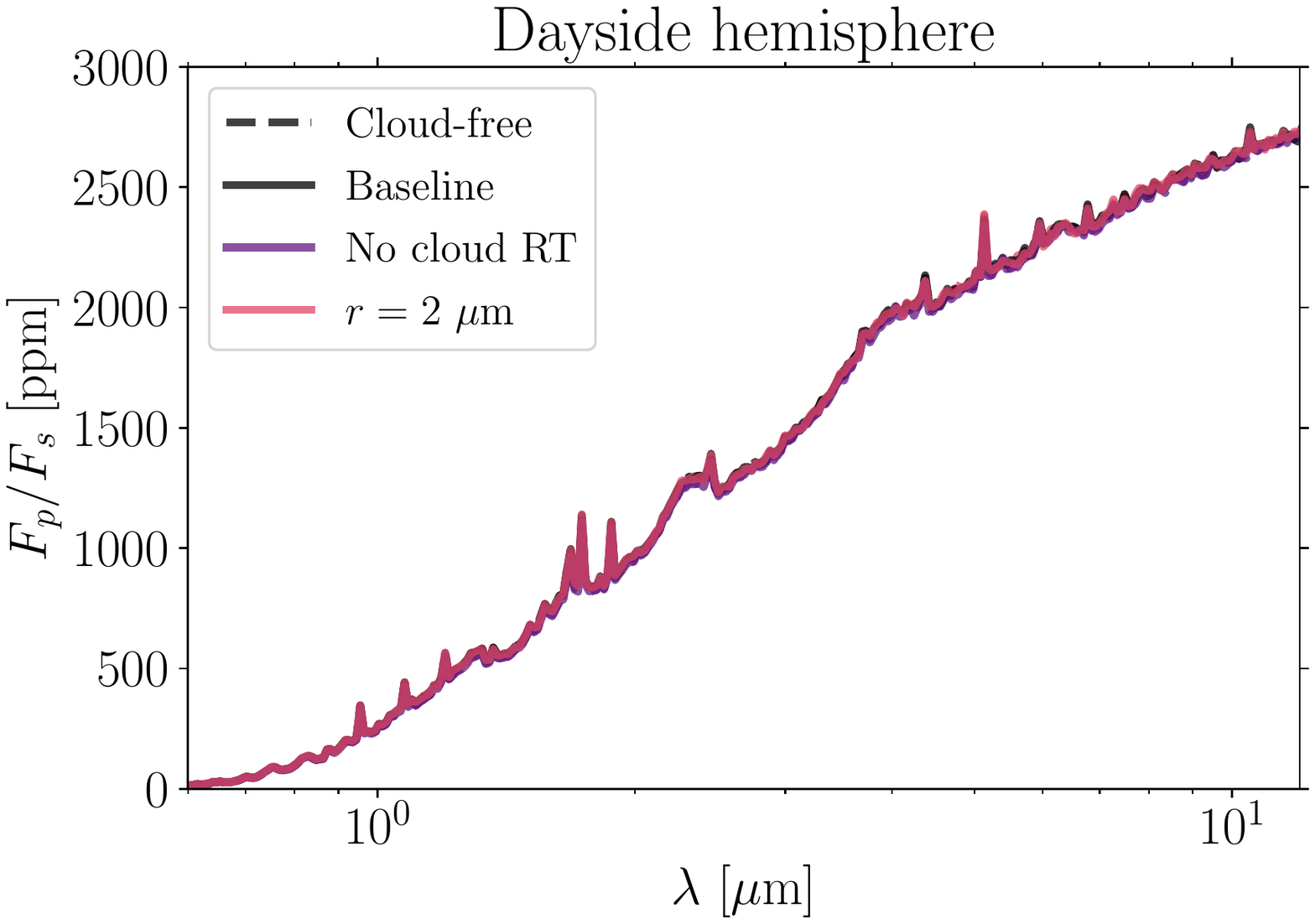}
    \includegraphics[width=0.45\textwidth]{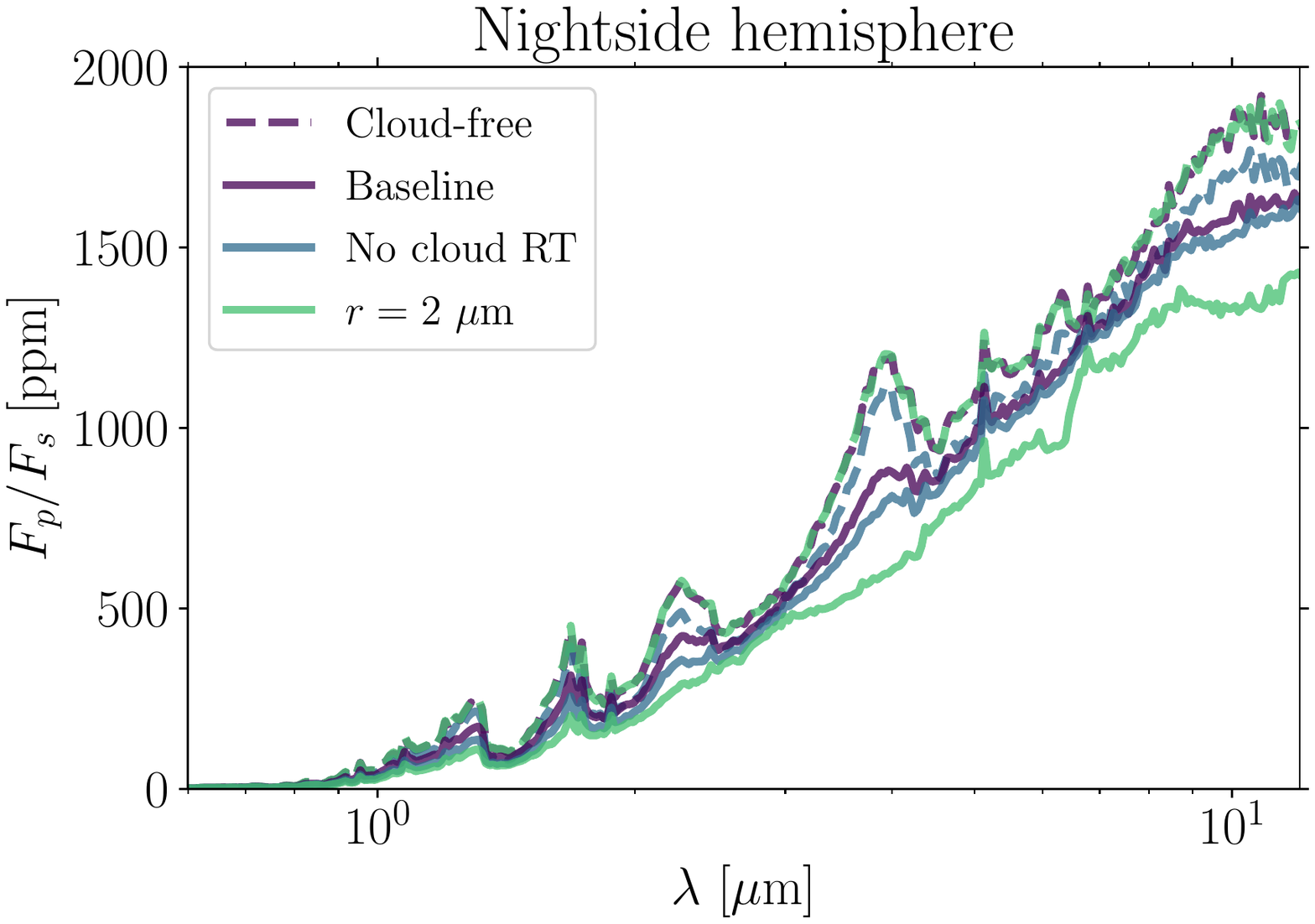}
    \caption{Simulated emission spectra for the dayside hemisphere (i.e., at secondary eclipse, top) and the nightside hemisphere (i.e., at transit, bottom) for cases with different cloud assumptions. Predictions that do not include clouds in the \texttt{gCMCRT} post-processing are shown with dashed lines, while cases that include clouds in both the GCM and post-processing are shown by solid lines. The dayside emission spectrum is unaffected by cloud assumptions, given the small amount of clouds on the dayside in our GCM simulations. The nightside emission spectrum is strongly affected by the inclusion of clouds, and in cases with clouds it is both affected by including significant cloud-radiative feedback in the GCM and varying the cloud particle size. The cases that do not include clouds in the post-processing have a higher emitted nightside flux than cloudy cases as emission from hotter regions at depth can more readily escape to space. The baseline case with radiatively active clouds has a higher nightside flux in the near- and mid-infrared than the reduced cloud opacity and no cloud radiative feedback cases. This is due to the patchy nightside cloud greenhouse effect warming the atmosphere below the cloud deck, which in turn increases the planetary longwave radiation that escapes to space in less cloudy regions near the anti-stellar point. }
    \label{fig:fpfs_daynight}
\end{figure}

Though the dayside spectra are similar for our cases with varying cloud assumptions, the nightside emergent spectrum is dependent on both the strength of cloud-radiative feedback and cloud particle size. \Fig{fig:fpfs_daynight} shows the dayside and nightside emission spectra from our \texttt{gCMCRT} calculations, assuming a PHOENIX stellar spectrum \citep{Allard2012} appropriate for TOI-1431 (T$_{\rm eff}$ = 7690 K, logg = 4.15, [Fe/H] = 0.43, \citet{Addison:2021aa}), interpolated using the pysnphot package \citep{pysnphot}. The dayside emission spectra are independent of the inclusion of clouds in the post-processing along with cloud microphysical and radiative assumptions, given that clouds only cover a thin slice of the dayside hemisphere near the western limb. 

For nightside emission spectra shown in \Fig{fig:fpfs_daynight}, cloud assumptions cause significant differences in the planet-to-star flux ratio in the near-to-mid infrared. Notably, the planet-to-star flux ratio on the nightside is larger in the baseline simulation with strongly radiatively active cloud tracers than in those with weak or zero cloud-radiative feedback. 
Interestingly, this increase in planet-to-star flux ratio with increasing cloud-radiative feedback is opposite to the effect that cloud-radiative feedback is expected to have on the nightside spectrum of cooler hot Jupiters, where the cold temperature of the cloud top causes the outgoing longwave radiation on the nightside to decrease \citep{Gao:2021vp}. As a result, we find a change in the sign of the impact of the nightside cloud greenhouse on the flux emerging from the nightside hemisphere between the regime of hot to that of ultra-hot Jupiters: for hot Jupiters, the cloud greenhouse leads to a decrease in the planetary nightside thermal emission, while in our ultra-hot Jupiter simulations the cloud greenhouse leads to an increase in the outgoing thermal flux from the nightside hemisphere. This change in the behavior of the nightside flux with cloud-radiative feedback occurs because the cloud tracer coverage in our GCMs is patchy rather than uniform, enabling radiation to escape to space from the warmer deep levels in cloud-free regions.  

Note that when we do not include clouds in the \texttt{gCMCRT} post-processing, the nightside emitted flux shown in the bottom panel of \Fig{fig:fpfs_daynight} is larger due to the increased photospheric pressure probing hotter layers. This increase in nightside planet-to-star flux ratio is especially large in continuum regions with reduced gas opacity. The increase in nightside flux when removing clouds from the post-processing implies that the change in the sign of the nightside cloud greenhouse effect is not due to clouds pushing the photosphere to lower pressures. Rather, because the cloud deck is sequestered beneath the thermal inversion layer the patchy cloud deck allows for enhanced emission in relatively cloud-free regions, analogous to the increased thermal emission from Jupiter's dry $5~\mu\mathrm{m}$ hot spots \citep{Seiff98,Showman:2000,depater05}.



\subsubsection{Phase curves}
\label{sec:phasecurve}

\begin{figure}
    \centering
    \includegraphics[width=0.45\textwidth]{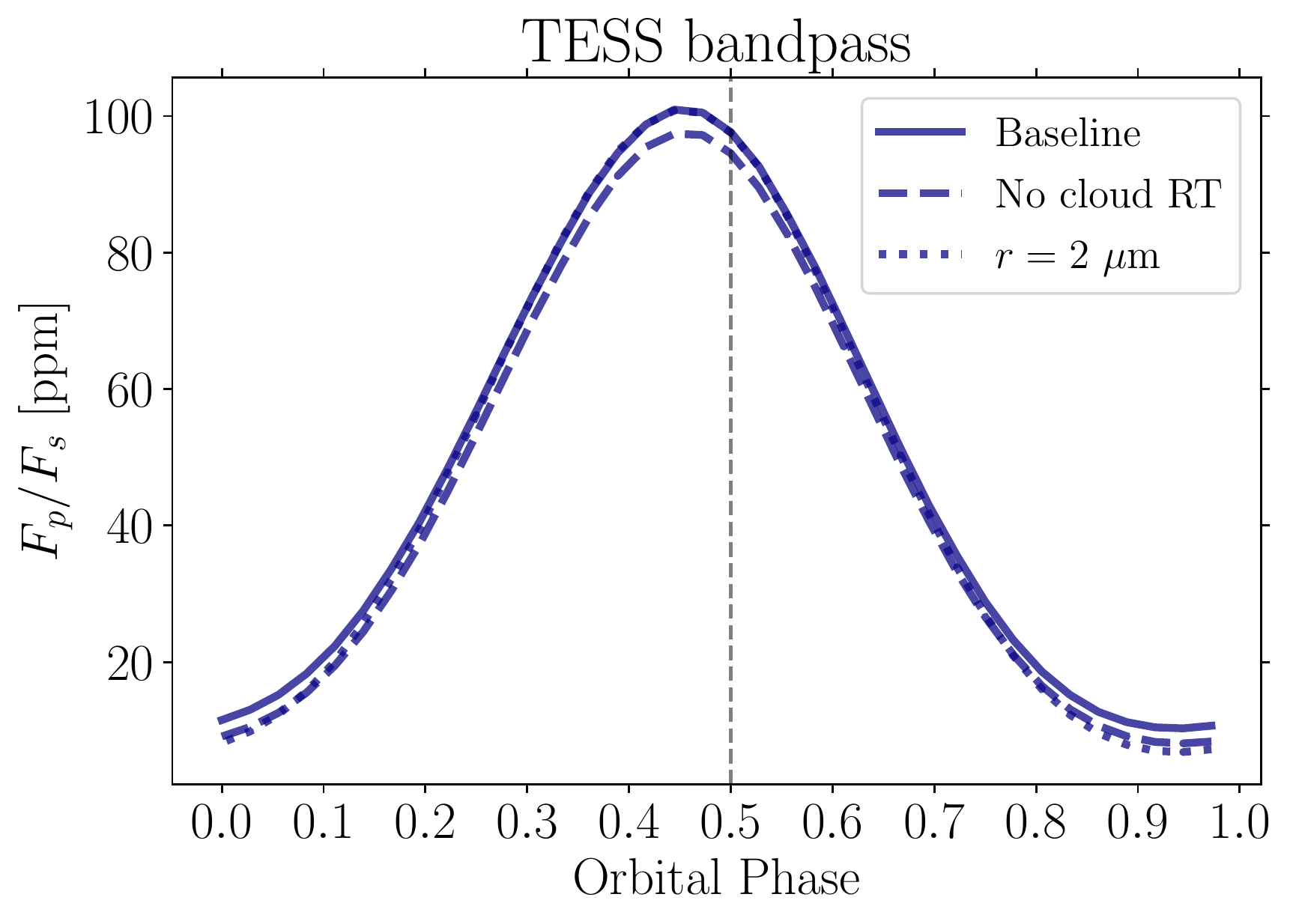}
    \includegraphics[width=0.455\textwidth]{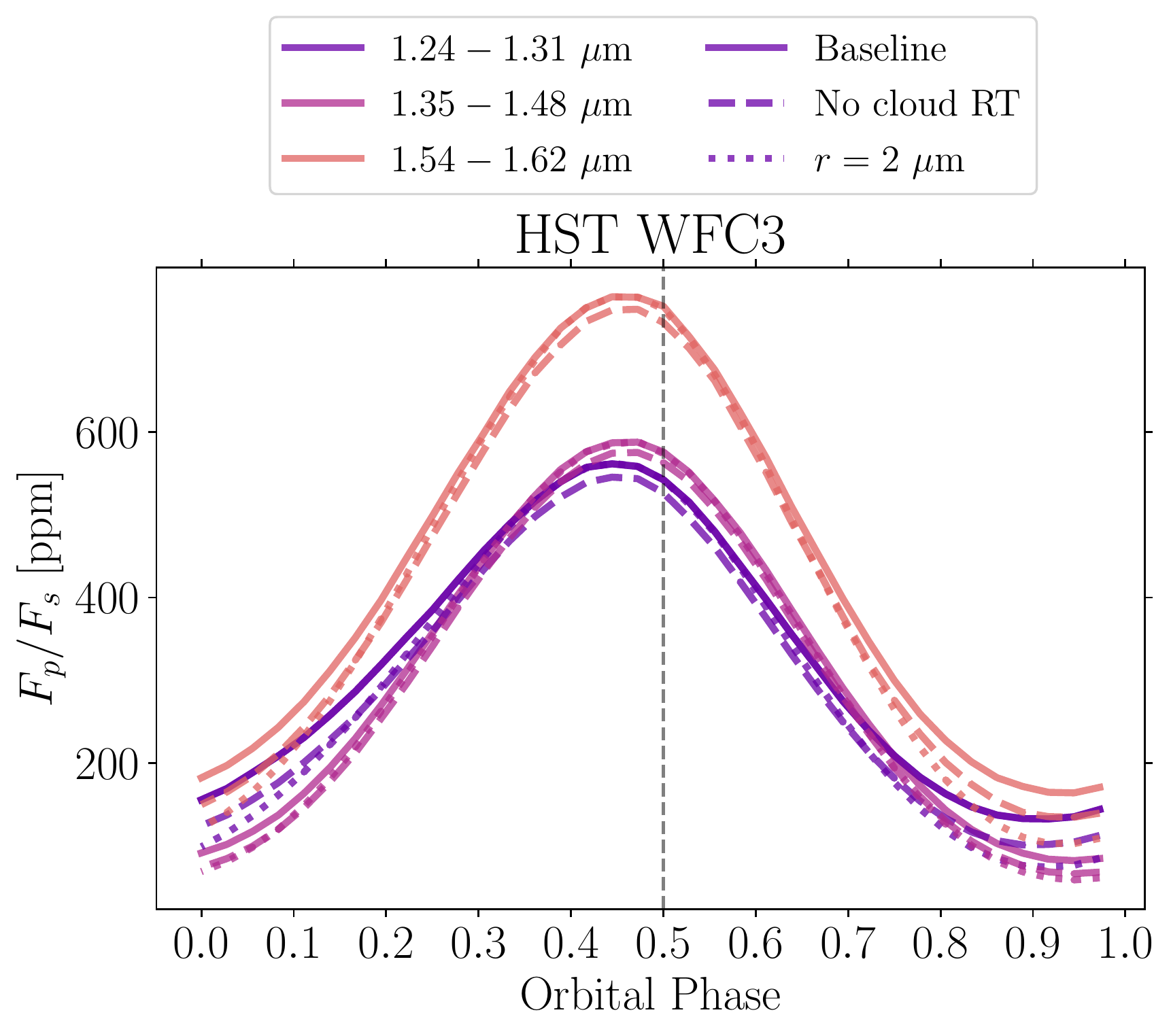}
    \caption{Simulated phase curves for varying cloud assumptions. The top panel shows the phase curves in the TESS bandpass, while the bottom panel shows phase curves at three different wavelength bands within the HST WFC3 bandpass. Secondary eclipse at an orbital phase of 0.5 is shown by the vertical dashed line. We find that the inclusion of significant cloud-radiative feedback in the GCM increases the peak flux, and that cases with a smaller characteristic cloud particle size have a reduced minimum level of nightside flux.}
    \label{fig:phasecurves}
\end{figure}

The patchy cloud greenhouse effect can potentially impact visible and near-infrared phase curves of ultra-hot Jupiters. \Fig{fig:phasecurves} shows phase curves including both planetary emission and reflected light due to clouds calculated with \texttt{gCMCRT} in the TESS and HST WFC3 bandpass from GCMs with varying cloud assumptions. All cases show an eastward phase curve offset and large phase curve amplitude, as is typical in GCMs of hot Jupiters \citep{Parmentier:2017}. There are only minor differences in the TESS bandpass, with a slight increase in dayside flux in cases with a stronger cloud greenhouse effect and a slight decrease in minimum flux in cases with a reduced cloud particle size. Due to the reduced settling velocity of the case with smaller characteristic cloud particle sizes, the greater cloud mass at low pressures reduces the nightside cloud top temperature in the case with a reduced particle size (see \Fig{fig:cldmassapp}).  

Note that none of the GCMs in our model grid can match the observed TESS phase curve of TOI-1431b, as the observations of \cite{Addison:2021aa} found a nightside planet-to-star flux ratio of $49^{+5}_{-4}~\mathrm{ppm}$ while our baseline simulation predicts a nightside planet-to-star flux ratio of just $10.7$ ppm. This may suggest that another process besides the cloud greenhouse effect and hydrogen dissociation and recombination are required to explain the hot nightside temperature of TOI-1431b (and the similar HAT-P-7b, \citealp{Bell:2021aa}). However, we caution that our GCMs are idealized in order to provide a framework to understand the processes that set the cloud distributions of ultra-hot Jupiters, and as a result more sophisticated non-gray GCMs may better match the observed phase curve (see \Sec{sec:limit}).  Additionally, spectrophotometric observations at longer wavelengths can provide additional context for the TESS phase curve observation of TOI-1431b. Notably, the predicted phase curves from our GCMs in the WFC3 bandpass show larger differences than in the TESS bandpass on the nightside hemisphere of up to tens of ppm between the cases with strong and weak cloud-radiative feedback. As for the nightside emergent flux shown above, this is due to the effect of the patchy cloud greenhouse warming the deep atmosphere and increasing the flux that escapes to space from the nightside hemisphere centered on the anti-stellar point. 


\subsubsection{Transmission spectra}
\label{sec:transmission}

\begin{figure}
    \centering
    \includegraphics[width=0.45\textwidth]{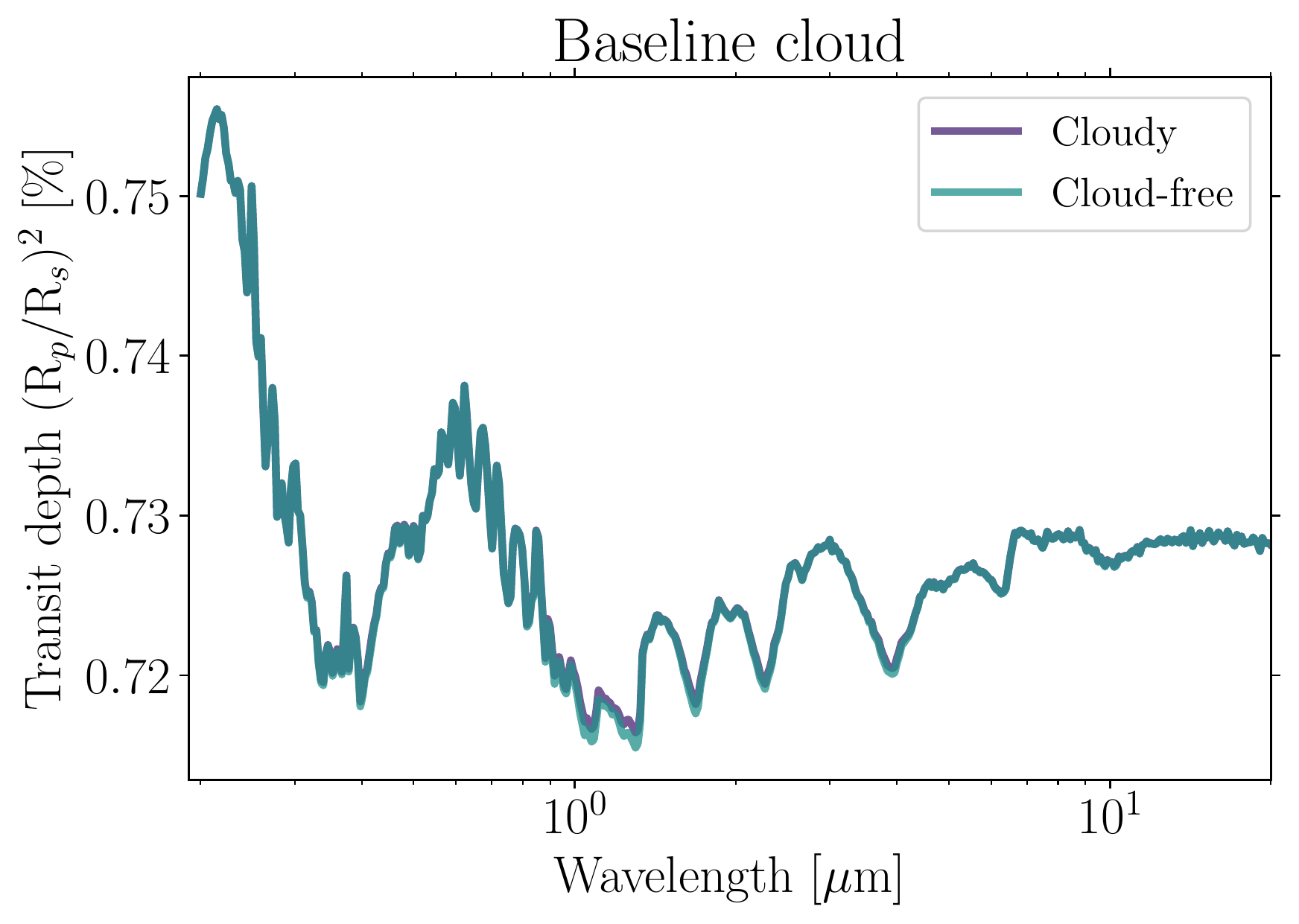}
    \includegraphics[width=0.45\textwidth]{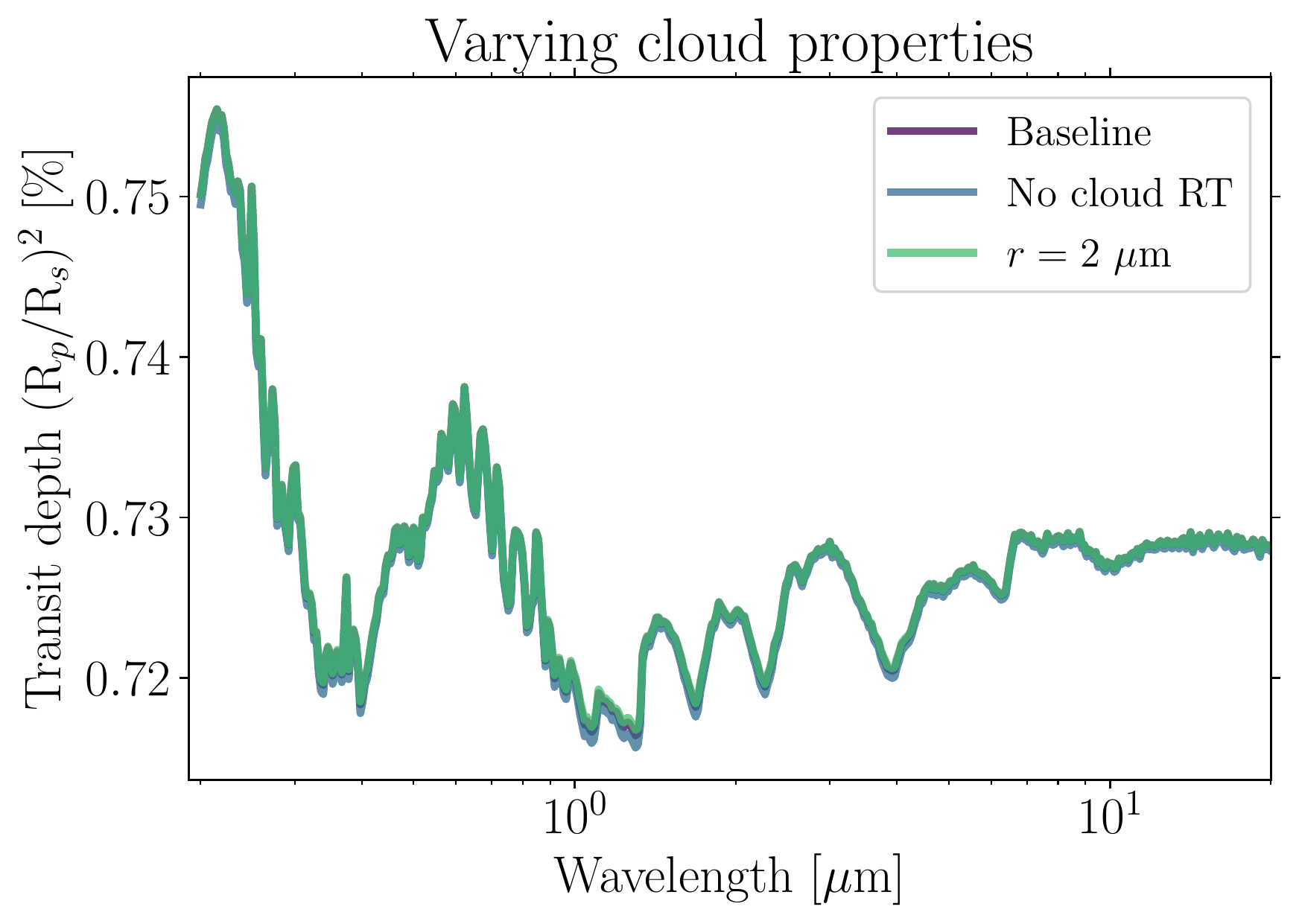}
    \caption{Simulated transmission spectra for varying cloud assumptions. The top panel shows simulated transmission spectra from our post-processed baseline GCM with active clouds, both including and ignoring the effect of clouds in the post-processing. The bottom panel shows simulated transmission spectra including the impact of clouds from our suite of models with varying cloud properties. We find that clouds do not have a significant impact on the continuum level of transmission spectra of ultra-hot Jupiters. The effect of clouds on transmission spectra is largest in cases with significant cloud-radiative feedback due to their higher altitude cloud decks.}
    \label{fig:transmission}
\end{figure}

We find that condensate clouds likely do not have a significant impact on the transmission spectra of ultra-hot Jupiters with equilibrium temperatures comparable to or greater than TOI-1431b ($T_\mathrm{eq} \gtrsim 2370~\mathrm{K}$). The top panel of \Fig{fig:transmission} shows predicted transmission spectra from the baseline case both including and removing the scattering and absorption by clouds. We find that the inclusion of clouds slightly raises the continuum level of absorption, especially blueward of the $1.4~\mu\mathrm{m}$ water absorption feature accessible with HST/WFC3. However, even the high temperature condensate corundum clouds considered here do not have a significant impact on the amplitudes of spectral features in transmission. The bottom panel of \Fig{fig:transmission} shows the impact of cloud microphysical and radiative properties on our simulated transmission spectra. The transit depth in the near and mid-infrared is somewhat larger in cases with a significant cloud radiative feedback than those with a weak cloud-radiative feedback. This is because the thermal impact of the cloud greenhouse shifts the cloud deck to lower pressures in cases with a strong cloud-radiative feedback, enhancing the scattering and absorption of transmitted stellar light at high altitudes.

In the cases with a significant cloud-radiative feedback, condensate cloud tracers in the GCM are lofted to low pressures of a few $\mathrm{mbar}$ on the western limb (see \Fig{fig:cloudp}). This cloud top pressure is similar to that expected from GCMs of WASP-121b that were post-processed to include clouds (\citealp{Parmentier:2018aa}, see their Figure 11), however here we include the dynamic transport of and radiative feedback due to clouds in the GCM itself. Our finding that the western limb is cloudy to low pressures and the eastern limb is cloud free further implies strong limb-to-limb contrasts in the cloud coverage of ultra-hot Jupiters, as expected from previous microphysical modeling \citep{Powell:2019aa,Helling:2021aa}. However, as found in the case of WASP-178b \citep{Lothringer:2022aa}, we predict that SiO vapor can persist in the gas phase and produce strong transmission spectral features in the NUV.


\section{Discussion}
\label{sec:disc}
\subsection{Effect of an enhanced visible opacity and hydrogen dissociation and recombination on thermal structure}
\label{sec:Hdisc}
\begin{figure}
    \centering
    \includegraphics[width=0.45\textwidth]{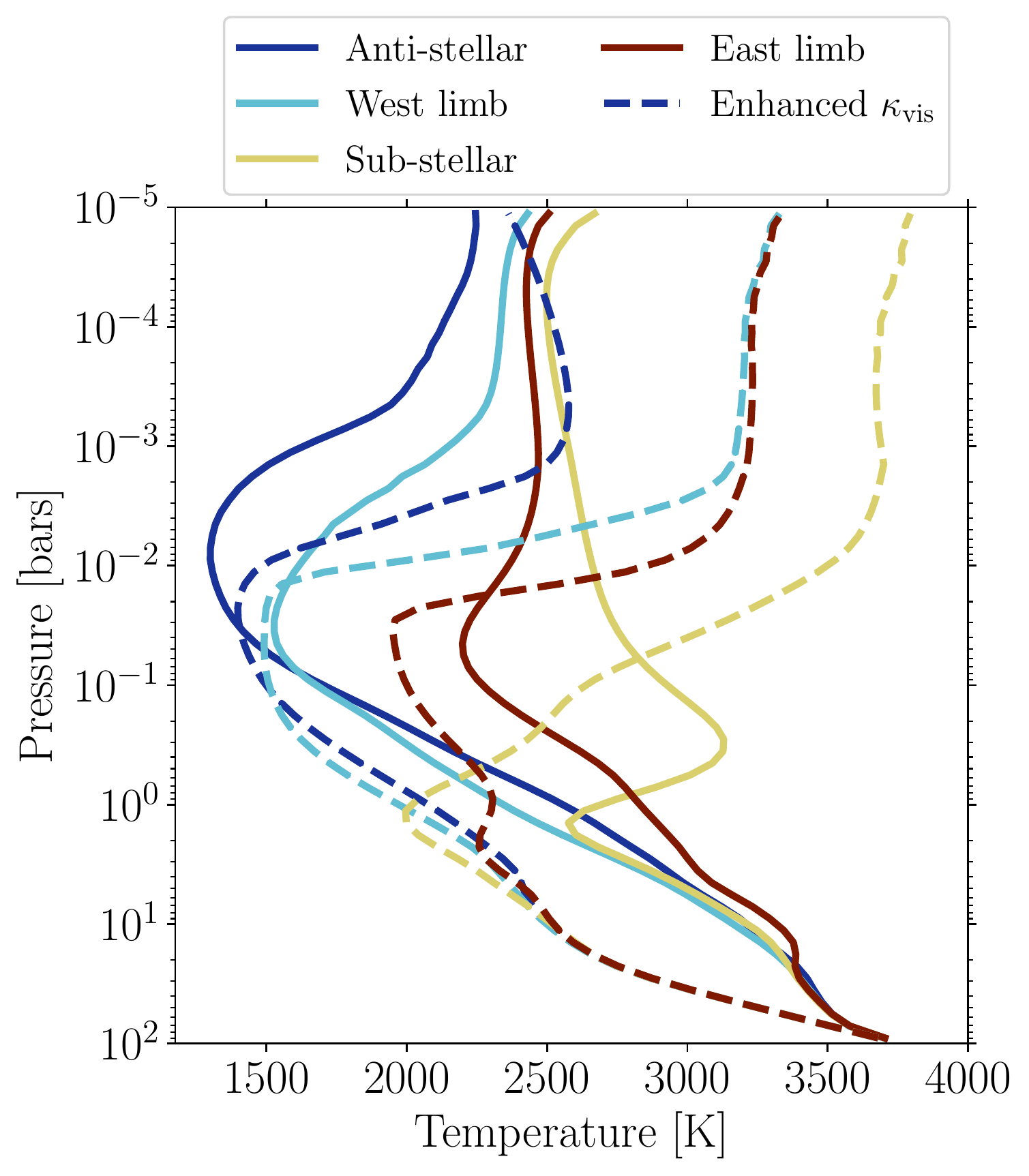}
    \caption{Impact of an enhanced visible opacity on the simulated thermal structure. Shown are meridional-mean temperature profiles from the baseline cloud case (solid lines) and a case with an enhanced visible opacity (dashed lines). A large ($\gtrsim 1500~\mathrm{K}$) thermal inversion on the dayside in the enhanced visible opacity case leads to a dynamically induced $\sim 1000~\mathrm{K}$ thermal inversion on the nightside that occurs at higher pressures than in the baseline case. This deeper and stronger nightside thermal inversion in the enhanced visible opacity run prevents significant cloud formation on the nightside.}
    \label{fig:ev_dissnodiss}
\end{figure}

We included a case with a significantly enhanced visible opacity in our suite of models in order to mimic the potential radiative impact of absorption of incident starlight by atomic metals in the atmospheres of ultra-hot Jupiters \citep{Lothringer:2018aa}. Another rationale for studying the impact of enhanced visible opacity is to determine the effect of a dayside inversion on the global dynamics and resulting three-dimensional temperature structure. In the double-gray or band-gray framework, an enhanced visible wavelength opacity warms the dayside upper atmosphere, causing deeper layers to cool to maintain global radiative equilibrium \citep{Guillot:2010,Parmentier:2014,Parmentier:2014a}. Additionally, with a thermal inversion the hotter upper layers on the dayside have a shorter radiative cooling timescale, which would naively imply reduced day-to-night heat transport at low pressures \citep{Perez-Becker:2013fv,Komacek:2015,Zhang:2016}. However, a thermal inversion also causes an increased abundance of hydrogen in atomic form at low pressures due to thermal dissociation (see Figure \ref{fig:tempwindp_enhancedvis}). As this atomic hydrogen is transported from dayside to nightside, it will recombine and warm the surrounding atmosphere \citep{Bell:2018aa,Tan:2019aa}, affecting the global temperature structure.

\Fig{fig:ev_dissnodiss} compares meridional mean temperature-pressure profiles at the anti-stellar point, west limb, sub-stellar point, and east limb between the baseline case and the case with an enhanced visible opacity. The case with an enhanced visible opacity has a strong dayside thermal inversion, while the baseline case is non-inverted on the dayside at pressures $\lesssim 300~\mathrm{mbar}$. However, \textit{both} cases have a significant thermal inversion on the nightside and western limb. This nightside inversion has a larger amplitude in the case with an enhanced visible opacity, but begins at a pressure of $\sim 10~\mathrm{mbar}$ in both cases. As a result, the deepest levels at which there is a nightside inversion coincide with pressures where the atmospheric flow is characterized by a strong eastward equatorial jet. Additionally, as shown below, the nightside inversion is significantly weakened with the removal of the thermodynamic impact of hydrogen dissociation and recombination. As a result, we hypothesize that on planets with strong ($\sim 1000~\mathrm{K}$) dayside thermal inversions, nightside thermal inversions can be sustained at low pressures (at much lower pressure levels than the cloud-free thermal photosphere, see \Tab{table:params}) due to the effective additional day-night heat transport from the conversion of atomic to molecular hydrogen. 


\begin{figure}
    \centering
    \includegraphics[width=0.45\textwidth]{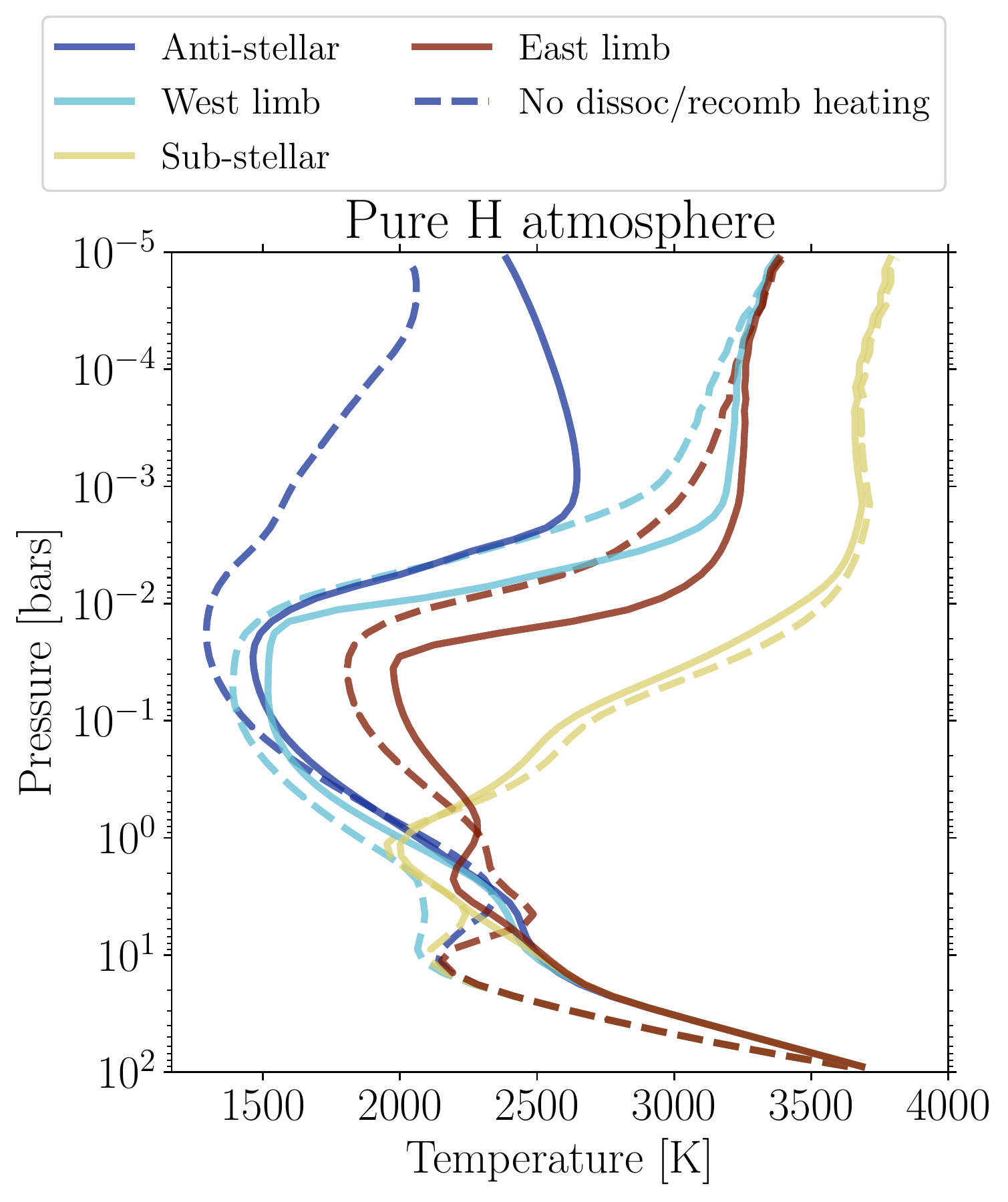}
    \caption{Impact of hydrogen dissociation and recombination on the thermal structure of ultra-hot Jupiters. 
    Shown are profiles of the meridional-mean temperature from cases with an enhanced visible opacity and a pure hydrogen atmosphere with and without the heating/cooling from hydrogen recombination/dissociation, resulting in a maximal effect of hydrogen recombination on the thermal structure. 
    The removal of the thermal effect of hydrogen dissociation and recombination reduces the strength of the nightside thermal inversion.}
    \label{fig:ev_dissnodiss_honlycomp}
\end{figure}

To isolate the effect of hydrogen dissociation and recombination on the atmospheric circulation and temperature structure, we conducted additional GCMs with the thermodynamic heating and cooling term from hydrogen dissociation and combination in \Eq{eq:thermo} set to zero. These models retain the same atomic hydrogen tracer formalism as the other GCMs conducted, and as a result retain the effect of hydrogen dissociation and recombination on the mean molecular weight, specific gas constant, and heat capacity. We conducted two additional simulations:
a case maximizing the impact of hydrogen dissociation and recombination by considering an atmosphere comprised entirely of hydrogen (i.e., no helium, with $x_{He/H} = 0$), with the same visible band and infrared band opacity choices as in our ``enhanced visible opacity'' simulation, and without radiatively active cloud tracers; and an equivalent case to that just described but without the thermal effect of hydrogen dissociation and recombination. 


\Fig{fig:ev_dissnodiss_honlycomp} compares the temperature structure between the case with a maximal impact of hydrogen dissociation and recombination on the thermal structure 
and an equivalent case without the thermal effect of hydrogen dissociation and recombination.
Due to the high visible opacity in this case, the dayside has a strong and deep thermal inversion similar to that in the equivalent case in our full model suite with enhanced visible opacity. 
The dayside thermal inversion is slightly stronger in the simulation without the thermal effect of hydrogen dissociation leading to cooling on the dayside. The heating due to hydrogen recombination causes the temperature on the limb at pressures of $100~\mu\mathrm{bar} \lesssim p \lesssim 1~\mathrm{bar}$ to be higher in the case with the thermal effect of hydrogen dissociation and recombination, although both cases have thermal inversions on each limb. Further, the temperature increase due to the nightside inversion is much larger in the case with the thermal effect of hydrogen dissociation and recombination\footnote{Note that there is still a slight thermal inversion at low pressures at the anti-stellar longitude in the case without the thermal effect of hydrogen dissociation. This occurs because the day-to-night winds increase in speed with decreasing pressure, while the radiative timescale is relatively independent of pressure in the upper regions of double-gray models. Given that the substellar longitude is nearly isothermal at pressures $\lesssim 1~\mathrm{mbar}$, this leads to a relative increase in substellar to anti-stellar heat transport with decreasing pressure in the upper regions of the model.}, leading to nightside temperatures that can be over $1,000~\mathrm{K}$ warmer than in the case without heating due to hydrogen recombination. 

The strong change in the three-dimensional thermal structure at low pressures due to the thermodynamic effects of hydrogen dissociation and recombination may have implications for phase-resolved observations of ultra-hot Jupiters. Notably, such a nightside inversion has not been found in recent low-resolution phase curves of the ultra-hot Jupiters WASP-76b and WASP-121b, which both show evidence for a dayside thermal inversion but a non-inverted nightside temperature profile \citep{May:2021ab,Mikal-Evans:2022wq}. However, because the photosphere of tidally locked gas giants is horizontally non-uniform \citep{Dobbs-Dixon:2017aa}, these low-resolution observations probe higher pressure levels on the nightside than on the dayside. The HST phase curve of WASP-121b probes pressures of $\sim 10 - 100~\mathrm{mbar}$ on the nightside, deeper than the nightside temperature inversion that occurs in our GCMs due to the thermal impact of hydrogen recombination that lies at pressures of $1-10~\mathrm{mbar}$. However, as discussed in \Sec{sec:limit}, we caution that our GCM is double-gray and may not adequately model radiative cooling from recombined molecules on the nightside. Further observational constraints on the presence of dayside and nightside inversions from both low-resolution phase curves and high-resolution phase-dependent emission spectra would help determine how the interplay between atmospheric dynamics and molecular dissociation and recombination affect the thermal structure of ultra-hot Jupiters.




\subsection{The impact of patchy clouds on emergent properties of ultra-hot Jupiters}
Our GCMs with radiatively active cloud tracers support the findings of previous cloud microphysical and GCM models that predict a transition in cloud coverage between hot and ultra-hot Jupiters, with hot Jupiters having uniformly cloudy nightsides and ultra-hot Jupiters having non-uniform cloud distributions 
\citep{Gao:2021vp,Helling:2021aa,Parmentier:2021tt,Roman:2021wl}. 
We find that the patchy cloud coverage on ultra-hot Jupiters causes their nightside cloud greenhouse to have a different effect on observable properties than cooler hot Jupiters. For hot Jupiters, the uniform nightside cloud deck constrains the source of outgoing longwave radiation to the region near the cloud top, reducing the outgoing flux compared to a cloud-free nightside \citep{Gao:2021vp}. In contrast, the patchy nightside clouds of ultra-hot Jupiters can locally enhance the thermal flux that escapes to space. This occurs because, as with a uniform cloud deck, the cloud greenhouse effect warms the deep atmosphere. However, unlike a uniform cloud deck, the patchiness of nightside clouds on ultra-hot Jupiters enables thermal emission from the warmer levels at depth to escape to space in regions with reduced cloud opacity. As a result, we find a change in the sign of the cloud greenhouse effect on the outgoing thermal flux from the nightside hemisphere of ultra-hot Jupiters as compared to hot Jupiters. This suggests that the radiative feedback due to the patchy cloud deck plays a role in the break of the flat nightside temperature trend between the hot and ultra-hot Jupiters inferred from the sample of Spitzer phase curves \citep{Beatty19,Keating:2019aa,Bell:2021aa}. 

Additionally, the patchy cloud greenhouse effect in our GCM simulations implies that there will be a trade-off between the patchy cloud coverage and cloud greenhouse effect that controls how the nightside temperature depends on equilibrium temperature. This is because in isolation, cloud dissipation at hotter equilibrium temperatures (i.e., higher incident stellar flux) leads to an increase in nightside flux as the hot deep levels can more easily radiate to space. However, at the same time the reduction in cloud coverage leads to slightly cooler temperatures at depth due to the reduced cloud greenhouse effect. The competition between patchy cloud coverage and the cloud greenhouse effect itself depends on the three-dimensional cloud distribution, necessitating a suite of three-dimensional models of ultra-hot Jupiters with radiatively active cloud tracers covering a broad range of equilibrium temperature in order to determine the resulting dependence of outgoing longwave radiation on planetary irradiation. 

We find that in our simulations with radiatively active cloud tracers, high-temperature condensate clouds can be mixed to low ($\sim 5~\mathrm{mbar}$) pressures, but only have a slight effect on transmission spectra. This agrees with the expectation from both GCMs with post-processed clouds \citep{Parmentier:2018aa} and cloud microphysics models \citep{Helling:2021aa} that condensate clouds can persist at $\sim\mathrm{mbar}$ pressures, however we do not find that clouds have a significant impact on transmission spectra of ultra-hot Jupiters with equilibrium temperatures comparable to or greater than TOI-1431b. Additionally, we find that strong limb-to-limb asymmetries in cloud coverage due to the large temperature contrast between the eastern and western limb \citep{Powell:2019aa} extend into the ultra-hot Jupiter regime, as potentially implied by recent time-resolved high-resolution transmission spectra of WASP-76b \citep{Ehrenreich:2020aa,Kesseli:2021ab}. 

We expect that the transition point between cloudy and cloud-free limbs as probed in transmission will occur within the ultra-hot Jupiter regime. From an emerging sample of NUV spectra of ultra-hot Jupiters, there is tentative evidence that such a transition occurs between equilibrium temperatures of $1950 - 2450~\mathrm{K}$ \citep{Lothringer:2022aa}. For WASP-121b ($T_\mathrm{eq} \approx 2350~\mathrm{K}$), \cite{Mikal-Evans:2022wq} found sufficiently cold nightside temperatures from HST/WFC3 phase curve observations for cloud condensation, while previous transmission spectra of WASP-121b point toward a cloud-free limb -- implying that atmospheric motions keep condensate clouds aloft in order to prevent rain-out to deeper atmospheric levels.  Our GCM simulations with an equilibrium temperature of $2368~\mathrm{K}$ show nightside cloud coverage that extends only slightly past the western limb, which lies at a similar planetary irradiation to the tentative observational transition in limb cloud coverage. 
However, a variety of planetary properties along with equilibrium temperature, including gravity and atmospheric metallicity, conspire to affect cloud microphysics and horizontal and vertical transport. This motivates observational surveys of both NUV and high-resolution transmission spectra as well as thermal phase curves to probe the transition in global climate and condensate cloud coverage between the hot and ultra-hot Jupiters. In addition, time-resolved high-resolution transmission and emission spectroscopy will be a powerful tool to test the expected patchy cloud configuration for individual ultra-hot Jupiters.



\subsection{Limitations}
\label{sec:limit}
Our current modeling framework is simplified in order to lend physical understanding of how a coupling between atmospheric dynamics and cloud-radiative feedback sets the cloud coverage in ultra-hot Jupiter atmospheres. Due to its idealized nature, our model setup has a variety of limitations that must be addressed before detailed comparison with observations of ultra-hot Jupiters. Most notably, as in many recent GCM studies of exoplanet atmospheric dynamics and its impact on observations (e.g., \citealp{Dietrick:2020aa,May:2020vr,Mendonca:2020aa,Roman:2021wl,Beltz:2021aa,Harada:2021tc,May:2021ab,Beltz:2022aa}), our model utilizes a double-gray radiative transfer scheme. Though double-gray schemes provide a simplified yet realistic framework within a GCM to gain physical insight into the processes that regulate atmospheric dynamics and heat transport, \cite{Lee:2021vo} recently demonstrated that GCMs with double-gray radiative transfer provide an inadequate representation of the thermal structure compared to those with band-gray or fully non-gray radiative transfer schemes. As a result, we caution that the quantitative predictions for thermal structure and cloud coverage may differ between our idealized GCMs and models with more realistic radiative transfer schemes. 

Importantly, non-gray models are required to fully understand the provenance of nightside inversions in ultra-hot Jupiters. Notably, the upper layers of the nightside will be cold enough for at least partial recombination of molecules (e.g., H$_2$O). Low pressures on the nightside will then undergo enhanced infrared cooling to space from molecular lines that may offset the thermodynamic warming impact of hydrogen recombination. Results from our idealized models, including patchy nightside clouds and a dynamically induced nightside thermal inversion, motivate future work with non-gray radiative transfer to fully determine how the thermal impact of hydrogen dissociation and recombination affects the temperature structure of ultra-hot Jupiters at low pressures. 

In this work, we have isolated the cloud-radiative feedback from a single high-temperature condensate, corundum, in order to study how the localized nature of vertical mixing and cloud-radiative feedback sets the patchy cloud coverage on the nightsides of ultra-hot Jupiters. Though our simulated ultra-hot Jupiter has an equilibrium temperature above the condensation temperature of even high-temperature condensates at pressures $\lesssim 1~\mathrm{bar}$, the simulated temperatures on the nightside in our GCM suite can be cold enough that a large variety of cloud species could locally condense, including silicates, sulfates, iron, and high-temperature condensates (e.g., perovskite). Additionally, microphysical models of cloud formation across the hot to ultra-hot Jupiter transition predict that a range of cloud condensate species can locally form on the nightside and western limb of ultra-hot Jupiters \citep{Helling:2021aa,Gao:2021vp}. Notably, the high optical thickness of silicate clouds (especially forsterite) may regulate the cooling from a patchy nightside cloud deck. Future work  including multiple types of condensate cloud tracers is required to ascertain how the formation, dynamical mixing, and radiative feedback of the range of possible cloud condensates affects the atmospheric dynamics and observable properties of ultra-hot Jupiters. 

Though the cloud particle size distribution and cloud-radiative parameters (e.g., asymmetry parameter, single scattering albedo) used for the GCM simulations conducted in this work are informed by \texttt{CARMA} cloud microphysics simulations \citep{Gao:2020aa,Gao:2021vp}, they neglect the dynamic three-dimensional interaction between the atmospheric dynamics and cloud microphysics. In reality, the cloud particle size distribution will evolve significantly due to the competition between cloud nucleation, vertical mixing, and settling, which will in turn affect the cloud optical thickness and the radiative scattering and absorption due to the cloud deck. The radiative feedback of clouds will then affect the dynamics, resulting in a coupled interplay between cloud microphysics, atmospheric circulation, and radiative transfer. Though we have included the interplay between the latter two in this work, we expect that a self-consistent inclusion of cloud microphysical processes would impact the resulting patchy cloud distribution. Previous GCMs with self-consistently coupled cloud microphysics (e.g., \citealp{Lee:2016,Lines:2018}) have found, albeit at great computational expense, that this coupling has a significant impact on the atmospheric dynamics and emergent properties of hot Jupiters.

Lastly, we only considered a single planetary parameter regime in this work in order to isolate the impact of cloud microphysical and radiative assumptions on the atmospheric circulation and cloud coverage of ultra-hot Jupiters. Future work is required to understand the coupled impact of cloud mixing and radiative feedback on the cloud coverage of hot gas giants across the transition from hot Jupiters to ultra-hot Jupiters. This broad parameter sweep would be necessary to test the prediction from one-dimensional models that the dissipation of the forsterite cloud deck leads to the increased nightside brightness temperatures of ultra-hot Jupiters \citep{Gao:2021vp}, as well as to further study how the coupling between mixing and radiative feedback of clouds impacts the phase curves of hot and ultra-hot Jupiters \citep{Parmentier:2021tt,Roman:2021wl}. 





\section{Conclusions}
\label{sec:conc}
Ultra-hot Jupiters have been found to be a unique class of exoplanet because of observational evidence for molecular dissociation, atomic ionization, and heavy metal vapor on their daysides. In this work, we have demonstrated using three-dimensional general circulation models with radiatively active high-temperature condensate cloud tracers that ultra-hot Jupiters are also unique because their cloud coverage is inherently patchy, as cloud condensation and vertical transport of condensed aerosol is only possible on the nightside and western limb of the planet. This is in broad agreement with previous theoretical work that has predicted a transition in cloud coverage between hot and ultra-hot Jupiters \citep{Powell:2019aa,Gao:2020aa,Helling:2021aa,Parmentier:2021tt,Roman:2021wl}. This also agrees with observational work that has found evidence for the clearing of a uniform nightside cloud deck at the transition between hot and ultra-hot Jupiters by studying the dependence of nightside outgoing longwave radiation with incident stellar irradiation \citep{Beatty19,Keating:2019aa}. Below we outline our key conclusions from this work, along with future avenues for further exploration of cloud-radiative feedback on the dynamics of ultra-hot substellar atmospheres. 
\begin{enumerate}
    \item The majority of ultra-hot Jupiters detected to date represent a transitional state in cloud coverage between hot Jupiters and cloud-free substellar objects (e.g., KELT-9b). While hot Jupiters are expected to have a nearly uniform blanket of nightside cloud coverage that acts to absorb and re-emit thermal radiation at nearly uniform cool cloud-top temperatures, there are only certain regions where cloud condensation is favorable on the nightsides of ultra-hot Jupiters -- even for the highest-temperature condensates. This causes the nightside and limb cloud distributions of ultra-hot Jupiters to be patchy and dependent on the nature of the atmospheric circulation and temperature structure. 
    \item We find that the patchy spatial distribution of clouds in ultra-hot Jupiter atmospheres does not directly track the local temperature conditions in the atmosphere. Notably, we find a lack of clouds in the coldest mid-latitude regions of the planet in every simulation in our suite of GCMs with varying cloud microphysical and radiative properties. This is because net vertical mixing of aerosols in a stably stratified atmosphere requires a positive correlation between tracer abundance and upward vertical winds \citep{parmentier_2013,Zhang:2018tp,Zhang:2018te}. In these mid-latitude regions, downwelling due to convergence at the flanks of the superrotating equatorial jet leads to net downward transport of cloud condensate, further enhancing the already patchy nature of the cloud deck on a given isobar. 
    \item We find in all of our GCMs with radiatively active or inactive cloud tracers that the patchy nightside cloud decks of ultra-hot Jupiters are sequestered at depth. The cloud deck does not reach pressures significantly below $1~\mathrm{mbar}$ at any location for any model in our suite of GCMs. This deep sequestration of condensate clouds is due to a dynamically induced thermal inversion on the nightside that is caused by the local heat deposition by hydrogen recombination, as a significant fraction of the atomic hydrogen transported from the dayside at low pressures recombines into molecular hydrogen. 
    \item The combination of spatially patchy cloud coverage and sequestration of clouds at depth in our simulations causes high-temperature condensate clouds to have a 
    greenhouse effect on the atmosphere below.
    Post-processing of our GCMs with the state-of-the-art Monte Carlo radiative transfer code \texttt{gCMCRT} shows that this patchy cloud greenhouse can have a potentially significant effect on the nightside emission spectra of ultra-hot Jupiters. Notably, this cloud greenhouse effect causes an increase in the emergent flux from the nightside hemisphere of ultra-hot Jupiters, in contrast to the expected decrease in outgoing longwave radiation due to the cloud greenhouse inferred for cooler hot Jupiters \citep{Gao:2021vp}. This is because the patchy nature of the cloud deck enables radiation to escape to space from relatively clear regions warmed by the cloud greenhouse. Additionally, we predict that the deep patchy cloud condensate coverage on the limbs of ultra-hot Jupiters with equilibrium temperatures of $\gtrsim 2370~\mathrm{K}$ does not greatly impact transmission spectra from the NUV to mid-infrared.
    \item In this work, we have only modeled the coupled dynamical transport of and radiative feedback by a single cloud condensate, corundum. We did so in order to cleanly investigate the transport of high-temperature condensates in ultra-hot Jupiter atmospheres that persist up to the irradiation level where ultra-hot Jupiter atmospheres become cloud-free. Future work is required to study the coupled impact of corundum along with other potentially abundant cloud species, especially silicates, along with other high-temperature cloud condensates (e.g., perovskite). Further work could investigate how the combined radiative and dynamical coupling of a combination of cloud condensates in GCMs impacts predictions for the transition in nightside and limb cloud coverage from hot to ultra-hot Jupiters along with the impact of these dynamically and radiatively active clouds on observable properties of the range of tidally locked gas giant atmospheres. 
    
\end{enumerate}

\acknowledgments
A significant portion of this work was performed on land that is the traditional homeland of the Piscataway and Nacotchtank peoples. We thank the anonymous referee for an insightful report that improved this work. X.T. is supported by the  European Research Council  advanced grant EXOCONDENSE (PI: R.T. Pierrehumbert). This work was completed with resources provided by the University of Chicago Research Computing Center (PI: Jacob Bean). The authors acknowledge the University of Maryland supercomputing resources (\url{http://hpcc.umd.edu}) made available for conducting the research reported in this paper.

\appendix

\section{Effect of cloud-radiative assumptions on atmospheric dynamics and cloud distribution}
\subsection{Temperature, winds, and atomic hydrogen mass mixing ratio}
\label{sec:appendixtemp}

\begin{figure*}
    \centering
    \includegraphics[height=1\textheight]{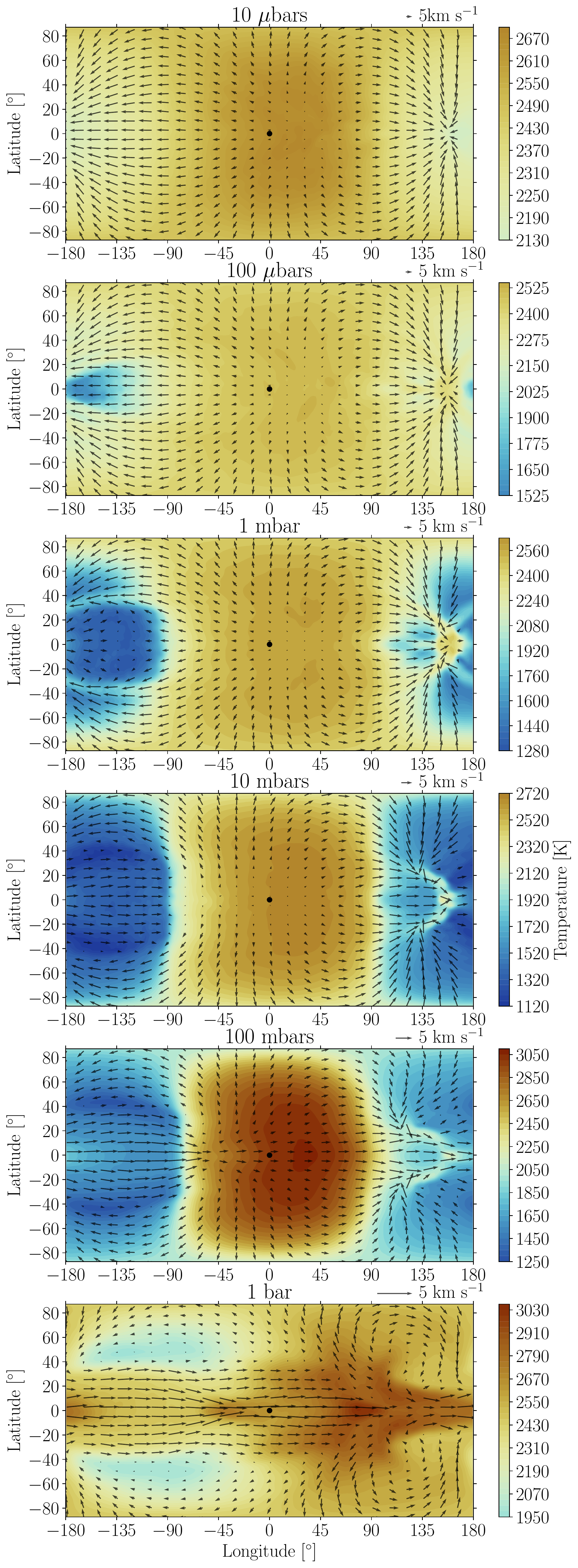}
    \includegraphics[height=1\textheight]{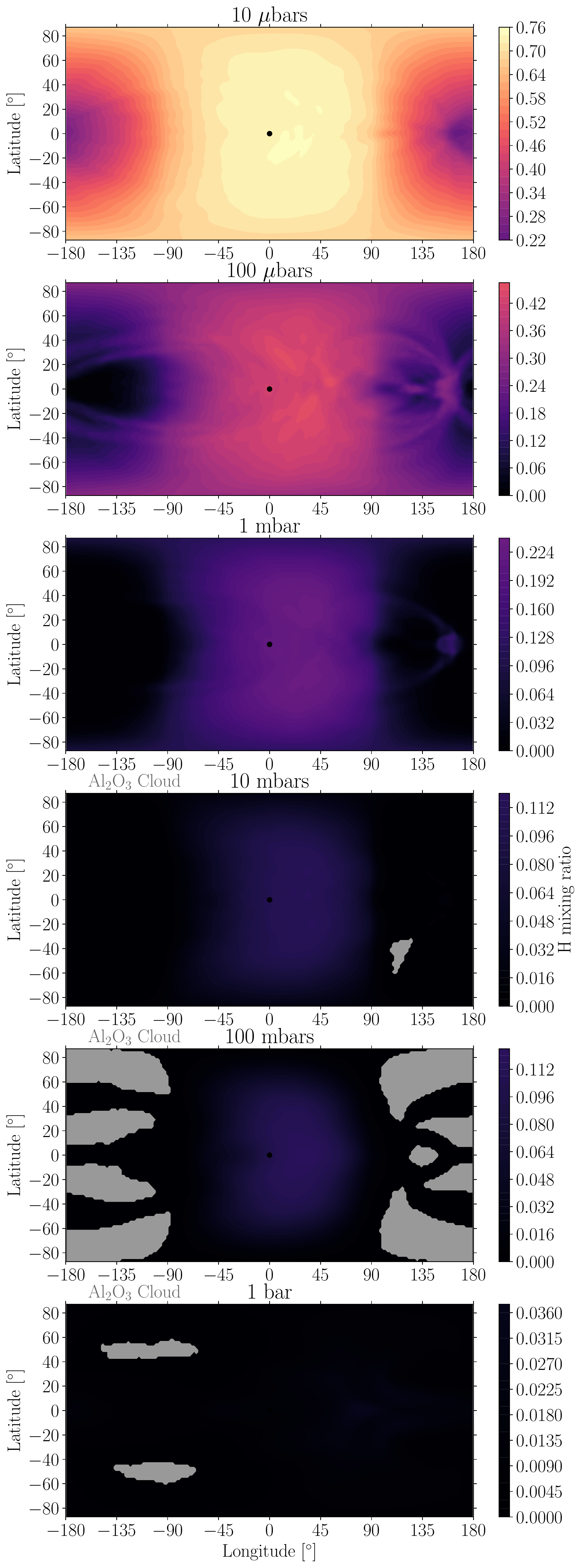}
    \caption{Temperature maps with overlaid wind arrows (left) and atomic hydrogen mass mixing ratio (right, colors) with overlaid cloud tracer distributions (gray regions show where the cloud mass mixing ratio is $\ge 5 \times 10^{-5}~\mathrm{kg}~\mathrm{kg}^{-1}$) plotted on isobars logarithmically spaced from 10 $\mu\mathrm{bars}$ to $1~\mathrm{bar}$ from the simulation with radiatively inactive clouds.}
    \label{fig:tempwindp_nocldrt}
\end{figure*}

\begin{figure*}
    \centering
    \includegraphics[height=1\textheight]{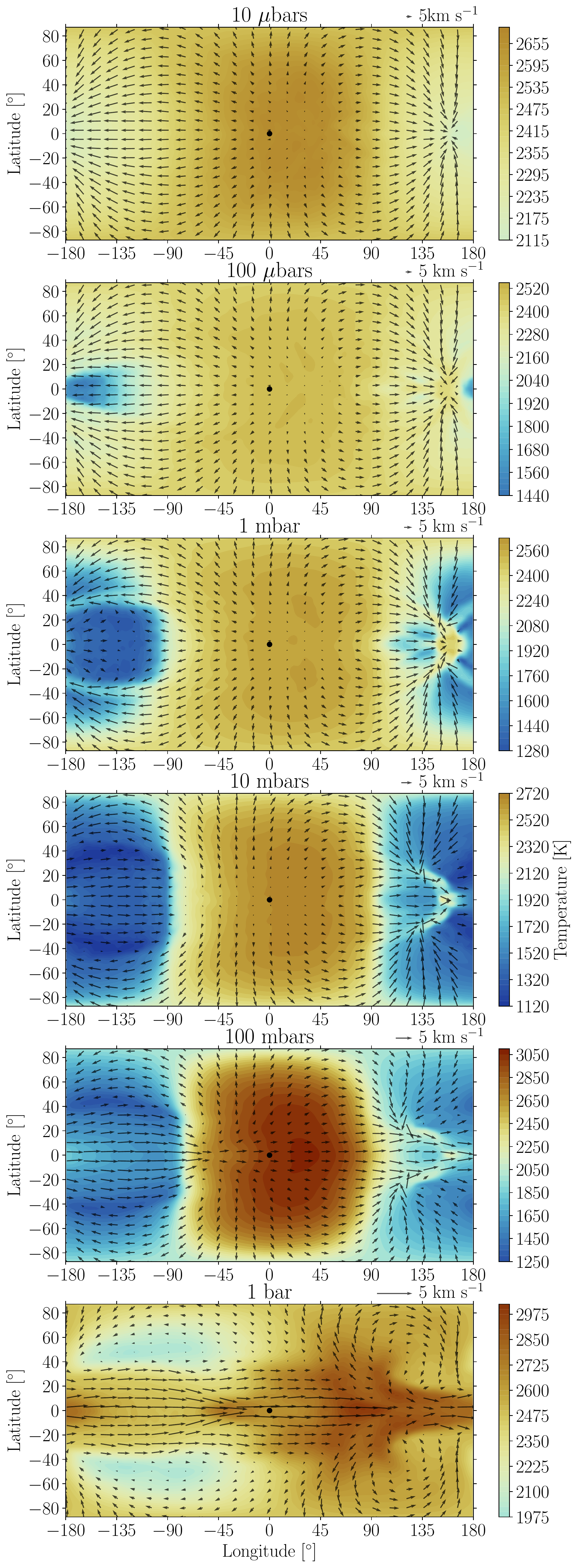}
    \includegraphics[height=1\textheight]{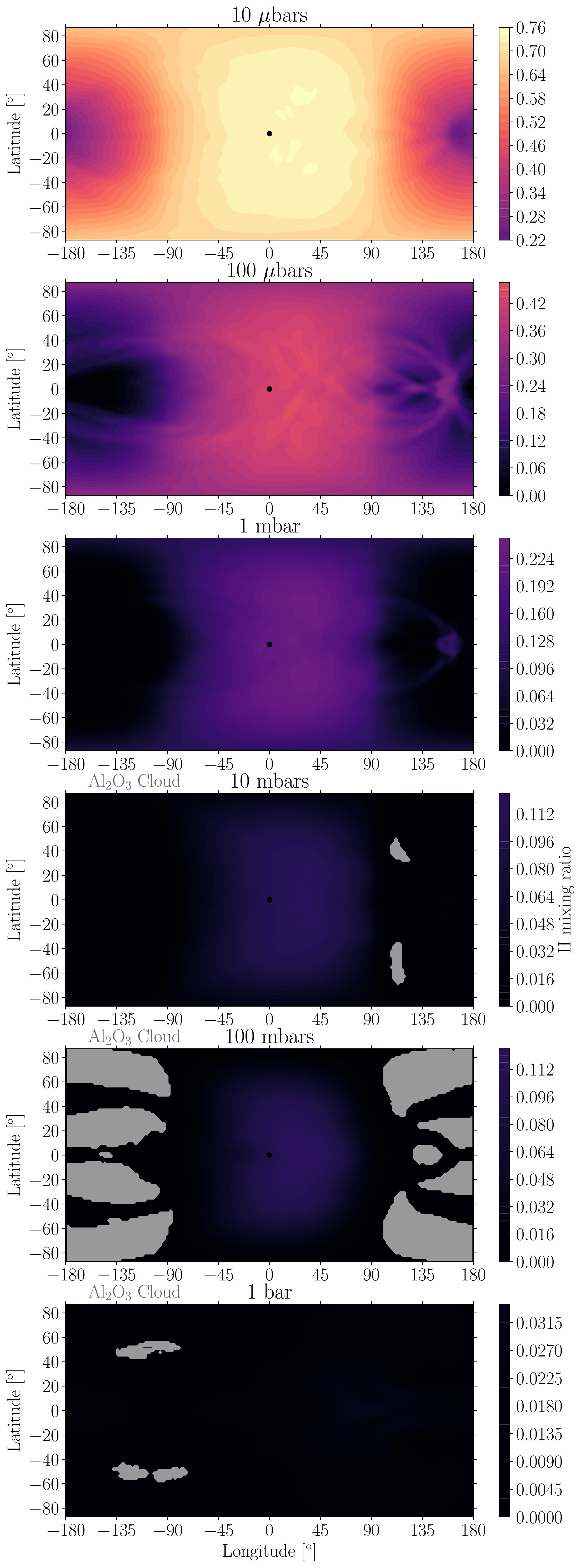}
    \caption{Temperature maps with overlaid wind arrows (left) and atomic hydrogen mass mixing ratio (right, colors) with overlaid cloud tracer distributions (gray regions show where the cloud mass mixing ratio is $\ge 5 \times 10^{-5}~\mathrm{kg}~\mathrm{kg}^{-1}$) plotted on isobars logarithmically spaced from 10 $\mu\mathrm{bars}$ to $1~\mathrm{bar}$ from the simulation with reduced cloud opacity.}
    \label{fig:tempwindp_redkappacld}
\end{figure*}

\begin{figure*}
    \centering
    \includegraphics[height=1\textheight]{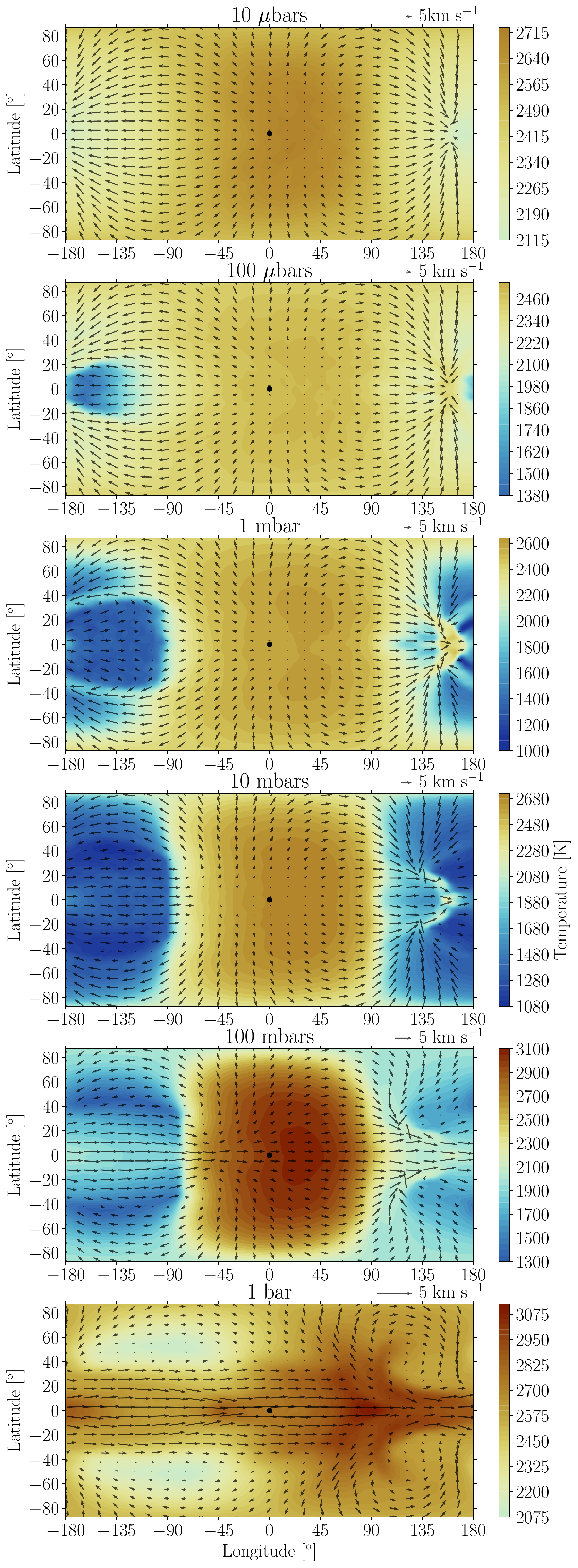}
    \includegraphics[height=1\textheight]{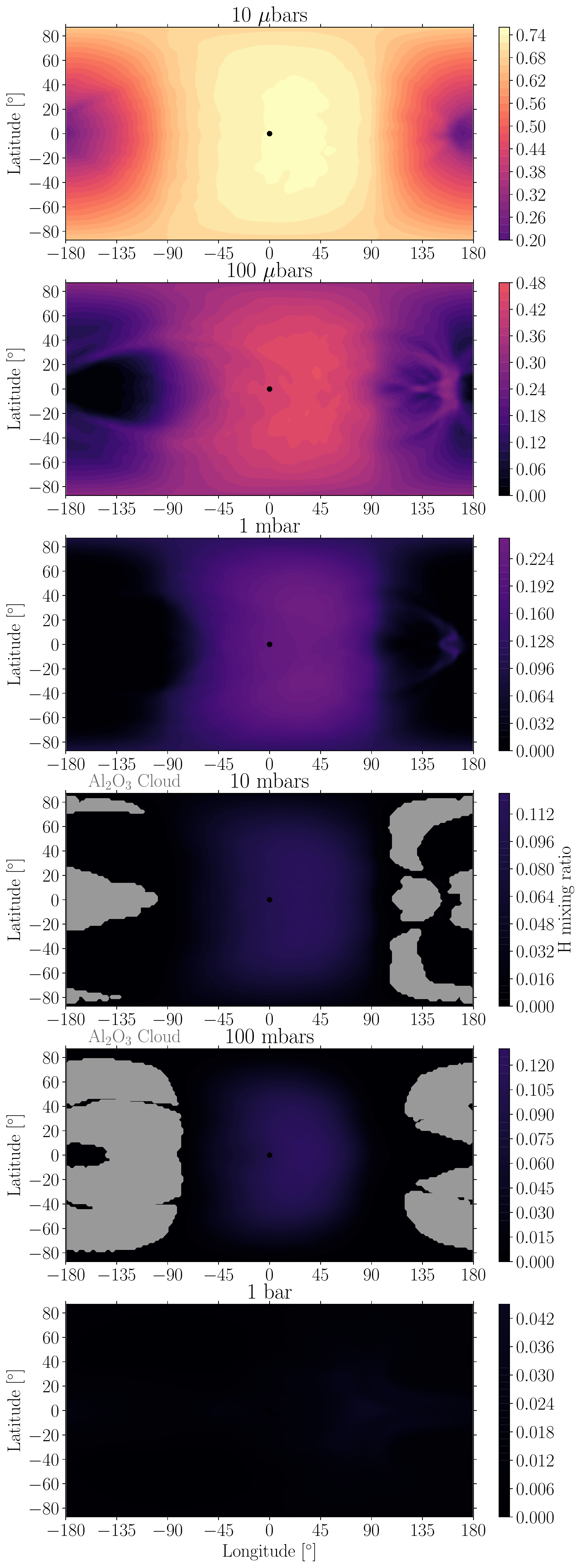}
    \caption{Temperature maps with overlaid wind arrows (left) and atomic hydrogen mass mixing ratio (right, colors) with overlaid cloud tracer distributions (gray regions show where the cloud mass mixing ratio is $\ge 5 \times 10^{-5}~\mathrm{kg}~\mathrm{kg}^{-1}$) plotted on isobars logarithmically spaced from 10 $\mu\mathrm{bars}$ to $1~\mathrm{bar}$ from the simulation with a mean cloud particle size of $r = 2~\mu\mathrm{m}$.}
    \label{fig:tempwindp_redr}
\end{figure*}

\begin{figure*}
    \centering
    \includegraphics[height=1\textheight]{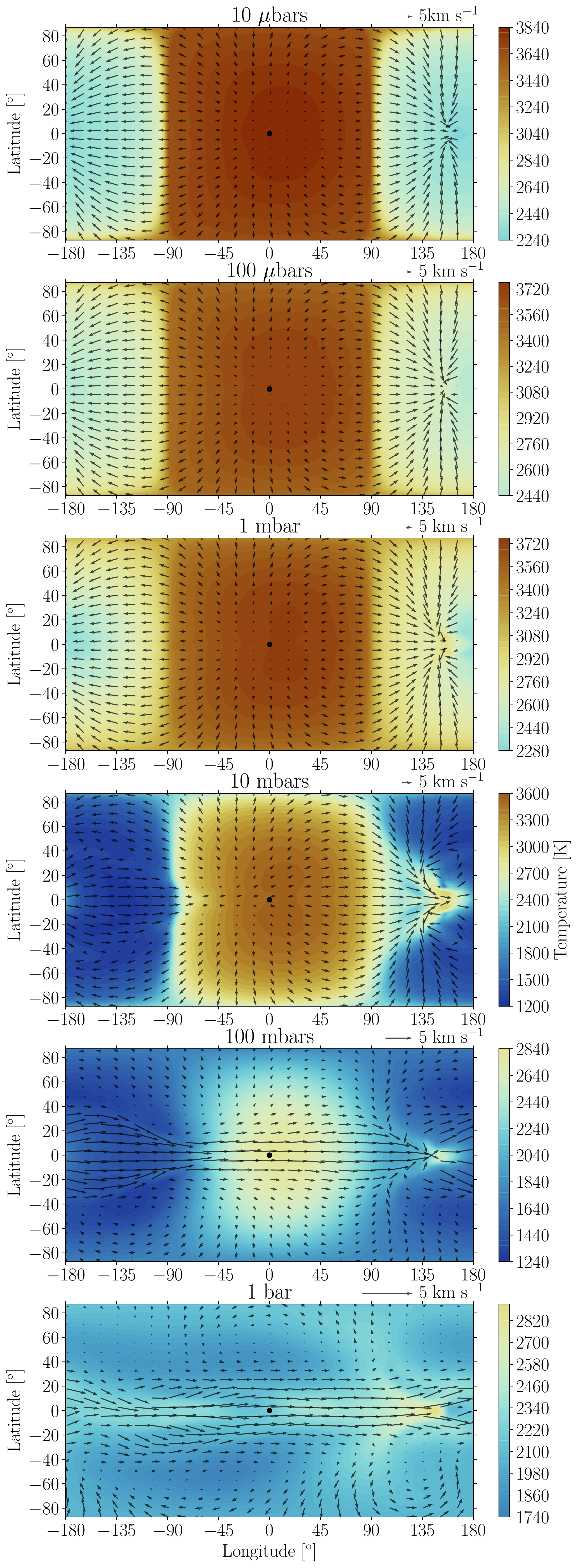}
    \includegraphics[height=1\textheight]{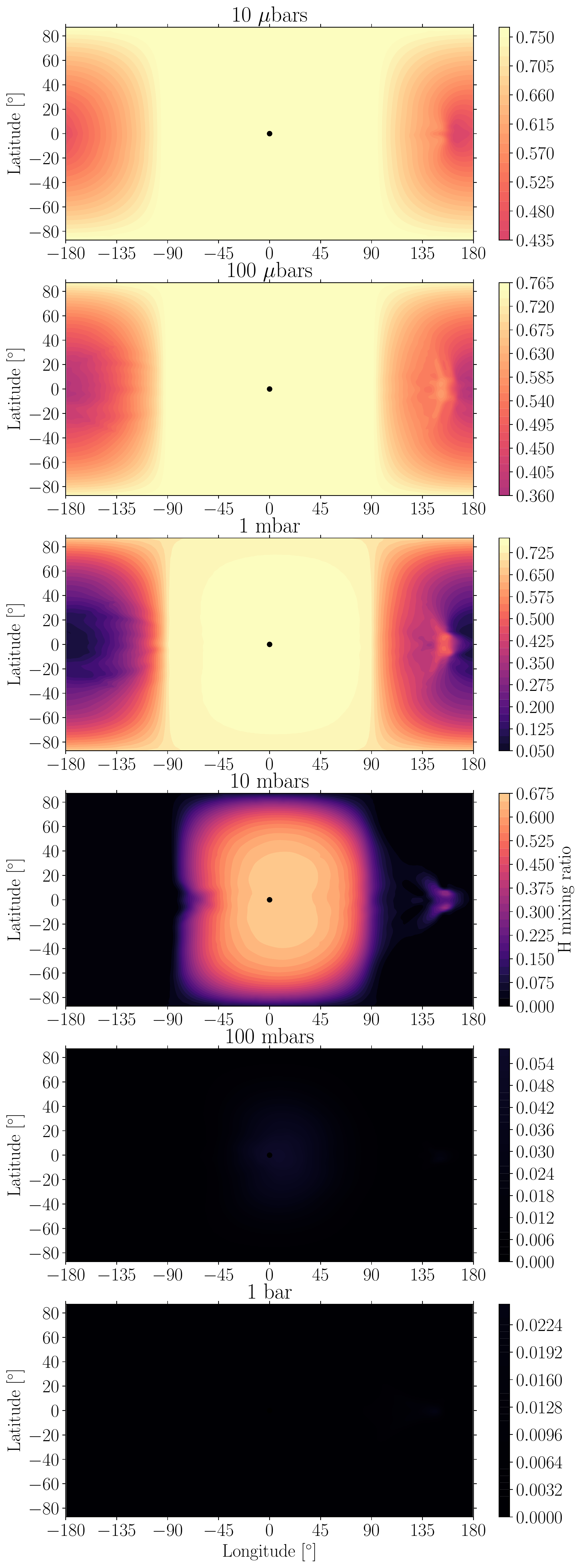}
    \caption{Temperature maps with overlaid wind arrows (left) and atomic hydrogen mass mixing ratio (right, colors) plotted on isobars logarithmically spaced from 10 $\mu\mathrm{bars}$ to $1~\mathrm{bar}$ from the simulation with an enhanced visible band opacity. Clouds with a local mass mixing ratio $\ge 5 \times 10^{-5}~\mathrm{kg}~\mathrm{kg}^{-1}$ do not form at any location in this simulation.}
    \label{fig:tempwindp_enhancedvis}
\end{figure*}

Equivalent maps to those for the baseline case in \Fig{fig:tempwindp} displaying the temperature and wind patterns as well as the hydrogen mass mixing ratio and regions with significant cloud mass mixing ratio on isobars are shown for the remaining cases from the full suite of GCMs in Figures \ref{fig:tempwindp_nocldrt}-\ref{fig:tempwindp_enhancedvis}. The largest differences between the baseline case and the cases with weak cloud-radiative feedback (i.e., the case no cloud radiative feedback shown in \Fig{fig:tempwindp_nocldrt} and the case with a weakened cloud-radiative feedback in \Fig{fig:tempwindp_redkappacld}) is the cooler deep atmosphere in the cases with a weak or non-existent cloud-radiative feedback due to the lack of a cloud greenhouse effect. This causes the cloud deck to move to higher pressures, allowing cloud condensate to persist on the mid-latitude nightside to pressures of a bar. There are only minor differences in the temperature and cloud mixing ratio pattern between the baseline case and the case with a reduced cloud particle size shown in \Fig{fig:tempwindp_redr}. This is because the cloud particle size affects just the vertical settling of clouds in our numerical scheme, while the cloud-radiative feedback and temperature structure has a larger control on the resulting cloud distribution. Regardless, the deep atmosphere is slightly warmer in the case with a reduced particle size than in the baseline case due to the thicker cloud deck caused by the reduced settling velocity with smaller characteristic cloud particle sizes.

\Fig{fig:tempwindp_enhancedvis} shows the temperature and wind as well as the  atomic hydrogen mass mixing ratio on isobars from the case with an enhanced visible opacity. The enhanced absorption of incoming stellar radiation leads to significant differences in temperature and winds relative to the other four cases in our model suite. The model with an enhanced visible opacity has a global thermal inversion at low pressures (see also \Sec{sec:Hdisc}), with a larger-amplitude thermal inversion on the dayside than nightside leading to significant day-to-night temperature contrasts at all pressures. This thermal inversion causes almost the entire dayside to have its hydrogen in atomic state at pressures $\lesssim 1~\mathrm{mbar}$. The thermal inversion also prevents significant cloud formation, as no clouds form with a mixing ratio $\ge 10^{-6}~\mathrm{kg}~\mathrm{kg}^{-1}$ in the case with an enhanced visible opacity. 

\subsection{Cloud and condensible vapor mass mixing ratio}
\label{sec:cldapp}
\begin{figure*}
    \centering
    \includegraphics[width=1\textwidth]{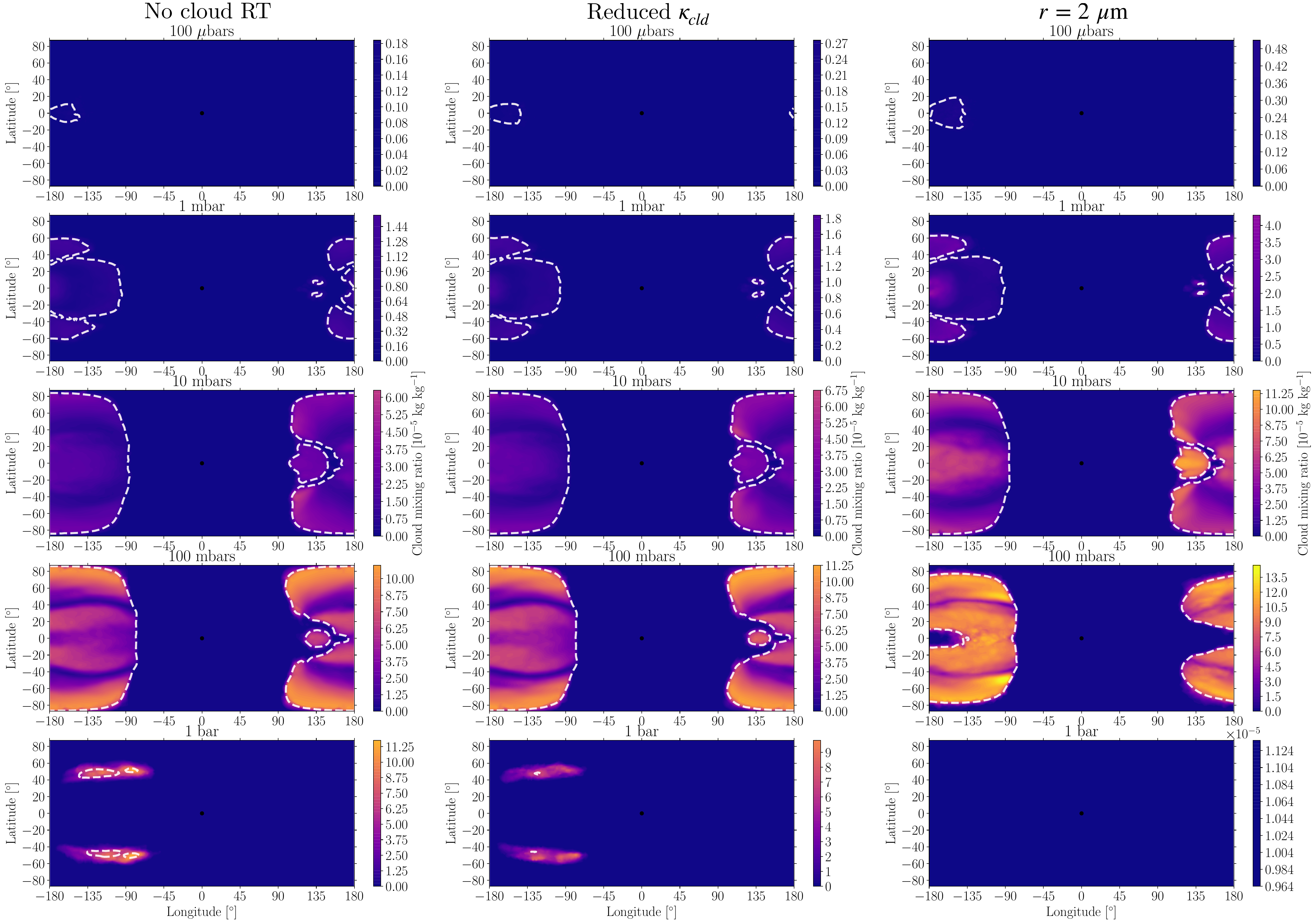}
    \caption{Maps of the cloud mass mixing ratio on isobars logarithmically spaced from $100~\mu\mathrm{bars}$ to 1 bar from three cases with varying cloud and radiative transfer assumptions: no cloud radiative feedback (No cloud RT), reduced cloud opacity (Reduced $\kappa_\mathrm{cld}$), and reduced mean cloud particle size ($r = 2~\mu\mathrm{m}$). All panels share a color scale, and the cloud mixing ratio is shown in units of $10^{-5}~\mathrm{kg}~\mathrm{kg}^{-1}$. The snow white dashed contour displays where the gas temperature is equal to the corundum condensation temperature on each isobar.}
    \label{fig:cldmassapp}
\end{figure*}

\begin{figure*}
    \centering
    \includegraphics[width=1\textwidth]{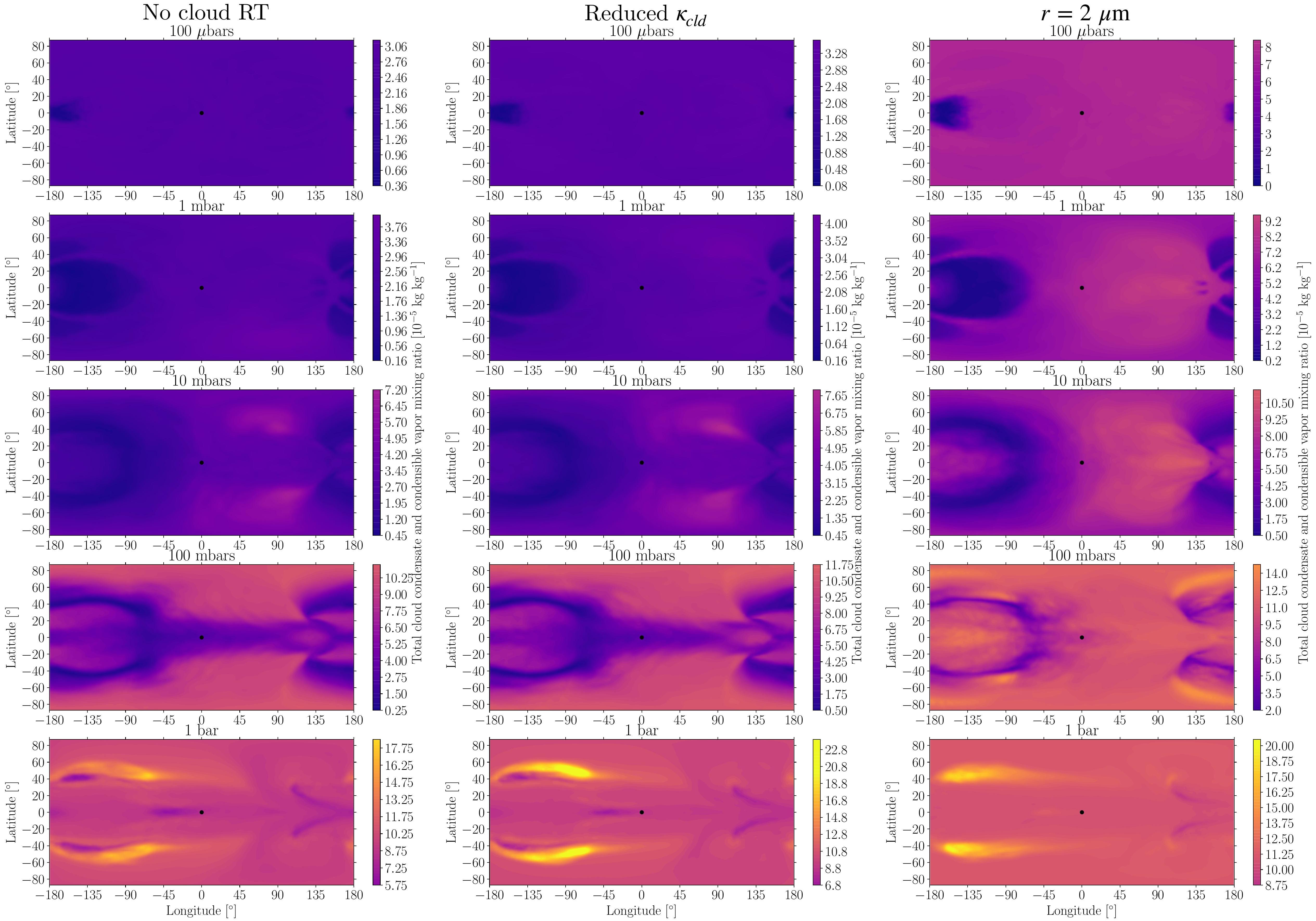}
    \caption{Maps of the total cloud condensate and condensible vapor mass mixing ratio ($q_c + q_v$) on isobars logarithmically spaced from $100~\mu\mathrm{bars}$ to 1 bar from three cases with varying cloud and radiative transfer assumptions: no cloud radiative feedback (No cloud RT), reduced cloud opacity (Reduced $\kappa_\mathrm{cld}$), and reduced mean cloud particle size ($r = 2~\mu\mathrm{m}$). All panels share a color scale, and the total cloud condensate and condensible vapor mixing ratio is shown in units of $10^{-5}~\mathrm{kg}~\mathrm{kg}^{-1}$.}
    \label{fig:totcldvapmassapp}
\end{figure*}

\Fig{fig:cldmassapp} shows maps of the cloud mass mixing ratio ($q_c$) on isobars, while \Fig{fig:totcldvapmassapp} shows maps of the total cloud condensate and condensible vapor mass mixing ratio ($q_c + q_v$) on isboars. Both figures are equivalent to those from the baseline case shown in \Fig{fig:cloudbaseline} but are from the cases with varying cloud microphysical and radiative properties. The no cloud radiative feedback and reduced cloud opacity cases have the largest differences in both cloud coverage and combined cloud plus vapor tracer distributions from the baseline case. This is due to their lack of a strong cloud-radiative feedback, which enables cloud formation at depth and shifts the cloud deck towards higher pressures. The models without strong cloud-radiative feedback also have a significant equatorial depletion at depth in total tracer, similar to that found in previous hot Jupiter GCMs \citep{parmentier_2013,Lines:2018,Komacek:2019aa}, as shown in Figure 13 of \citealp{Showman:2020rev}. The case with a reduced cloud particle size has a similar three-dimensional cloud distribution to the baseline case due to their comparable levels of cloud-radiative feedback and thus similar temperature structure and wind patterns, but with a slight increase in the maximum cloud mass mixing ratio due to the reduced settling flux of smaller cloud particles. 

\subsection{Zonal-mean zonal wind}
\label{sec:appendixwind}
\begin{figure*}
    \centering
    \includegraphics[width=1\textwidth]{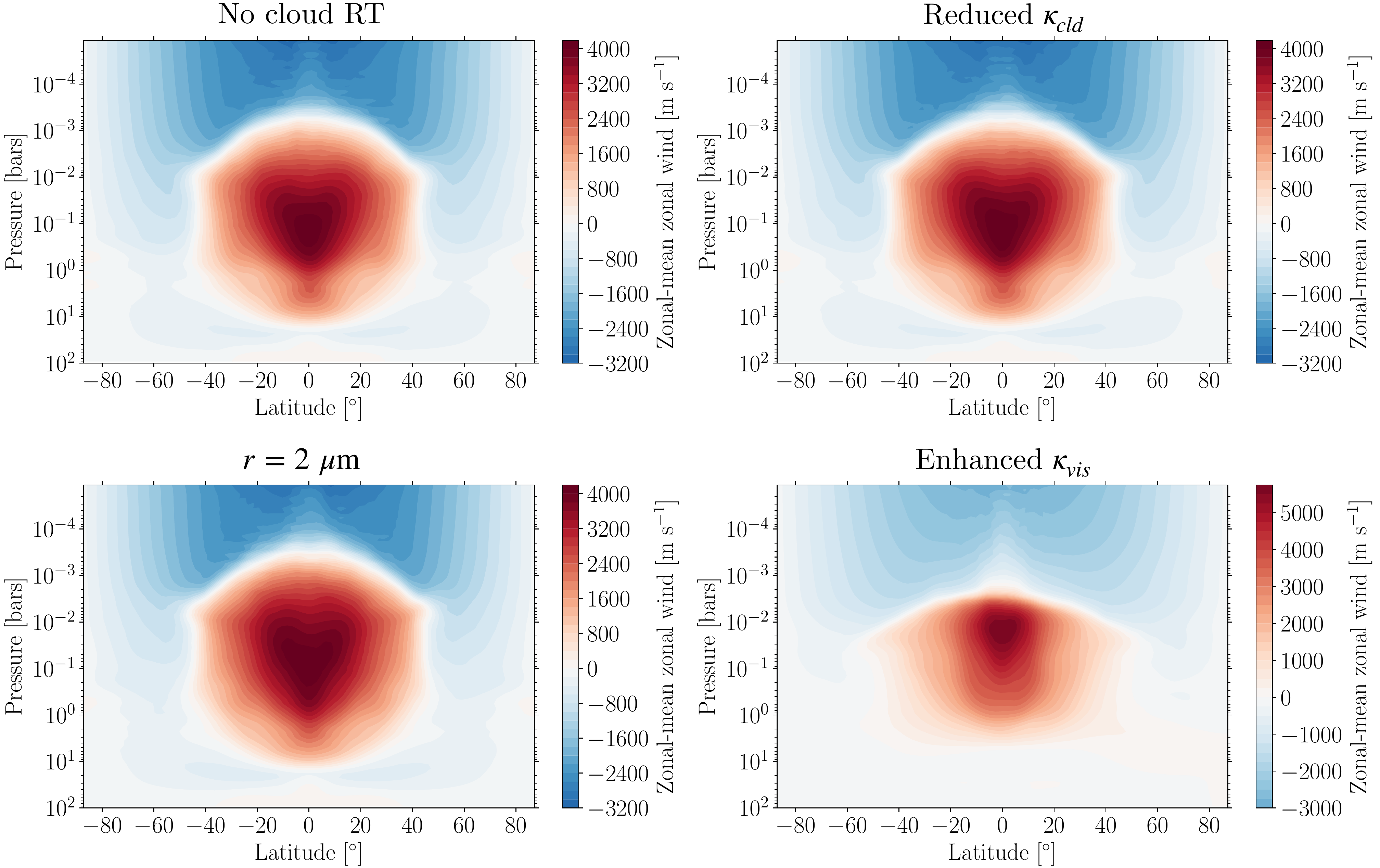}
    \caption{Zonal-mean zonal wind from four cases with varying cloud and radiative transfer assumptions: no cloud radiative feedback (No cloud RT), reduced cloud opacity (Reduced $\kappa_\mathrm{cld}$), reduced mean cloud particle size ($r = 2~\mu\mathrm{m}$), and enhanced visible band opacity (Enhanced $\kappa_\mathrm{vis}$).}
    \label{fig:uzonalapp}
\end{figure*}

\Fig{fig:uzonalapp} shows latitude-pressure profiles of the zonal-mean zonal wind equivalent to those from the baseline case shown in \Fig{fig:uzonal} from the remainder of the GCM suite. The peak zonal-mean zonal wind speeds for all cases with varying cloud parameters are similarly just above $4~\mathrm{km}~\mathrm{s}^{-1}$. The no cloud radiative transfer and reduced cloud opacity cases without a strong cloud-radiative feedback have a zonal wind maximum that is slightly more confined to high pressures than the baseline and reduced cloud particle size cases with significant cloud-radiative feedback. The largest difference in zonal-mean zonal wind structure is between the enhanced visible opacity case and all other cases, as this case has a faster peak wind speed of the equatorial jet associated with the hotter dayside and resulting larger day-to-night contrast. The eastward jet is also further confined to depth in the case with an enhanced visible opacity, showing that the transition from flow characterized by an eastward equatorial jet to predominantly day-to-night flow occurs at a higher pressure in the case with a strong thermal inversion.

\subsection{Vertical mixing}
\label{sec:appendixmix}
\begin{figure*}
    \centering
    \includegraphics[width=0.328\textwidth]{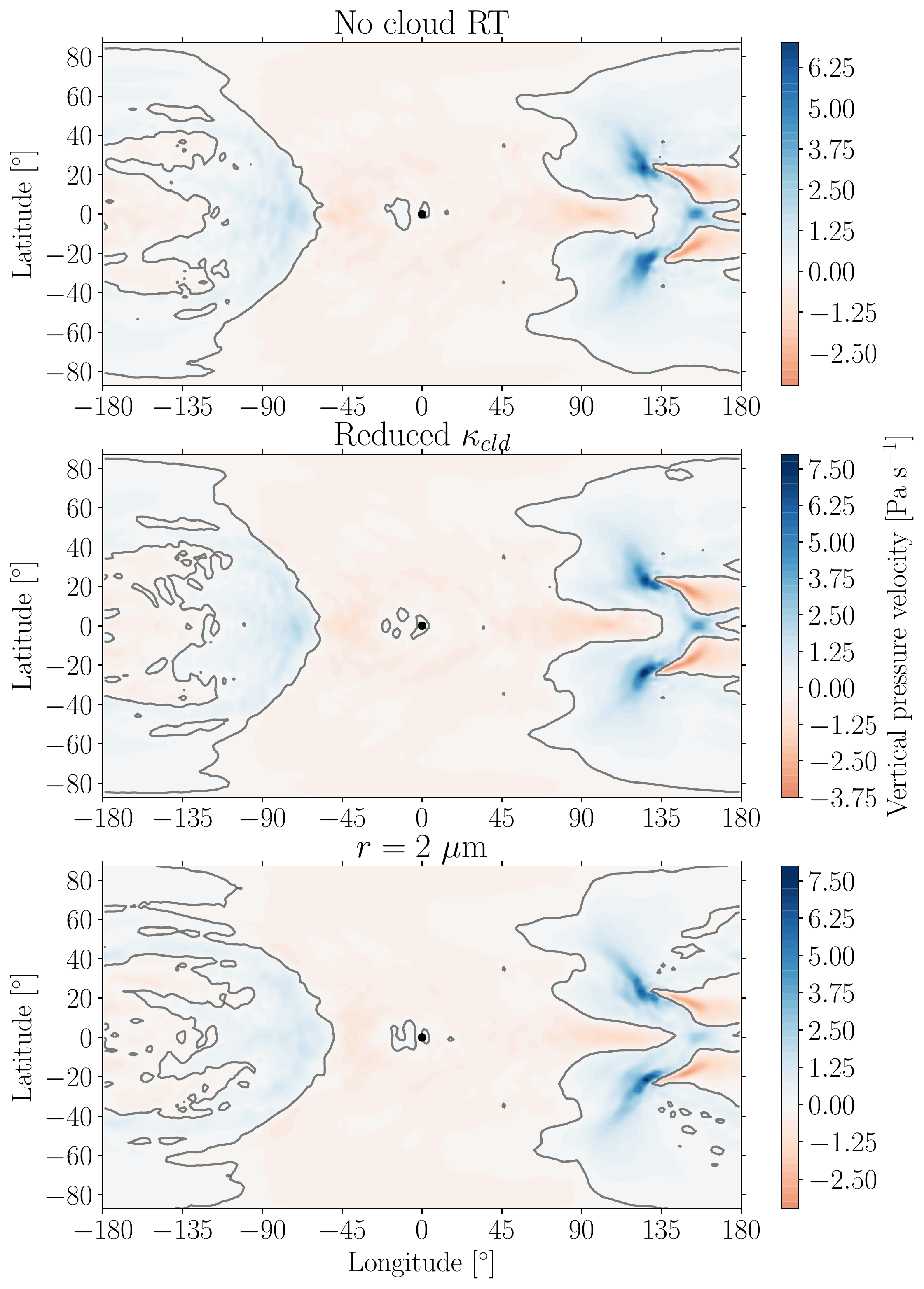}
    \includegraphics[width=0.32\textwidth]{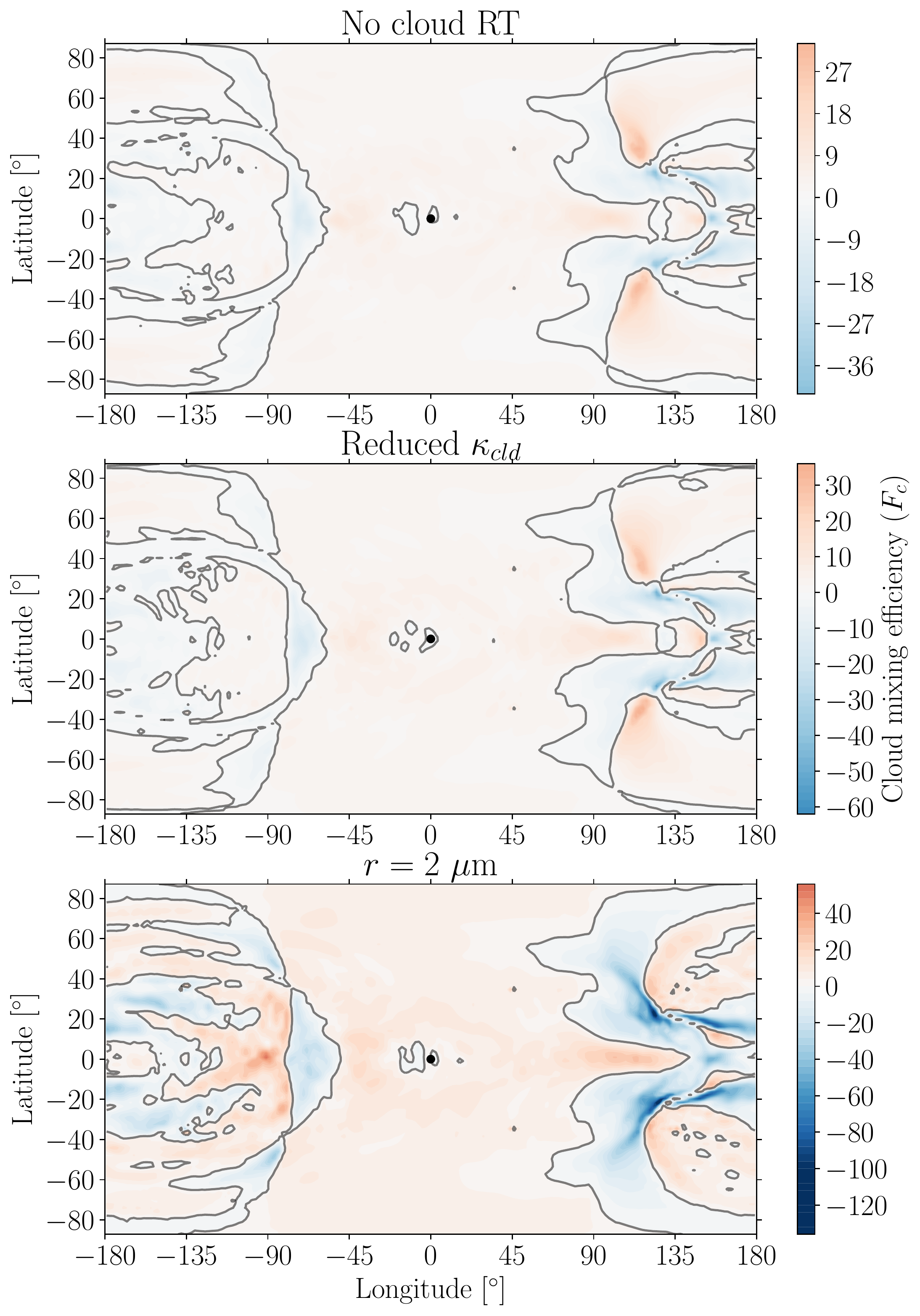}
    \includegraphics[width=0.32\textwidth]{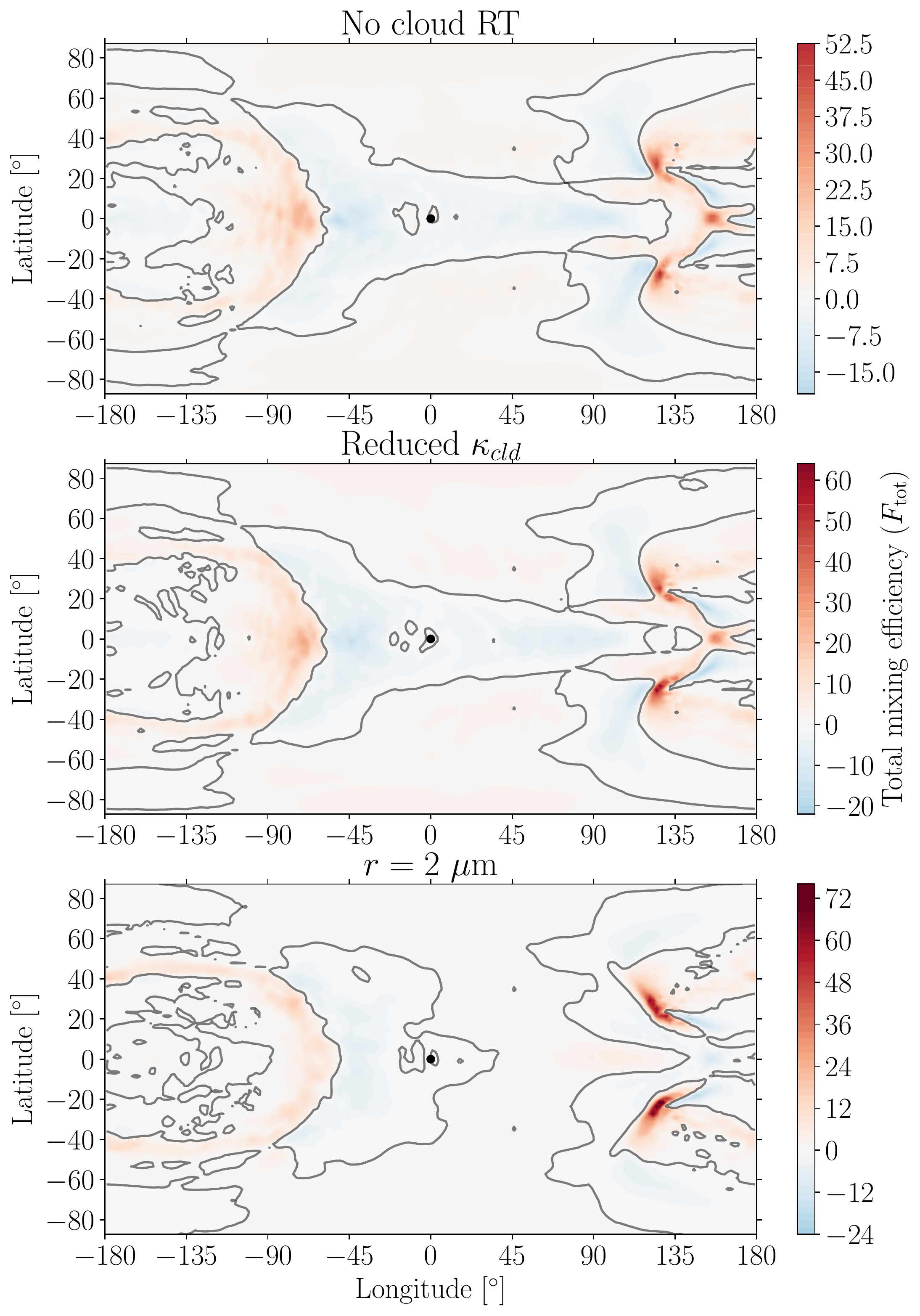}
    \caption{Vertical pressure velocity (left), cloud condensate mixing efficiency (middle), and total (cloud condensate plus condensible vapor tracer) mixing efficiency (right) at a pressure of 100 mbar for three cases with varying cloud and radiative transfer assumptions: no cloud radiative feedback (No cloud RT), reduced cloud opacity (Reduced $\kappa_\mathrm{cld}$), and reduced mean cloud particle size ($r = 2~\mu\mathrm{m}$). The gray contour denotes the zero velocity or mixing efficiency line. All panels for vertical velocity have separate color scales, while each column of mixing efficiency panels share a color scale. For both velocity and mixing efficiency, red corresponds to upward motion or mixing while blue corresponds to downward motion or mixing.}
    \label{fig:mixeffapp}
\end{figure*}

\Fig{fig:mixeffapp} shows maps of the vertical pressure velocity and both cloud condensate tracer and total (cloud condensate and condensible vapor) tracer mixing efficiency at the 100 mbar level from the cases with varying cloud parameters, equivalent to those for the baseline case shown in the bottom row of \Fig{fig:mixeffisobars}. The pattern of vertical velocity at 100 mbars is qualitatively similar for all cases, characterized by upwelling throughout much of the dayside and local upwelling and downwelling on the nightside associated with regions of horizontal divergence and convergence, respectively. However, the spatial pattern of both the cloud and total mixing efficiencies is different between the cases with a strong cloud-radiative feedback (baseline and reduced cloud particle size) and those with a weak or non-existent cloud radiative feedback (reduced cloud opacity and no cloud radiative feedback). This is due itself to the spatial differences in cloud cover, and as a result condensible vapor mixing ratio, between the cases with strong and weak cloud-radiative feedback. In the cases with stronger cloud-radiative feedback, the greater peak cloud mass mixing ratio at 100 mbars leads to larger peak absolute values of both cloud and total mixing efficiency across the 100 mbar isobar. The cases with weaker cloud radiative feedback have a relative equatorial depletion of total tracer (see \Fig{fig:totcldvapmassapp}), which in turn causes negative total mixing efficiency throughout much of the equatorial dayside. Additionally, the change in cloud pattern between cases with strong and weak cloud-radiative feedback causes the regions of positive and negative cloud mixing efficiency to differ between them, as vertical motions in the same direction can either drive a positive or negative cloud mixing efficiency depending on the local value of cloud mass mixing ratio relative to the horizontal mean on an isobar. As a result, the coupled nature of vertical winds and tracer abundance sets the amount of vertical mixing across isobars in our model suite.  

\bibliography{references,References_all,references_paste}





\end{document}